\documentclass[a4paper,12pt]{article}

\usepackage[T1]{fontenc}
\usepackage[authoryear]{natbib}
\usepackage{amsmath}
\usepackage{amsfonts}
\usepackage{amssymb}
\usepackage{graphicx}
\usepackage{latexsym}
\usepackage{array}
\usepackage{booktabs}
\usepackage{float}
\usepackage{multirow}
\usepackage{subfig}
\usepackage[
            top=25mm,
            bottom=25mm,
            left=30mm,
            right=20mm,
                        marginratio=5:6,
            includeheadfoot]{geometry}

\title{ \textbf{Bayesian Modeling and MCMC Computation in Linear Logistic Regression for Presence-only Data}}

\author
{Fabio Divino$^1$,Natalia Golini$^2$,Giovanna Jona Lasinio$^2$\and Antti Penttinen$^3$\\
{\small 1-Division of Physics, Computer Science and Mathematics,
University of Molise }\\
{\small 2-Department of Statistical Sciences,
University of Rome \textit{"La Sapienza"}}\\
{\small 3-Department of Mathematics and Statistics, 
University of Jyv\"{a}skyl\"{a}
}}
\begin{document}

\maketitle
\begin{abstract}
Presence-only data are referred to situations in which, given a censoring mechanism, a binary response can be observed only with respect to on outcome, usually called \textit{presence}. In this work we present a Bayesian approach to the problem of presence-only data based on a two levels scheme. A probability law and a case-control design are combined to handle the double source of uncertainty:  one due to the censoring and one due to the sampling. We propose a new formalization for the logistic model with presence-only data that allows  further insight into inferential issues related to the model. We concentrate on the case of the linear logistic regression and, in order to make inference on the parameters of interest, we present a Markov Chain Monte Carlo algorithm with data augmentation that does not require the a priori knowledge of the population prevalence. A simulation study concerning 24,000 simulated datasets related to different scenarios is presented comparing our proposal to  optimal benchmarks.
\end{abstract}
\section{Introduction}
There is a significant body of literature in statistics, econometrics and ecology dealing with the modeling of discrete responses under biased or preferential sampling designs. They are particularly popular in the natural sciences when species distributions are studied. Such sample schemes may reduce the survey cost especially when one of the responses is rare. A large part of statistical literature concerns the case-control design, retrospective, choice-based or response-based sampling \citep{lanc:imb:1996}. In the simplest case a sample of cases and a sample of controls are available and for each observation a set of ``attributes/covariates'' is observed in both samples. Then inference is carried out following standard statistical procedures \citep{armenian:2009}.\\
A case that has received increasing attention in the  literature is the situation where the sample of controls is a random sample from the whole population with information only on the attributes and not on the response \citep{lanc:imb:1996}. This situation is fairly common in ecological studies where only species' presence is recorded when field surveys are carried out. In the ecological literature, since the 1990's such data are called \emph{presence only data} \citep[see][and references therein]{ara:wil:2000}.  \citet{pea:boy:2006} define presence-only data as ``consisting only of observations of the organism but with no reliable data where the species was not found''. Atlases, museum and herbarium records, species lists, incidental observation databases and radio-tracking studies are examples of such data. \\
In recent years we find a considerably growing literature describing approaches to the  modeling of this type of data, among the many ecological papers we recall \citet{kea:che:2004}, \citet{pea:boy:2006}, \citet{elith:al:2006}, \citet{eli:lea:2009}, \citet{franklin:2010} and, most notably, in the statistical literature \citet{ward:al:2009}, \citet{war:she:2010}, \citet{chakraborty:al:2011}, \citet{dilorenzo:al:2011} and \citet{dorazio:2012}. While in \citet{war:she:2010} and \citet{chakraborty:al:2011} to model the presence-only data Poisson point processes are considered in the likelihood and  Bayesian framework respectively, in \citet{ward:al:2009} and \citet{dilorenzo:al:2011} a modified case-control logistic model is adopted  in the likelihood and Bayesian perspective respectively, in both papers there is no account for  possible dependence structure in the observations. In \citet{dorazio:2012} the asymptotic relations between the two approaches are discussed. \\ 
A different approach, MaxEnt, is based on the maximum entropy principle \citep{jaynes:1957}. In MaxEnt \citep{philips:al:2006,elith:al:2011} the relative entropy between the distribution of covariates at locations where the species is present and the unconditional background distribution of covariates is maximized subject to some constrains concerning empirical statistics (see Philips et al., 2006, for details). As pointed by \citet{dorazio:2012} ``the MaxEnt method requires knowledge of species' prevalence for its estimator of occurrence to be consistent''.\\
In what follows we are going to use the name \emph{presence-only data} when referring to the above sketched general problem of having information on the presence and covariates jointly on a sample from a  population, while information on only the covariates is available on any sample from the same population. This work is developed in the same discrete setting as in \citet{ward:al:2009} and \citet{dilorenzo:al:2011}, i.e., we have a population of independent units, no dependence structure, such as spatial correlation, is anticipated. We defer the treatment of this extension to a subsequent work. \\
The main contribution of the paper is a new rigorous formalization of the logistic regression model with presence-only data that allows further insight into the inferential issues. This leads us to an algorithmic procedure that, among other results, returns a MCMC approximation of the response prevalence under general knowledge of the process generating the data.
We also present  a large simulation study involving 24,000 simulated datasets and comparing our approach  to other two models representing optimal benchmarks.\\
The paper is organized as follows. Section \ref{sec:reglog} introduces a general framework for the presence-only data problems, Section \ref{sec:bayes} presents our Bayesian approach, Section \ref{sec:mcmc} describes the MCMC algorithm while results related to the simulation study are reported in Section \ref{sec:sim}. Finally in Section \ref{sec:conc} some conclusions are drawn and future developments briefly described.

\section{Linear logistic regression for presence-only data.}\label{sec:reglog}
The analysis of a binary response related to a set of explicative covariates is usually carried out through the use of the logistic regression where the logit of the conditional probability of occurrence is modeled as a function of covariates. In this section, we first introduce a general framework for the modeling  of presence-only data and then consider the case of the linear logistic regression. The approach proposed is built on two levels and we partially follow the formulation introduced by \citet{ward:al:2009} but adopting a Bayesian scheme  as in \citet{divino:al:2011}.

\subsection{A two level approach.}
Let  $Y$ be a binary variable informing on the presence ($Y=1$) or absence ($Y=0$) of a population's attribute and let $X=(X_1,...,X_k)$ denote a set of highly informative, on the same attribute, covariates which are available on the same population. Then, the presence-only problem can be formalized by  considering a censorship mechanism that acts when observing the response $Y$, so that part of the population units are not reachable. In particular, we refer to the situation in which we are able to detect only a partial set of units on which the attribute of interest is present while the information on the covariates  $X$ is available on the entire population. In this situation, we have to consider two types of uncertainty: the uncertainty due to the mechanism of censorship and the uncertainty due to the sampling procedure. Moreover, since we are not able to collect a random sample of observable data, we need to adjust for the sampling mechanism through the use of a case-control scheme \citep{bre:dey:1980, breslow:2005, armenian:2009} . \\
In order to build a Bayesian model, in this framework we adopt the following conceptual scheme in two levels.\\
\paragraph*{Level 1.} 
Given the population of interest $\mathcal{U}$ of size $N$, the binary responses $\mathbf{y}=(y_{1},...,y_{N})$ are generated independently by a probability law $\mathcal{M}$. \\
\paragraph*{Level 2.}
Let $\mathcal{U}_p$ be the subset of $\mathcal{U}$ where we observe $Y=1$. A modified case-control design is applied so that a sample of presences, considered as cases, is selected from $\mathcal{U}_p$ and a sample of ``contaminated'' controls \citep{lanc:imb:1996} is selected from the whole population $\mathcal{U}$, with all the covariates but no information on $Y$.
\paragraph*{}
Here, we cannot approach the model construction using only a finite population approach \citep{sarndal:1978} because of the censoring mechanism that ``masks'' distributional information on $Y$ already at the population  level. By the introduction of Level 1 we can describe the censored observations as random quantities generated by the model $\mathcal{M}$. Hence, the problem of presence-only data can be formalized as a problem of missing data \citep{rubin:1976, lit:rub:1987}.

\subsection{The model generating population data.}
At the first level, we assume that the law $\mathcal{M}$ is defined in terms of the conditional probability of occurrence $Pr(Y=1|x)$, denoted by $\pi^*(x)$, when the covariates are $X=x$. Moreover, we consider that the relation between $Y$ and $X$ is formalized through a regression function $\phi(x)$ on the logit scale 
\begin{equation}
\phi(x)=\mathrm{logit} \, \pi^*(x),
\label{eq:phi1}
\end{equation}
that is
\begin{equation}
\pi^*(x)=\frac{e^{\phi(x)}}{1+e^{\phi(x)}}.
\label{eq:prev1}
\end{equation}
When the data $\mathbf{y}=(y_{1},...,y_{N})$ are independently generated from $\mathcal{M}$, we denote by $\pi$ the empirical prevalence of the binary response $Y$ in $\mathcal{U}$, expressed as the ratio of the number of presences $N_1$ to the size of the population, that is 
\[\pi=\frac{N_1}{N}.\]
\subsection{The modified case-control design.}
At the second level, we adopt a case-control design modified for presence-only data \citep{lanc:imb:1996} in order to account for the specific sampling procedure considered. The use of the case-control scheme is necessary at all times when it is appropriate to select observations in fixed proportions with respect to the values of the response variable. This can occur when the attribute of interest represents a phenomenon that is rare among the units of the population as for example a rare disease or a rare exposure in epidemiological studies \citep{woodward:2005}.\\
Now, let $C$ be a binary indicator of inclusion into the sample ($C=1$ denotes that a unit is in the sample), let $\rho_0=Pr(C=1|Y=0)$ and $\rho_1=Pr(C=1|Y=1)$ be the inclusion probability of the absences and the presences, respectively. Under the assumption that, given $Y$, the sampling mechanism is independent from the covariates $X$, the conditional probability of occurrence is modified through the Bayes rule as 
\begin{equation}
Pr(Y=1|C=1,x)=\frac{\rho_1 e^{\phi(x)}}{\rho_0+\rho_1e^{\phi(x)}}.
\label{eq:prev2}
\end{equation}
Hence, the corresponding case-control regression function $\phi_{cc}(x)$ defined as the logit of \eqref{eq:prev2} is given by
\begin{equation}
\phi_{cc}(x)=\phi(x)+\mathrm{log}\,\frac{\rho_1}{\rho_0}. 
\label{eq:phi2}
\end{equation}
In particular, if the selection of cases ($n_1$) and controls ($n_0$) is made independently without replacement, the inclusion probabilities are given in terms of the empirical prevalence $\pi$ by
\[\rho_0=\frac{n_0}{(1-\pi)N}\] 
and 
\[\rho_1=\frac{n_1}{\pi N},\] 
so that the equation \eqref{eq:phi2} becomes
\begin{equation}
\phi_{cc}(x)=\phi(x)+\mathrm{log}\,\frac{n_1}{n_0}-\mathrm{log}\,\frac{\pi}{1-\pi}.
\label{eq:phi3}
\end{equation}
In our framework, since the response variable $Y$ is already censored at the population level, the standard case-control design cannot be adopted but it should be modified in such a way that a sample of presences is matched with an independent sample drawn from the entire population, named the \textit{background sample} \citep{zaniewski:al:2002, ward:al:2009}. Remark that in this sample the response variable is unobserved and only the covariates are available.\\
In this way, the complete sample $S$ is composed by a set $S_u$ of $n_u$ independent background data, where the response $Y$ is not observed, drawn from the entire $\mathcal{U}$ and by a set $S_p$ of $n_p$ independent observations selected from the sub-population of presences $\mathcal{U}_p$. This procedure implies that the reference population $\mathcal{U}$ is augmented with its subset $\mathcal{U}_p$ so that the total number of observations considered in the sampling scheme becomes $N+N_1$. To illustrate the sampling framework we are going to adopt here, let us consider the following situation: we can label population units of type $y=1$ only when they are isolated from units of type $y=0$. This can be formalized by introducing a binary stratum variable $Z$ such that $Z=0$ indicates when an observation is drawn from the entire population $\mathcal{U}$ while $Z=1$ denotes the sampling from the sub-population $\mathcal{U}_p$. Remark that $Z=1$ implies $Y=1$ whilst $Z=0$ implies that $Y$ is an unknown value $y \in \{0,1\}$. Moreover, by construction $Z$ is independent from the covariates $X$, given the response $Y$. The introduction of the variable $Z$ allows us to define the structure of the data at the population level and at the sample level in terms of presences/absences ($Y$) and known/unknown data ($Z$), as reported in Table \ref{tab1} and Table \ref{tab2}.
\begin{table}[H]
\centering
\begin{tabular}{cccc}
\midrule
Y/Z       & $Z=0$    & $Z=1$    & Total     \\
\midrule
$Y=0$     & $N_{0}$  & $0$      & $N_{0}$   \\
$Y=1$     & $N_{1}$  & $N_{1}$  & $2N_{1}$  \\
\midrule
Total     & $N    $  & $N_{1}$  & $N+N_{1}$ \\
\midrule
\end{tabular}
\caption{Data structure at the population level.}\label{tab1} 
\end{table} 
\begin{table}[H]
\centering
\begin{tabular}{cccc}
\midrule
Y/Z       & $Z=0$    & $Z=1$     & Total    \\
\midrule
$Y=0$     & $n_{0u}$ & $0$       & $n_{0}$  \\
$Y=1$     & $n_{1u}$ & $n_{p}$   & $n_{1}$  \\
\midrule
Total     & $n_{u}$  & $n_{p}$   & $n$      \\
\midrule
\end{tabular}
\caption{Data structure at the sample level.}\label{tab2}
\end{table} 
In Table \ref{tab1}, $N_0$ is  the number of absences in the population $\mathcal{U}$ while in Table \ref{tab2}, $n_{0u}$ and $n_{1u}$ respectively denote the unknown frequencies of absences and presences in the sub-sample $S_u$. Remark that, in the above described situation, the inclusion probability of units with or without the mentioned attribute changes. In fact, while an absence can be drawn only when sampling from $\mathcal{U}$, a presence can be selected when sampling both from $\mathcal{U}$ and from $\mathcal{U}_p$. Thus, one has 
\begin{equation}
\rho_{0}=\frac{n_{0}}{N_{0}}=\frac{n_{0u}}{(1-\pi)N},
\label{eq:rho0}
\end{equation}
and
\begin{equation}
\rho_{1}=\frac{n_{1}}{2N_{1}}=\frac{n_{1u}+n_{p}}{2\pi N}.
\label{eq:rho1}
\end{equation}
The introduction of the stratum variable $Z$ allows us also to exactly derive the logistic regression model under the case-control design modified for presence-only data. In fact, when we consider the population $\mathcal{U}$ augmented with its subset $\mathcal{U}_p$, the model $\pi^*(x)$ represents the conditional probability to mark a presence only when $Z=0$, that is $Pr(Y=1|Z=0,x)=\pi^*(x)$. On the other hand, when $Z=1$, we simply have $Pr(Y=1|Z=1,x)=1$. We can prove the following result.\\
\paragraph*{Proposition 1.} 
Under the assumption that $Z$ is independent from $X$ given $Y$, one has
\begin{equation}
Pr(Y=1|x)=\frac{2\pi^*(x)}{1+\pi^*(x)}.
\label{eq:prop1}
\end{equation}
\paragraph*{Proof.}
From the hypothesis of conditional independence it results
\[Pr(Z|Y,x)=Pr(Z|Y),\]
that can be express also as
\[\dfrac{Pr(Y|Z,x)Pr(Z|x)}{Pr(Y|x)}=\dfrac{Pr(Y|Z)Pr(Z)}{Pr(Y)}.\]
Let consider the case with $Y=1$ and $Z=0$, one has
\[\dfrac{Pr(Y=1|Z=0,x)Pr(Z=0|x)}{Pr(Y=1|x)}=\dfrac{Pr(Y=1|Z=0)Pr(Z=0)}{Pr(Y=1)}.\]
The probabilities enclosed in the second term can be derived from Table \ref{tab1} and one has
\begin{equation}
\dfrac{\pi^*(x)Pr(Z=0|x)}{Pr(Y=1|x)}=\dfrac{\frac{N_1}{N}\frac{N}{N+N_1}}{\frac{2N_1}{N+N_1}}=\frac{1}{2}.
\label{eq:prop1.1}
\end{equation}
In the case $Y=1$ and $Z=1$ one similarly obtains
\begin{equation}
\dfrac{Pr(Z=1|x)}{Pr(Y=1|x)}=\dfrac{\frac{N_1}{N_1}\frac{N_1}{N+N_1}}{\frac{2N_1}{N+N_1}}=\frac{1}{2}.
\label{eq:prop1.2}
\end{equation}
From \eqref{eq:prop1.2} it results $Pr(Y=1|x)=2\,Pr(Z=1|x)$ and by substituting in \eqref{eq:prop1.1}, one can derive that $Pr(Z=0|x)=\dfrac{1}{1+\pi^*(x)}$ and hence $Pr(Z=1|x)=\dfrac{\pi^*(x)}{1+\pi^*(x)}$. Now, it is simple to obtain that
\[Pr(Y=1|x)=\dfrac{2\pi^*(x)}{1+\pi^*(x)}.\]
$\square$\\
If we assume that, given $Y$, the inclusion into the sample ($C=1$) is independent from the covariates $X$, one has \footnote{see Appendix for the detailed proof.}
\begin{equation}
Pr(Y=0|C=1,x)\,Pr(C=1|x)=\frac{1-\pi^*(x)}{1+\pi^*(x)}\,\rho_0 
\label{eq:cor1}
\end{equation}
and
\begin{equation}
Pr(Y=1|C=1,x)\,Pr(C=1|x)=\frac{2\pi^*(x)}{1+\pi^*(x)}\,\rho_1. 
\label{eq:cor2}
\end{equation}
Then, from the ratio of \eqref{eq:cor2} to \eqref{eq:cor1}, it results
\begin{equation}
\nonumber
\frac{Pr(Y=1|C=1,x)}{Pr(Y=0|C=1,x)}= \frac{2\pi^*(x)}{1-\pi^*(x)} \,\frac{\rho_1}{\rho_0},
\end{equation}
and by plugging the quantities $\rho_0$ and $\rho_1$, as defined in \eqref{eq:rho0} and in \eqref{eq:rho1}, into the logit of $Pr(Y=1|C=1,x)$, one obtains the following relation 
\begin{eqnarray}
\nonumber
\mathrm{logit}\,Pr(Y=1|C=1,x)&=&\mathrm{log}\left[\frac{2\pi^*(x)}{1-\pi^*(x)}\,\frac{\rho_1}{\rho_0}\right] 
\\
\nonumber
&=&\mathrm{log}\left[\frac{2\pi^*(x)}{1-\pi^*(x)}\,\frac{n_{1u}+n_p}{n_{0u}}\,\frac{1-\pi}{2\pi}\right]
\\
\nonumber
&=&\mathrm{log}\left[\frac{\pi^*(x)}{1-\pi(x)}\,\frac{n_{1u}+n_p}{n_{0u}}\,\frac{1-\pi}{\pi}\right]
\\
\nonumber
&=&\mathrm{logit}\,\pi^*(x)+\mathrm{log}\,\frac{n_{1u}+n_p}{n_{0u}}-\mathrm{log}\,\frac{\pi}{1-\pi}
\\
&=&\phi(x)+\mathrm{log}\,\frac{n_{1u}+n_p}{n_{0u}}-\mathrm{log}\,\frac{\pi}{1-\pi},
\label{eq:phi4}
\end{eqnarray}
that represents the logistic regression model under the case-control design for presence-only data. As well, we can now formalize the presence-only data regression function $\phi_{pod}(x)$ as
\begin{equation}
\phi_{pod}(x)=\phi(x)+\mathrm{log}\,\frac{n_{1u}+n_p}{n_{0u}}-\mathrm{log}\,\frac{\pi}{1-\pi}.\label{eq:phi5}
\end{equation}
Although the derivation is substantially different, we end with the same formulation as in \citet{ward:al:2009}. Now, in order to make parameter estimation possible, we need to handle the ratio
\begin{equation}
\frac{\rho_1}{\rho_0}=\frac{n_{1u}+n_p}{n_{0u}} \, \frac{1-\pi}{2 \pi},
\label{eq:ratio1}
\end{equation}
where the quantities $\pi$ and $n_{1u}$ are unknown ($n_{0u}=n_u-n_{1u}$).\\ 
In the recent literature, two main approaches have been proposed. The first one by \citet{ward:al:2009} replace the ratio $\frac{n_{1u}+n_p}{n_{0u}}$ with the ratio of the expected numbers of presences and absences in the sample, that is 
\begin{equation}
\frac{\rho_1}{\rho_0}\approx\frac{E[n_{1u}+n_p]}{E[n_{0u}]}\,\frac{1-\pi}{2 \pi}
=\frac{\pi n_u+n_p}{(1-\pi)n_u}\,\frac{1-\pi}{2\pi}
=\frac{\pi n_u+n_p}{2 \pi n_u}.
\label{eq:ratio2}
\end{equation}
These authors adopt a likelihood approach and computation is carried out via the EM algorithm. As they underline, this approximation can be easily implemented if the empirical population prevalence $\pi$ is known a priori. They discuss also the possibility to estimate $\pi$ jointly with the regression function when the prevalence is identifiable, as for example in the linear logistic regression, and with respect to this case they present a simulation example. The difficulty in obtaining efficient joint estimates because of the correlation between $\pi$ and the intercept of the linear regression term is discussed. Notice that \citet{ward:al:2009} considers a slightly different representation of the ratio \eqref{eq:ratio2}, omitting the multiplier ``2'' in the denominator. \\
\citet{dilorenzo:al:2011}, dealing with a problem of abundance data, use the approximation \eqref{eq:ratio2}, but they adopt a Bayesian approach and consider the population prevalence $\pi$ as a further parameter in the model. They choose an informative \textit{Beta} prior for $\pi$, but their MCMC algorithm contains an unusual weakness since the simulation of $\pi$ is performed from its prior and not from the posterior that can be derived through the interaction between the parameter $\pi$  and the regression function $\phi(x)$.\\
A different approximation of the ratio \eqref{eq:ratio1} can be obtained by considering the sample prevalence in $S_u$ (the background sample)
\[\pi_u=\frac{n_{1u}}{n_u},\]
where
\[n_{1u}=\sum_{i \in S_u} y_i.\] 
Due to the censorship process, this quantity is unknown but it would be the maximum likelihood estimator for $\pi$ if the data  $\mathbf{y}_u=\{y_i, i \in S_u\}$ could be observed. Now, replacing $\pi$ by $\pi_u$ in \eqref{eq:ratio1} one obtains
\begin{eqnarray}
\frac{\rho_1}{\rho_0}
\approx\frac{n_{1u}+n_p}{n_{0u}}\,\frac{1-\pi_u}{2 \pi_u}
=\frac{n_{1u}+n_p}{n_u-n_{1u}}\,\frac{n_u-n_{1u}}{2n_{1u}}
=\frac{n_{1u}+n_p}{2n_{1u}},
\label{eq:ratio3}
\end{eqnarray}
that allows to formulate a computable version of the regression function for presence-only data as
\begin{equation}
\phi_{pod}(x) \approx \phi(x)+\mathrm{log}\,\frac{n_{1u}+n_p}{n_{1u}}.
\label{eq:phi6}
\end{equation}
This function depends on  the data $\mathbf{y}_u$ in $S_u$ which are not directly observable, but if $\mathbf{y}_u$ is treated as missing data one can enclose it into the estimation process and then obtain a consistent approximation for $\phi_{pod}(x)$. In particular, in a Bayesian framework, this idea can be performed by using a Markov Chain Monte Carlo computation with data augmentation. Moreover from the use of MCMC simulations we can also obtain an approximation of $\pi_u$ and therefore an estimate of the empirical population prevalence $\pi$. Details are given in Section 4.\\
The approximation \eqref{eq:ratio3} can, in principle, be always adopted, but some care must be used as identifiability issues are present. We follow the recommendation in \citet{ward:al:2009} to approach jointly estimates of $\phi(x)$ and $\pi$ only when the latter is identifiable with respect to the regression function, as for example in the linear regression case  \citep[see][for mathematical details]{ward:al:2009}. 

\subsection{The linear logistic regression.}
If we consider a linear regression function $\phi(x)=x\beta$, where $\beta=(\beta_1,...,\beta_k)$ is the vector of the regression parameters, a computable model for presence-only data can be defined through the following approximation
\begin{equation}
\phi_{pod}(x) \approx x \beta +\mathrm{log}\,\frac{n_{1u}+n_p}{n_{1u}},
\label{eq:phi7}
\end{equation}
or equivalently through the approximation of the conditional probability of occurrence at the sample level
\begin{eqnarray}
\nonumber
Pr(Y=1|C=1,x) &\approx & \frac{\mathrm{exp} \{ x\beta +\mathrm{log} \,\frac{n_{1u}+n_p}{n_{1u}} \} }
{1+\mathrm{exp} \{ x\beta +\mathrm{log} \,\frac{n_{1u}+n_p}{n_{1u}} \} } \\
&=& \frac{ \left( 1+ \frac{n_p}{n_{1u}} \right) \, \mathrm{exp} \{ x\beta \}}
{1+ \left(1+ \frac{n_p}{n_{1u}} \right) \mathrm{exp} \{ x\beta \}}.
\label{eq:prev3}
\end{eqnarray}
In this particular case, all the unknowns of the model are the linear parameters vector $\beta$ and the missing data $\mathbf{y}_u$ in the background sample $S_u$. 
\section{The hierarchical Bayesian model.}\label{sec:bayes}
\begin{figure}[H]
\centering
\includegraphics[scale=0.4]{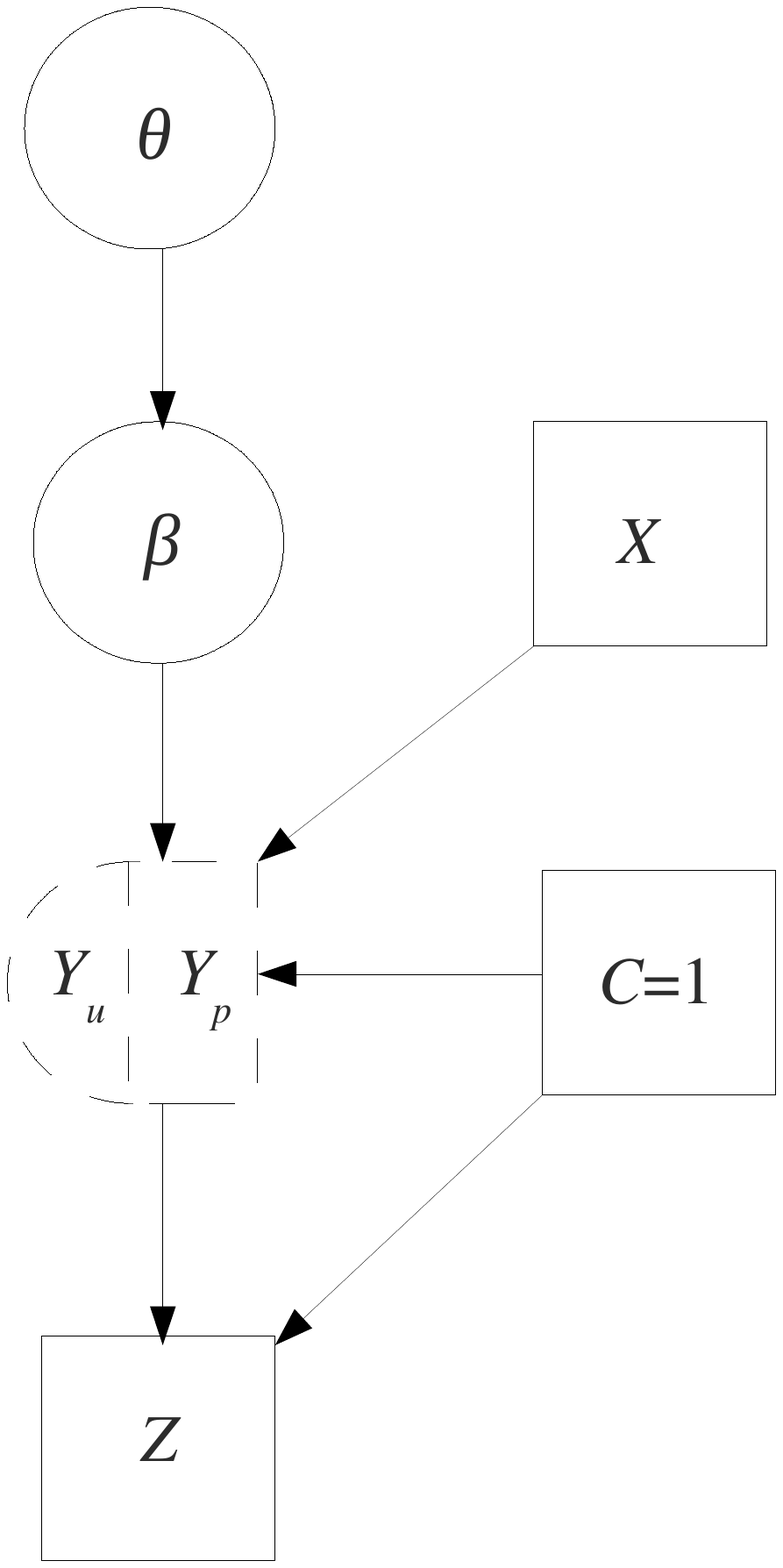}
\caption{Graphical representation of the hierarchical Bayesian model.}\label{fig:graph}
\end{figure}

Due to the censorship process affecting the data, we can acquire complete information only on the stratum variable $Z$ and not on the binary response $Y$. Then, it seems natural to model $Z$ as the observable variable. If we consider the conditional joint distribution of $Z$ and $Y$
\begin{equation}
Pr(Z,Y|C=1,x)=Pr(Z|Y,C=1,x)Pr(Y|C=1,x), 
\label{eq:bayes1}
\end{equation}
through the marginalization over $Y$, the probability $Pr(Z|C=1,x)$ can be obtained and we can  express the relation between presences and covariates in terms of regression of $Z$ respect to $X$. Notice that, while $Pr(Y|C=1,x)$ can be obtained from \eqref{eq:prev3}, the term $Pr(Z|Y,C=1,x)$, due to the conditional independence between $Z$ and $X$ given $Y$, simply reduce to be equal to $Pr(Z|Y,C=1)$ that can be derived from  Table \ref{tab2}. \\
We point out that, even if the response $Y$ does not play an explicit role after the marginalization step, we need to keep it in the model as a hidden variable in order to obtain the approximation for the quantity $n_{1u}=\sum_{i \in S_u} y_i$, necessary to correct the linear regression function for presence-only data.\\
Now, we can formalize the hierarchical Bayesian model to estimate the parameters of a linear logistic regression under the case-control scheme adjusted for presence-only data. In order to better explain the conditional relationship underlying the hierarchy, we introduce the graph in Figure \ref{fig:graph}. The dashed node indicates a variable hidden with respect to the conditional relationships.
\paragraph*{The priors.}
At the top of the hierarchy, we assume the hyper parameter $\theta$ distributed as $p(\theta)$. At the second level, we consider the prior probability distribution on $\beta$ depending on the hyper parameter $\theta$, that is $\beta|\theta \sim p(\beta|\theta)$. At the third level, the unobserved data $\mathbf{y}_u$ in $S_u$ are considered latent parameters with prior distribution  \textit{Bernoulli} (denoted by $Be$) with probability of occurrence given by the approximation in \eqref{eq:prev3}, that is 
\begin{equation}
\nonumber
y_i|C_i=1,x_i,\beta \sim Be\left(\frac{\left(1+\frac{n_p}{n_{1u}}\right)\,\mathrm{exp}\{x\beta\}}
{1+\left(1+\frac{n_p}{n_{1u}}\right)\mathrm{exp}\{x\beta\}}\right), \; i \in S_u.
\end{equation}
This point is important for deriving the predictive distribution of the unobserved data $\mathbf{y}_u$ necessary in the estimation algorithm.
\paragraph*{The likelihood.}
At the lowest level of the hierarchy, we have the likelihood, defined with respect to the observable stratum variable $Z$. Recalling that from the Table \ref{tab2} we have $Pr(Z=1|Y=0,C=1)=0$ and $Pr(Z=1|Y=1,C=1)=\dfrac{n_p}{n_{1u}+n_p}$, when \eqref{eq:bayes1} is marginalized over $Y$, one obtains the approximation
\begin{eqnarray}
\nonumber
Pr(Z=1|C=1,x,\beta) &\approx & \frac{n_p}{n_{1u}+n_p}\,\frac{\mathrm{exp}\{x\beta+\mathrm{log}\,\frac{n_{1u}+n_p}{n_{1u}}\}}
{1+\mathrm{exp}\{x\beta+\mathrm{log}\,\frac{n_{1u}+n_p}{n_{1u}}\}} \\
&=&\frac{\frac{n_p}{n_{1u}}\,\mathrm{exp}\{x\beta\}}{1+\left(1+\frac{n_p}{n_{1u}}\right)\mathrm{exp}\{x\beta\}}
\label{eq:lik1}
\end{eqnarray} 
and hence
\begin{eqnarray}
\nonumber
Pr(Z=0|C=1,x,\beta) &=& 1-Pr(Z=1|C=1,x,\beta)\\
&\approx & \frac{1+\mathrm{exp}\{x\beta\}}{1+\left(1+\frac{n_p}{n_{1u}}\right)\mathrm{exp}\{x\beta\}}.
\label{eq:lik2}
\end{eqnarray}
Thus, we can assume that for all $i \in S$ the conditional distribution of $Z_i$ is  \textit{Bernoulli} with probability of occurence given by \eqref{eq:lik1}, that is
\begin{equation}
\nonumber
Z_i|C_i=1,x_i,\beta \sim Be \left(\frac{\frac{n_p}{n_{1u}}\,\mathrm{exp}\{x\beta\}}{1+\left(1+\frac{n_p}{n_{1u}}\right)\mathrm{exp}\{x\beta\}}\right), \; i \in S. 
\end{equation}
Recalling that $Z_i=0$ for all $i \in S_u$ while $Z_i=1$ for all $i \in S_p$, the likelihood function  can be written as
\begin{equation}
\nonumber
L(\beta;\mathbf{z},\mathbf{x})= 
\prod_{i \in S_u} \frac{1+\mathrm{exp}\{x_i\beta\}}{1+\left(1+\frac{n_p}{n_{1u}}\right)\mathrm{exp}\{x_i\beta\}}\times 
\prod_{i \in S_p} \frac{\frac{n_p}{n_{1u}}\mathrm{exp}\{x_i\beta\}}{1+\left(1+\frac{n_p}{n_{1u}}\right)\mathrm{exp}\{x_i\beta\}}.
\end{equation}
\citet{ward:al:2009} defines this function as the \textit{observed likelihood} versus the \textit{full likelihood} that, instead, considers the distribution of the stratum variable $Z$ jointly with the response $Y$.
\paragraph*{The posterior.}
Now, through the Bayes rule we derive the full posterior
\begin{eqnarray}
p(\beta,\theta |\mathbf{z},\mathbf{x}) &\propto & p(\theta)p(\beta|\theta)L(\beta;\mathbf{z},\mathbf{x})
\label{eq:post}
\end{eqnarray}
that can be used to make inference on the  quantities of interest.
\section{The MCMC computation.}\label{sec:mcmc}
Samples from \eqref{eq:post} can be obtained via Markov Chain Monte Carlo simulation \citep{rob:cas:2004, liu:2008}. While it seems quite standard to implement a direct sampler for the vector $\beta$ and the hyper parameter $\theta$, we need to sample also the latent $\mathbf{y}_u$. For this reason we introduce a step of data augmentation \citep{tan:won:1987, tanner:1996} in the estimation procedure. The basic idea of the data augmentation technique is to augment the set of observed data to a set of completed data that follow a simpler distribution \citep{liu:wu:1999}. In our framework, we need to augment the observations of the stratum variable $\mathbf{z}$ with the missing values $\mathbf{y}_u$ in order to have, at each iteration, a consistent value of the quantity $n_{1u}$, necessary to adjust the regression function $\phi_{pod}(x) \approx x\beta +\mathrm{log}\,\frac{n_{1u}+n_p}{n_{1u}}$ for presence-only data. The following result allows for an easy implementation of the data augmentation step.\\

\paragraph*{Proposition 2.} 
Using the approximation \eqref{eq:ratio3} of the ratio \eqref{eq:ratio1}, the posterior predictive probability of occurrence for an unobserved response $y$ in the sub-sample $S_u$ is approximated by the model $\mathcal{M}$ that generates the data at the population level, that is
\begin{equation}
Pr(Y=1|Z=0,C=1,x) \approx \pi^*(x). 
\label{eq:prop2}
\end{equation}
\paragraph*{Proof.}
From the conditional independence between $Z$ and $X$ given $Y$, the predictive probability of occurrence in $S_u$ is given by
\[Pr(Y=1|Z=0,C=1,x)= \dfrac{Pr(Z=0|Y=1,C=1)Pr(Y=1|C=1,x)}{Pr(Z=0|C=1,x)}. \]
From Table \ref{tab2} we have that $Pr(Z=0|Y=1,C=1)= \frac{n_{1u}}{n_p+n_{1u}}$ and hence
\begin{equation}
Pr(Y=1|Z=0,C=1,x) = \dfrac{n_{1u}}{n_p+n_{1u}}\,\dfrac{Pr(Y=1|C=1,x)}{Pr(Z=0|C=1,x)}.
\label{eq:prop2.1}
\end{equation}
Now, recalling that in the general case one has
\begin{eqnarray}
Pr(Y=1|C=1,x) \approx \frac{\left(1+\frac{n_p}{n_{1u}}\right) \, \mathrm{exp}\{\phi(x)\}}
{1+\left(1+\frac{n_p}{n_{1u}}\right)\mathrm{exp}\{\phi(x)\}}
\label{eq:prop2.2}
\end{eqnarray}
and 
\begin{eqnarray}
Pr(Z=0|C=1,x) \approx \frac{1+\mathrm{exp}\{\phi(x)\}}{1+\left(1+\frac{n_p}{n_{1u}}\right)\mathrm{exp}\{\phi(x)\}},
\label{eq:prop2.3}
\end{eqnarray}
by substituting  \eqref{eq:prop2.2} and \eqref{eq:prop2.3} in \eqref{eq:prop2.1}, one obtains
\begin{eqnarray}
\nonumber
Pr(Y=1|Z=0,C=1,x) &\approx & \dfrac{n_{1u}}{n_p+n_{1u}} \,\dfrac{\left(1+\frac{n_p}{n_{1u}}\right) \, \mathrm{exp} \{ \phi(x) \}} {1+ \mathrm{exp} \{ \phi(x) \}} \\
\nonumber
&=& \dfrac {\mathrm{exp}\{\phi(x)\}}{1+\mathrm{exp}\{\phi(x)\}}\\
\nonumber
&=& \pi^*(x).
\end{eqnarray}
$\square$\\
\subsection{The data augmentation algorithm.}
A general MCMC scheme to perform inference on a linear regression model for presence-only data can be defined as follow.
\bigskip
\begin{quote}
\begin{description}
\hrule
\bigskip
\item[Step 0.] Initialize $\theta$, $\beta$ and $\mathbf{y}_u$
\item[Step 1.] Set $n_{1u}=\sum_{i \in S_u} y_i$
\item[Step 2.] Sample $\theta$ from $p(\theta|\mathbf{z},\mathbf{x},\beta)$
\item[Step 3.] Sample $\beta$ from $p(\beta |\mathbf{z},\mathbf{x},\theta)$
\item[Step 4.] Sample $y_i$ from  $p(y_i|Z_i=0,C_i=1,x_i,\beta)$ for all $i \in S_u$ 
\item[] Goto Step 1
\bigskip
\hrule
\end{description}
\end{quote}
\bigskip
After the initialization of  all the arrays (Step 0), Step 1 sets a current value for the quantity $n_{1u}$ to adjust the regression function $\phi_{pod}(x)$. Step 2 and Step 3 consider the sampling from the posterior of the hyper parameter $\theta$ and the regression parameter $\beta$, respectively, and they can be performed by Metropolis-Hasting schemes \citep{rob:cas:2004}. Step 4 concerns the data augmentation for the unobserved $\mathbf{y}_u$ in order to update consistently the quantity $n_{1u}$ at the following iteration. From the result \eqref{eq:prop2}, this simulation can be obtained by a Gibbs sampler  \citep{rob:cas:2004} since the posterior predictive distribution for all $i \in S_u$ is approximated by \textit{Bernoulli} with parameter of occurrence $\pi(x_i)=\frac {\mathrm{exp}\{x_i\beta\}}{1+\mathrm{exp}\{x_i\beta\}}$.

\subsection{The estimation of the prevalence $\mathbf{\pi}$.}\label{sec:prev}
From the data augmentation algorithm we can  obtain a MCMC estimate of the population prevalence $\pi$. In fact, if at each iteration $t$, after the Markov chain has reached the  equilibrium, we save the current value $n_{1u}^{(t)}$, we can obtain a consistent MCMC approximation of the sample prevalence $\pi_u$ in $S_u$ by
\begin{equation}
\hat{\pi}_{mcmc}= \dfrac{\bar{n}_{1u}}{n_u}
\label{eq:mcmc}
\end{equation}
where $\bar{n}_{1u}$ is the ergodic mean of the augmentations $n_{1u}^{(t)}$ over the Markov chain, that is
\[ \bar{n}_{1u}=\dfrac{\sum_{t=1}^T  n_{1u}^{(t)}}{T}. \]
Therefore, since $\pi_u$ would be a consistent estimator for $\pi$, $\hat{\pi}_{mcmc}$ represents also a consistent estimation of the empirical population prevalence.
\section{A comparative simulation study.}\label{sec:sim}
We present a simulation experiment to evaluate the performances of the model \eqref{eq:prev3}. To this aim we generate several datasets in the way described below and we compare our proposal with respect to two models acting in two different situations: (a) the censorship process does not act on the population $\mathcal{U}$ so that the data $\mathbf{y}$ are completely observed; (b) the censorship is present, but we assume known the population prevalence so that  approximation \eqref{eq:ratio2} can be used. In (a) we are able to estimate a linear logistic model (denoted by $M_{0}$), no correction is required and $\phi_{0}(x)=x\beta$. In (b) we consider a linear logistic model for presence-only data, denoted by $M_{1}$, with regression function $\phi_{1}(x)=x\beta+\frac{\pi n_{u}+n_{p}}{\pi n_{u}}$. Model \eqref{eq:prev3} (denoted by $M_2$) is estimated when the censorship process acts on the data and no information is available on the population prevalence. In this case, the regression function is given by $\phi_{2}(x)=x\beta+\frac{n_{1u}+n_{p}}{n_{1u}}$. Remark that model $M_{2}$ can be estimated when the least amount of information is available, $M_1$ requires less information than $M_0$ but more than $M_2$ and $M_0$ can be used only in the ideal situation of complete information. We assume $M_{1}$ as  benchmark model in the case of presence-only data.

\paragraph*{The generation of data.}
In order to set the simulation study, we need to generate the covariates $X$ and the binary response $Y$. In particular, we consider two covariates: $X_{1}$, giving strong information on the distribution of the response $Y$, and $X_{2}$, representing a term of noise, not available in the estimation step. We assume $X_{1}$ distributed as a mixture of two \textit{Gaussian} densities (denoted by \textit{N}),  centred in $\mu_{a}=4.0$ and $\mu_{b}=-4.0$ respectively, and with equal variances $\sigma^{2}=4.0$, that is
\[X_{1}\sim w N_{a}(\mu_{a};\sigma^{2})+(1-w)N_{b}(\mu_{b};\sigma^{2}).\]
The weight $w$ is a realization of a \textit{Bernoulli} random variable with probability of occurrence fixed to $p=0.165$. $X_{2}$ has standard \textit{Gaussian} distribution $N(0,1)$. Finally, the binary response $Y$, given the covariates $X_{1}$ and $X_{2}$, is \textit{Bernoulli} distributed with probability of occurrence  
\[\pi(x)=\dfrac{exp\{\beta_{0}+\beta_{1}x_{1}+\beta_{2}x_{2}\}}{1+exp\{\beta_{0}+\beta_{1}x_{1}+\beta_{2}x_{2}\}}.\]
We generate covariates and binary response  with respect to a population $\mathcal{U}$ of size $N=10000$. Three general scenarios with different level of complexity have been considered:
\begin{quote}
\begin{description}
\item[\textit{(i)}] $\beta_{0}=0$, $\beta_{1}=1$, $\beta_{2}=0$ : only the informative covariate $X_1$ generates the data;
\item[\textit{(ii)}] $\beta_{0}=0$, $\beta_{1}=1$, $\beta_{2}=1$ : a term of noise $X_2$ is added to the informative covariate;
\item[\textit{(iii)}] $\beta_{0}=1$, $\beta_{1}=1$, $\beta_{2}=1$ : $X_{1}$, $X_{2}$ and a constant effect generate the data.
\end{description}
\end{quote}

\begin{figure}[H]
\centering
\begin{tabular}{cccc}
\toprule
$n$ & $\beta_0$ & $\beta_1$ & $\pi$\\
\toprule
\raisebox{1cm} {50} &\includegraphics[height=2cm, width=3.5cm]{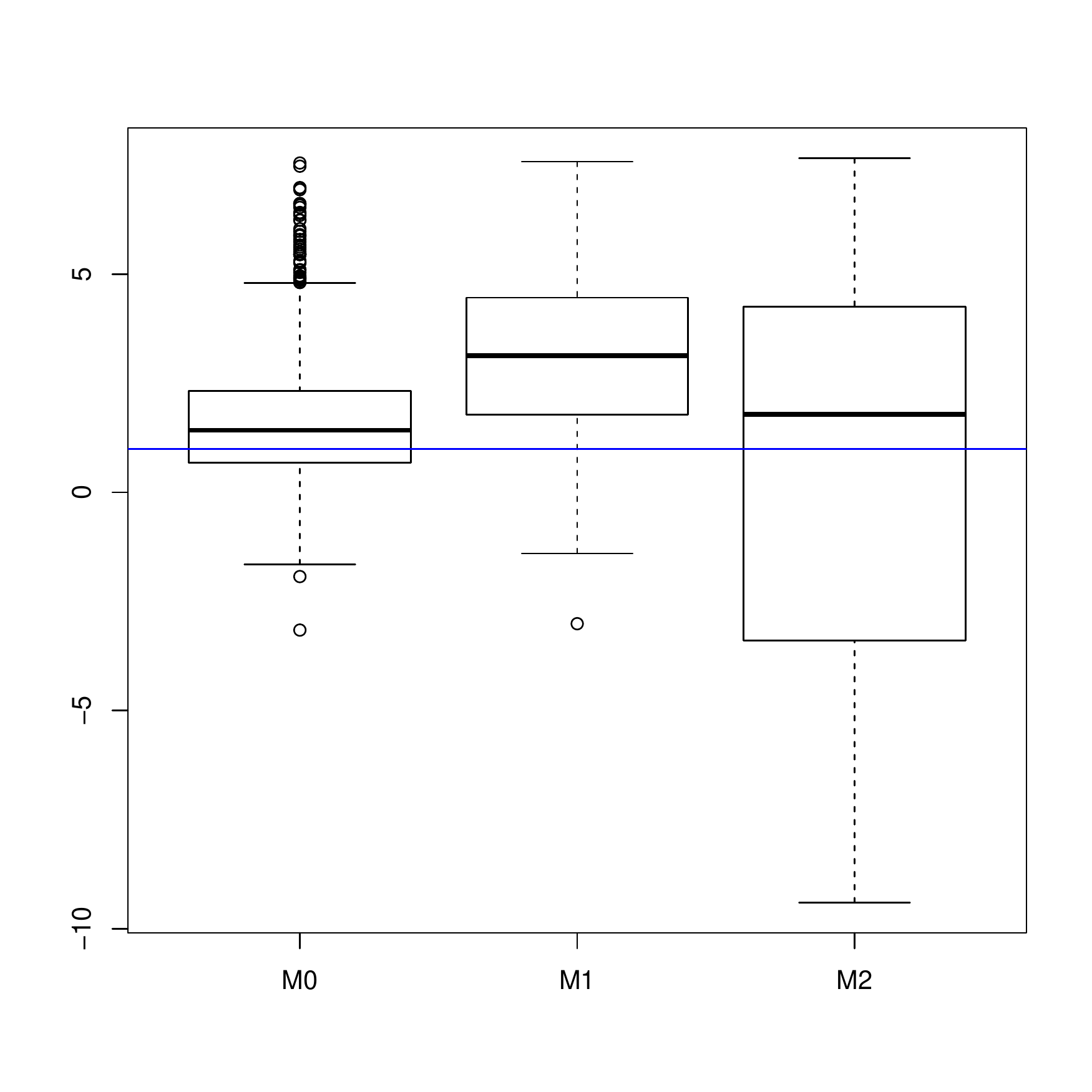}
&\includegraphics[height=2cm, width=3.5cm]{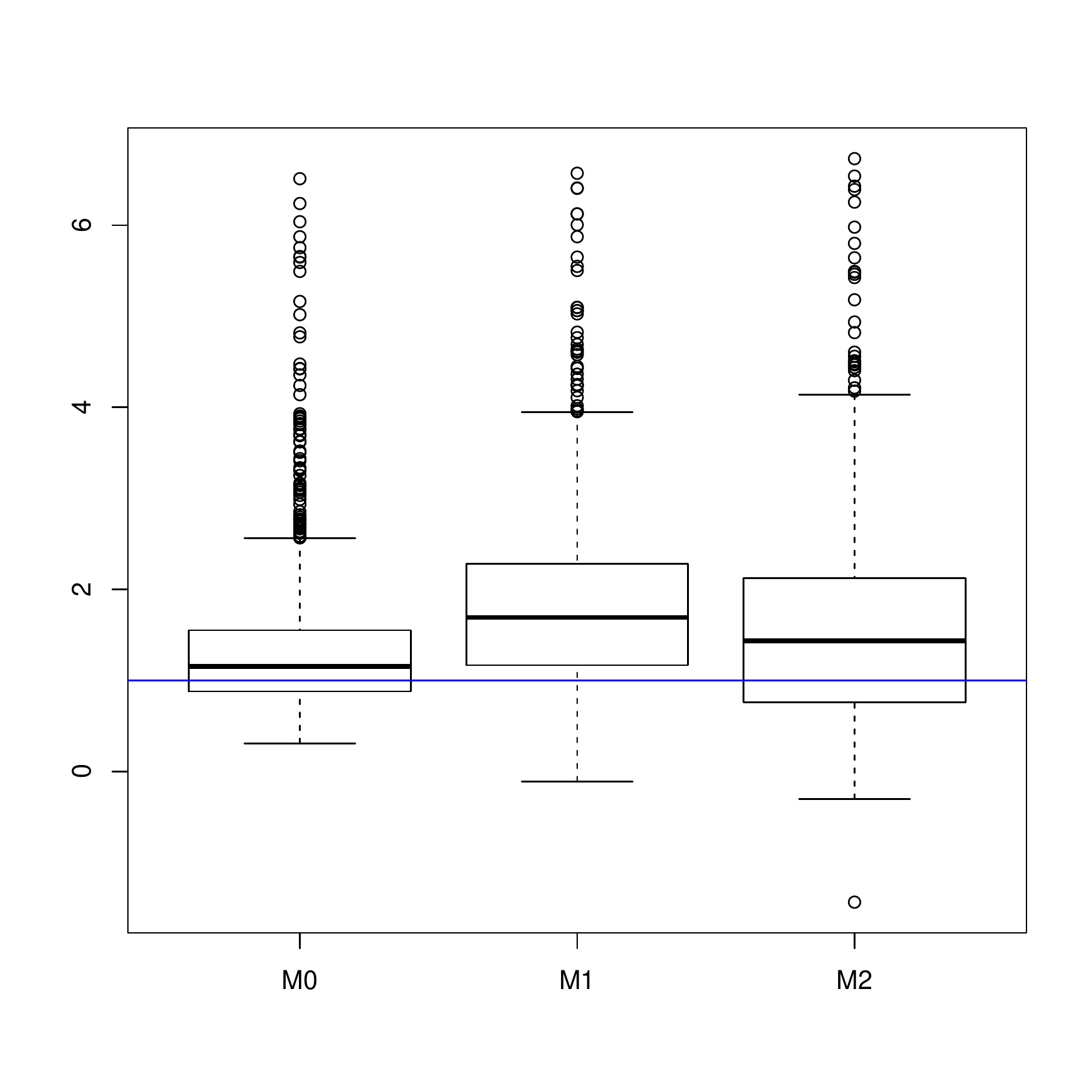}&\includegraphics[height=2cm, width=3.5cm]{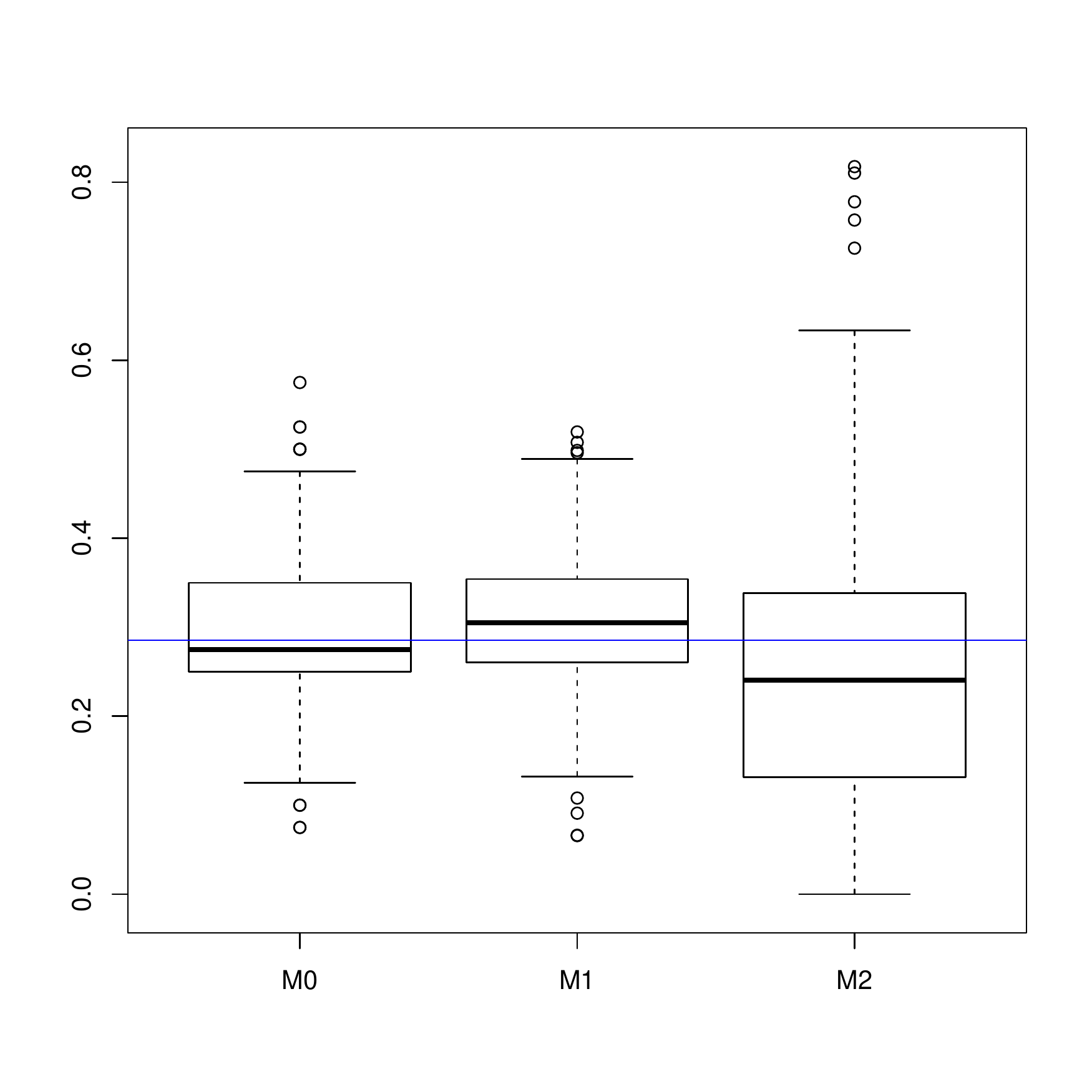}\\
\midrule
\raisebox{1cm} {100}&\includegraphics[height=2cm, width=3.5cm]{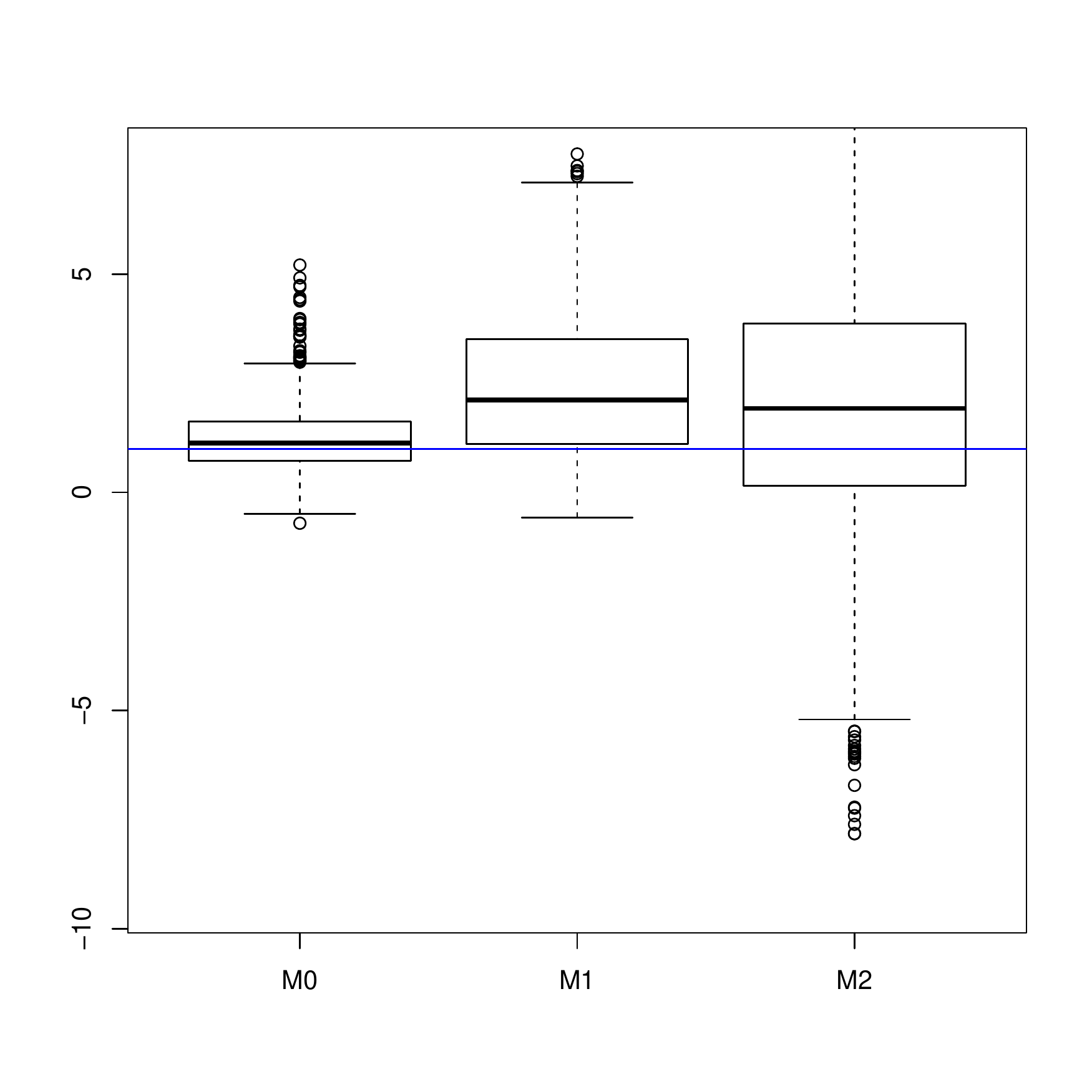}
&\includegraphics[height=2cm, width=3.5cm]{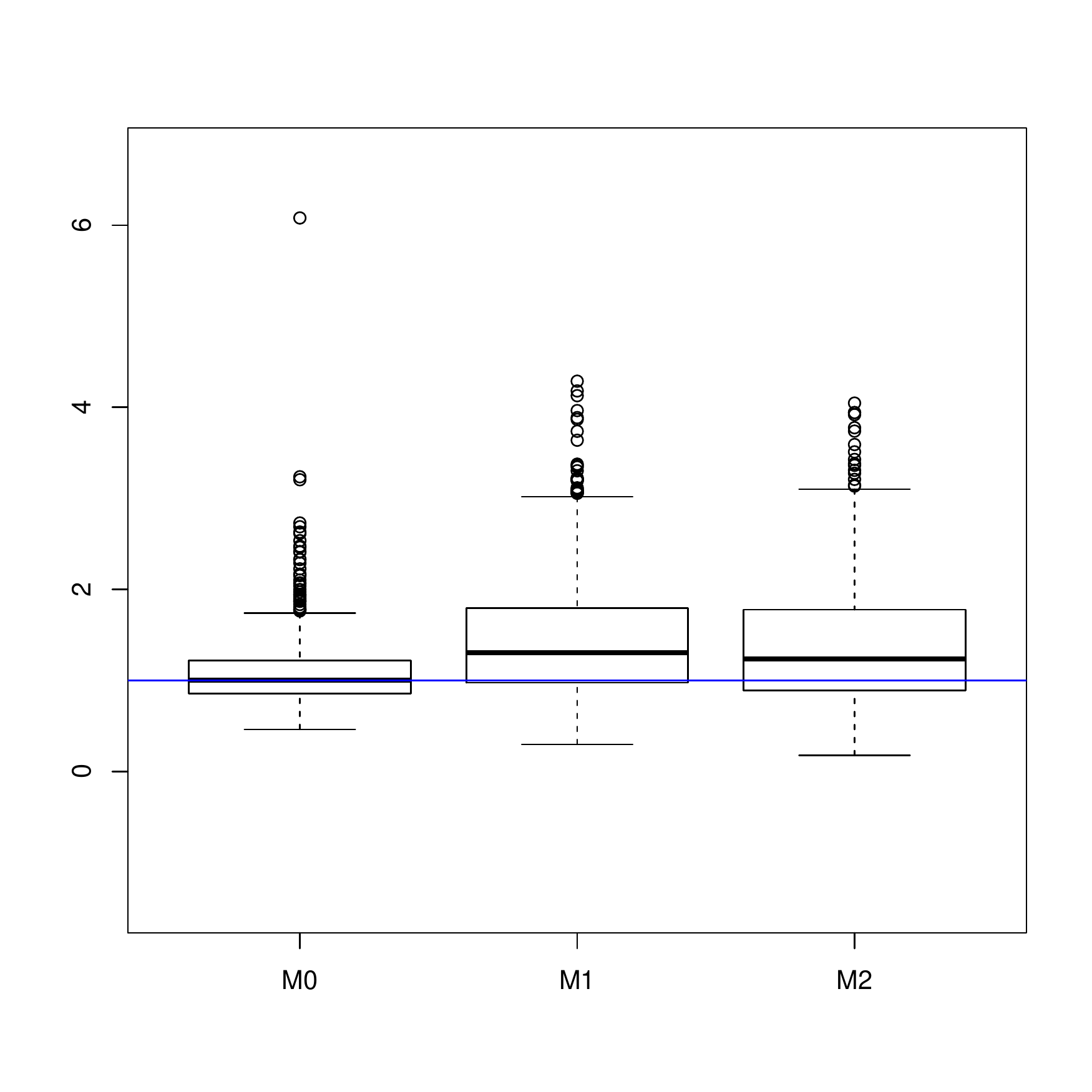}&\includegraphics[height=2cm, width=3.5cm]{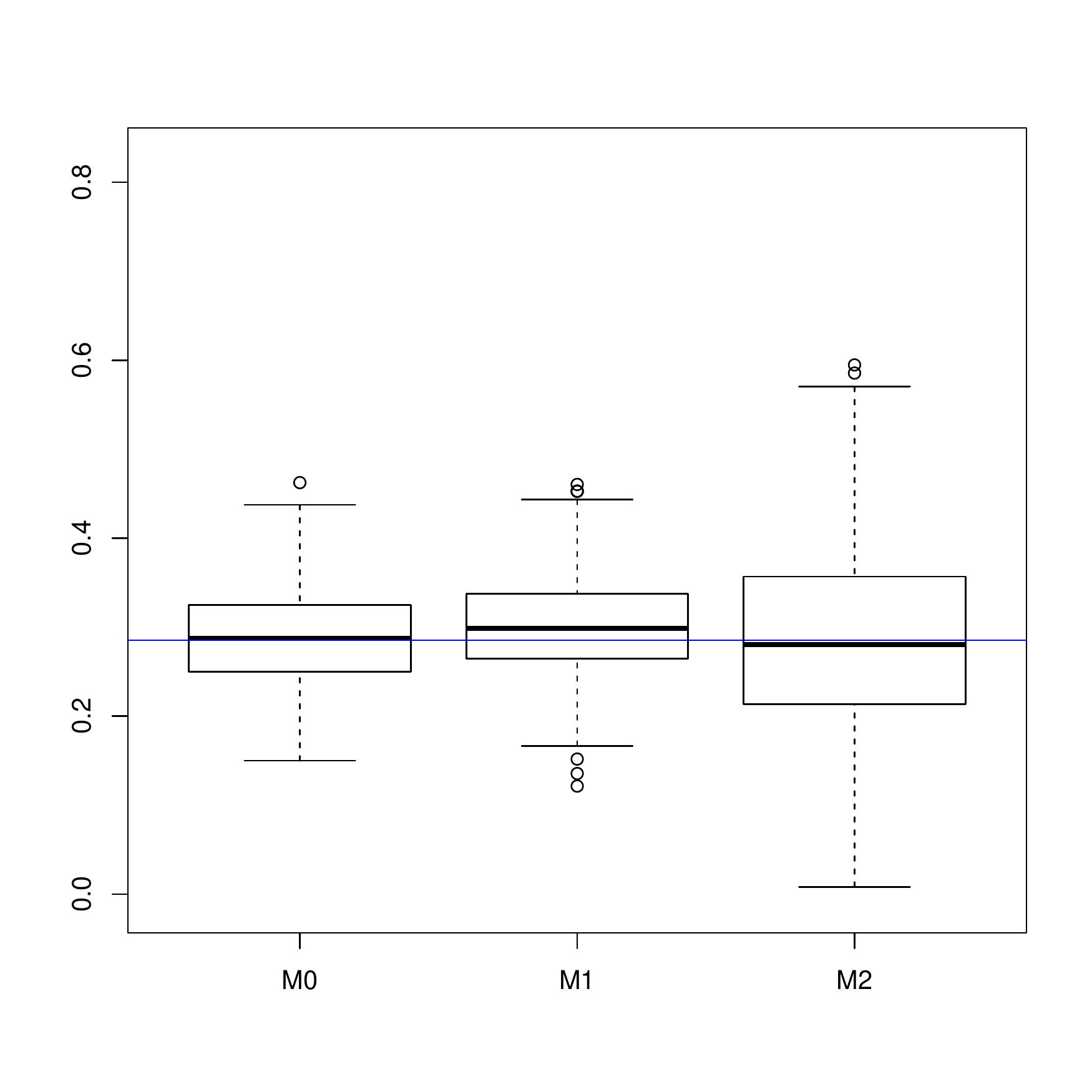}\\
\midrule
\raisebox{1cm} {200}&\includegraphics[height=2cm, width=3.5cm]{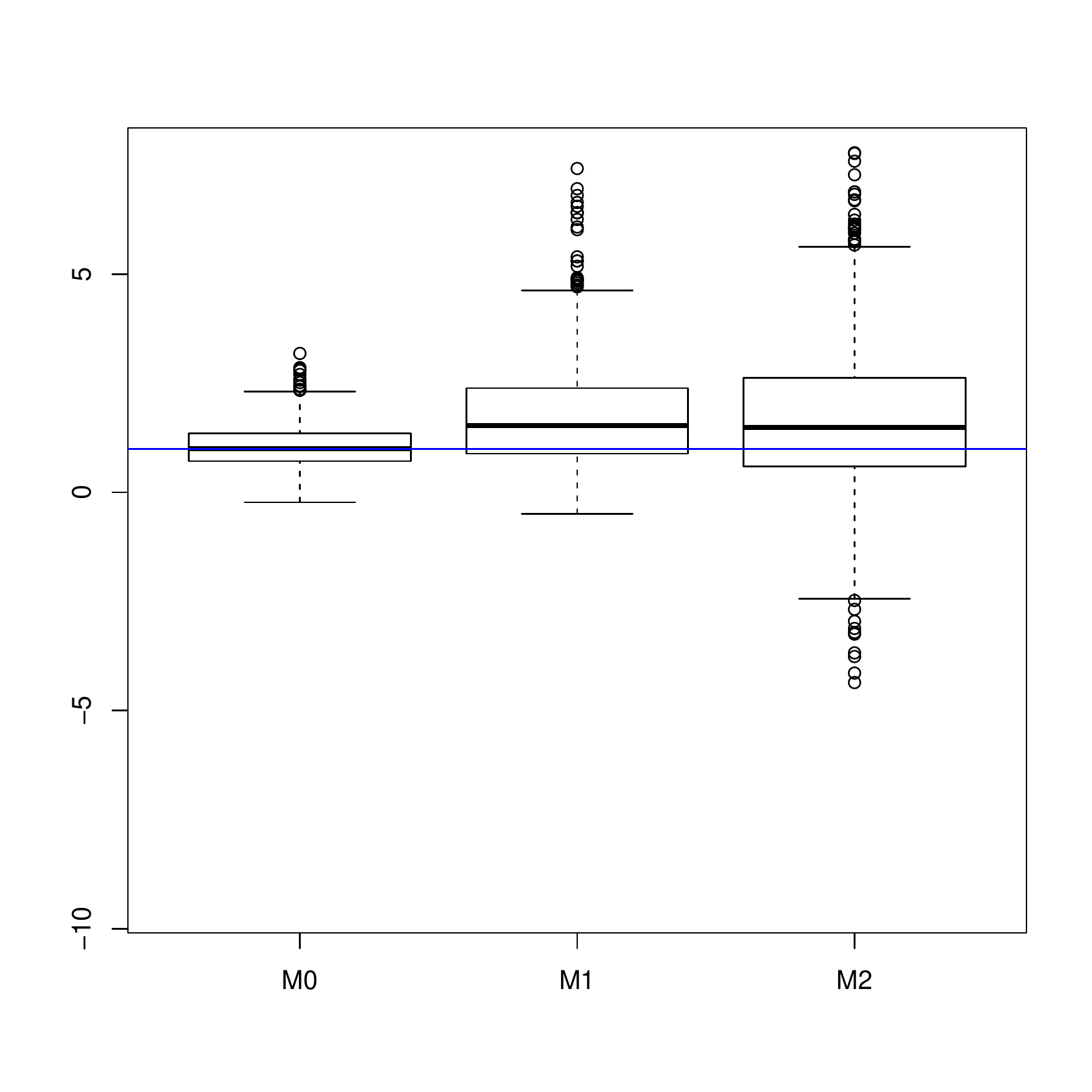}
&\includegraphics[height=2cm, width=3.5cm]{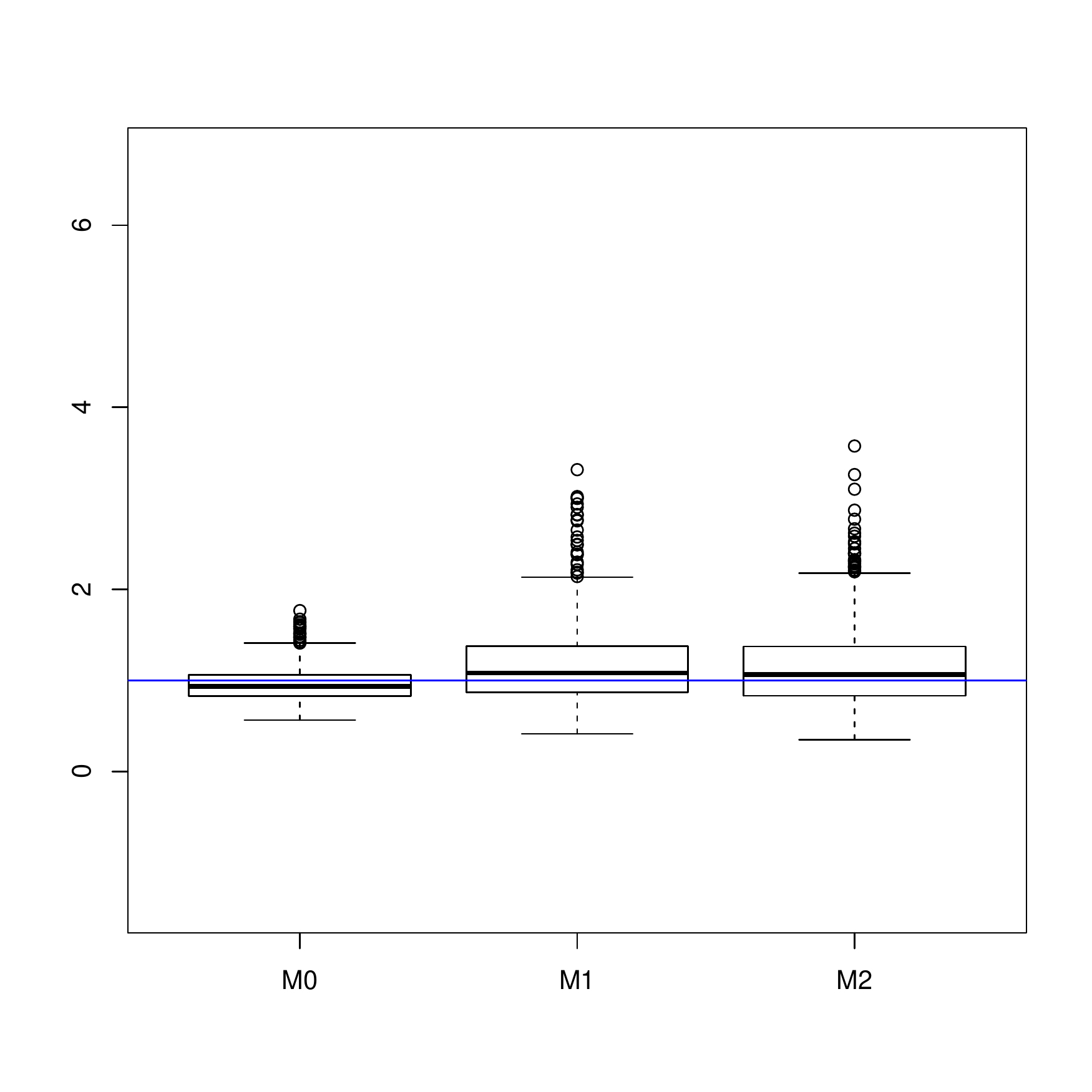}&\includegraphics[height=2cm, width=3.5cm]{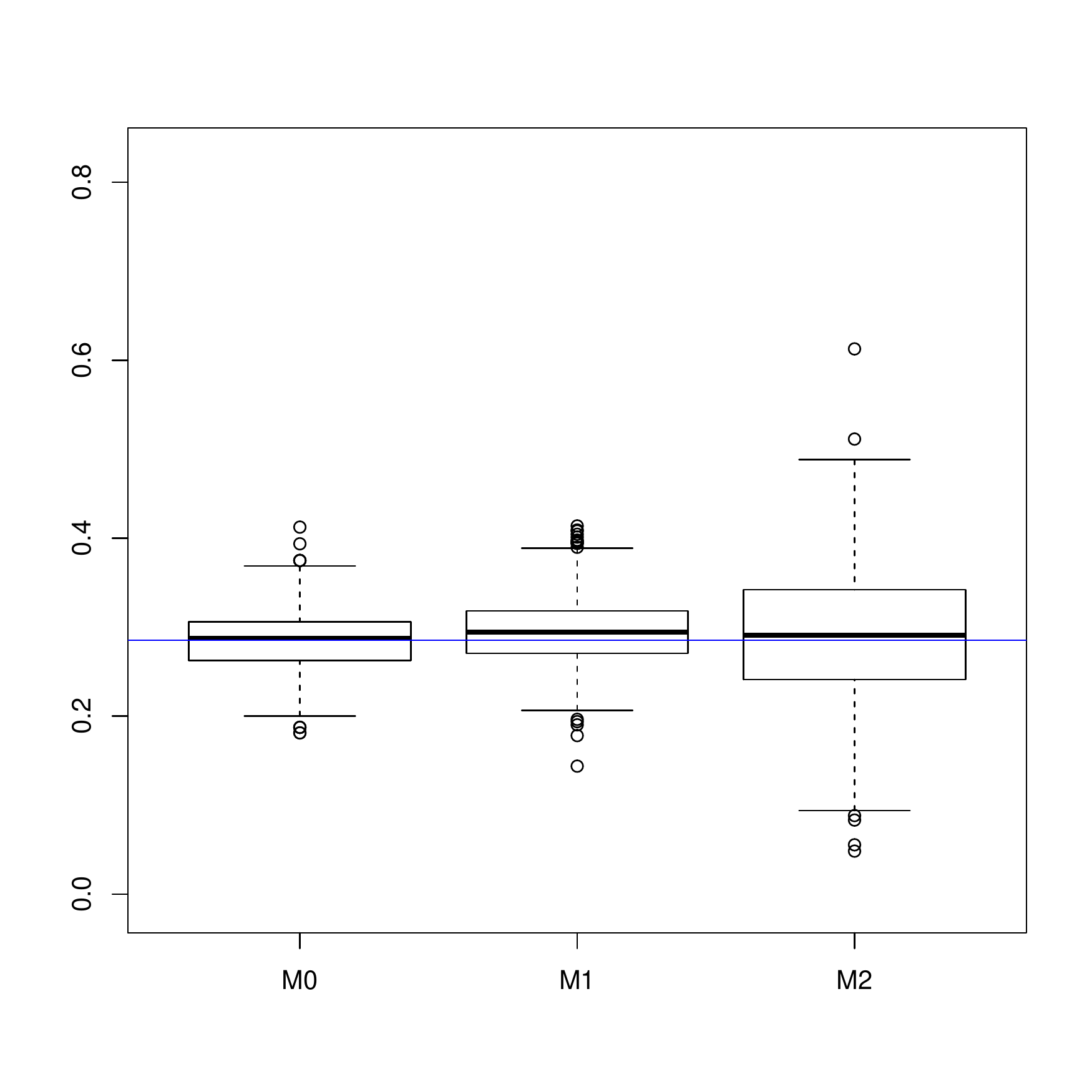}\\
\midrule
\raisebox{1cm} {500}&\includegraphics[height=2cm, width=3.5cm]{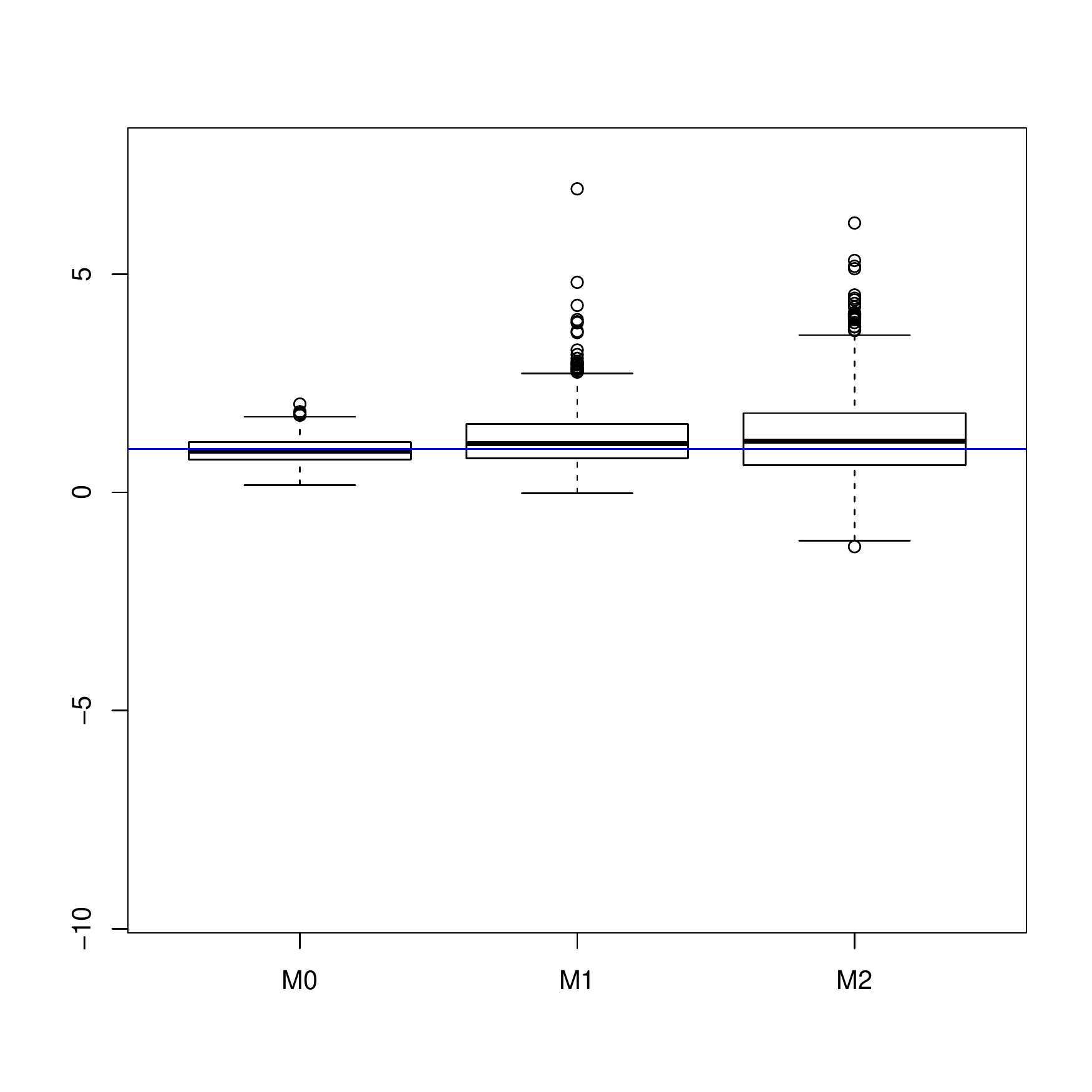}
&\includegraphics[height=2cm, width=3.5cm]{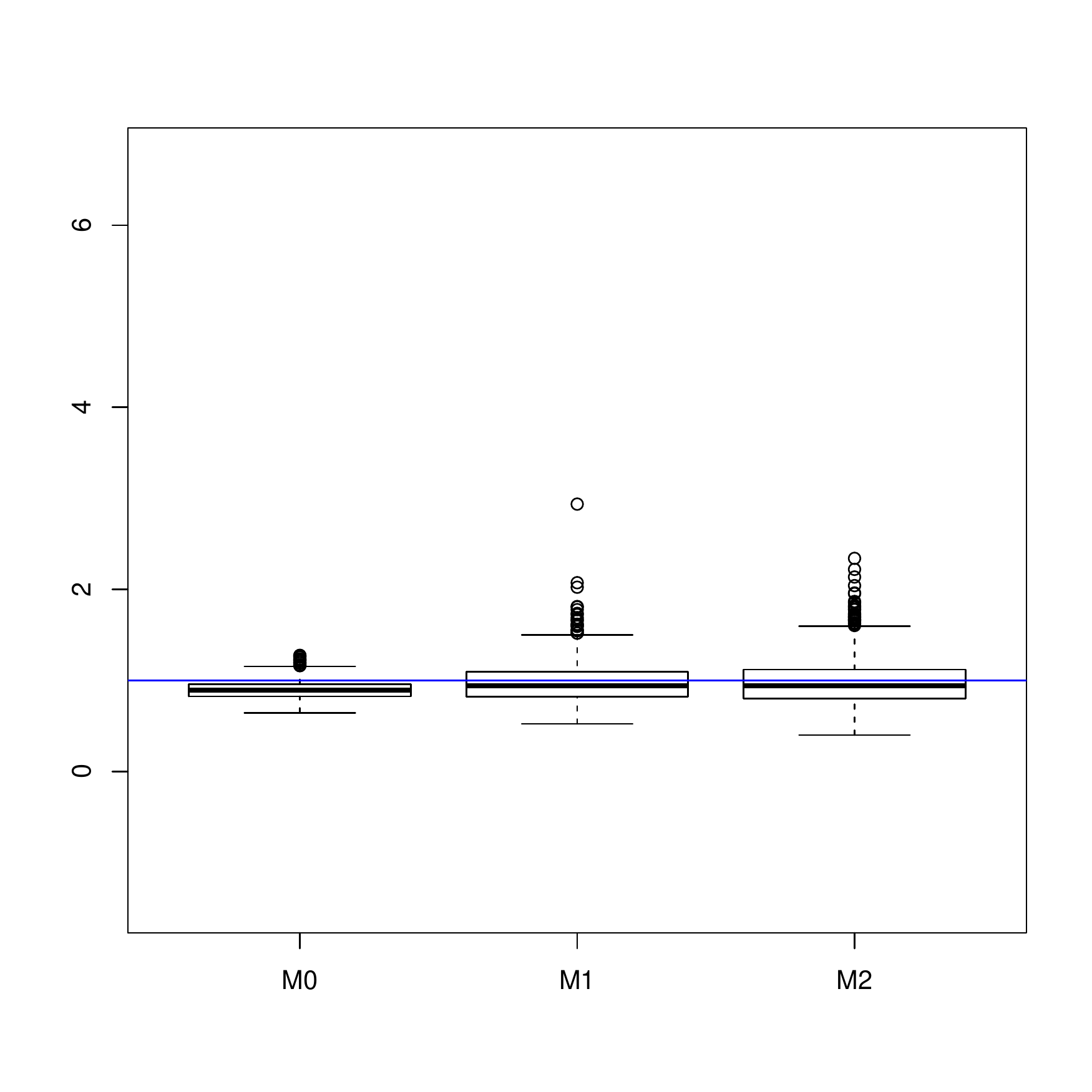}&\includegraphics[height=2cm, width=3.5cm]{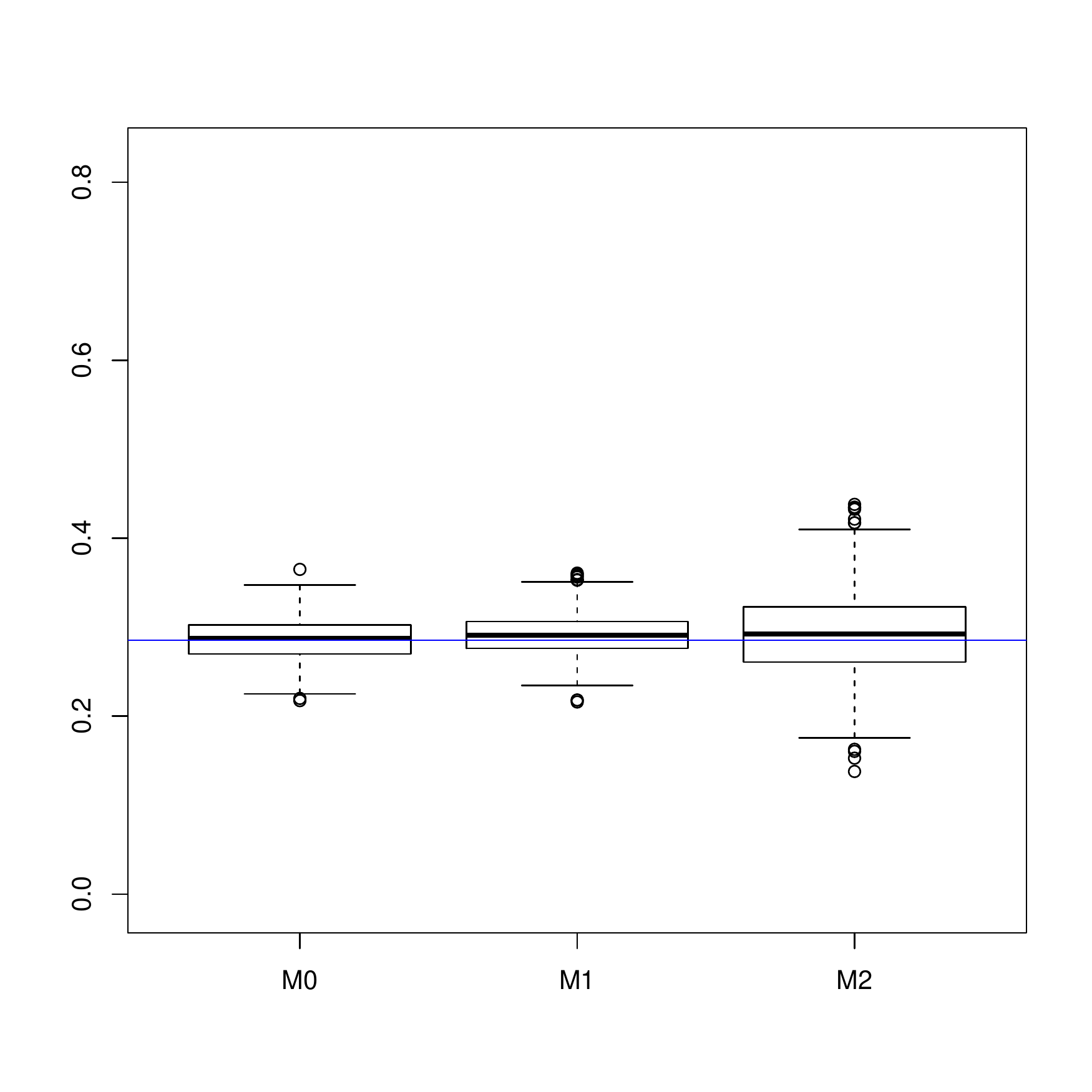}\\
\midrule
\raisebox{1cm}{1000}&\includegraphics[height=2cm, width=3.5cm]{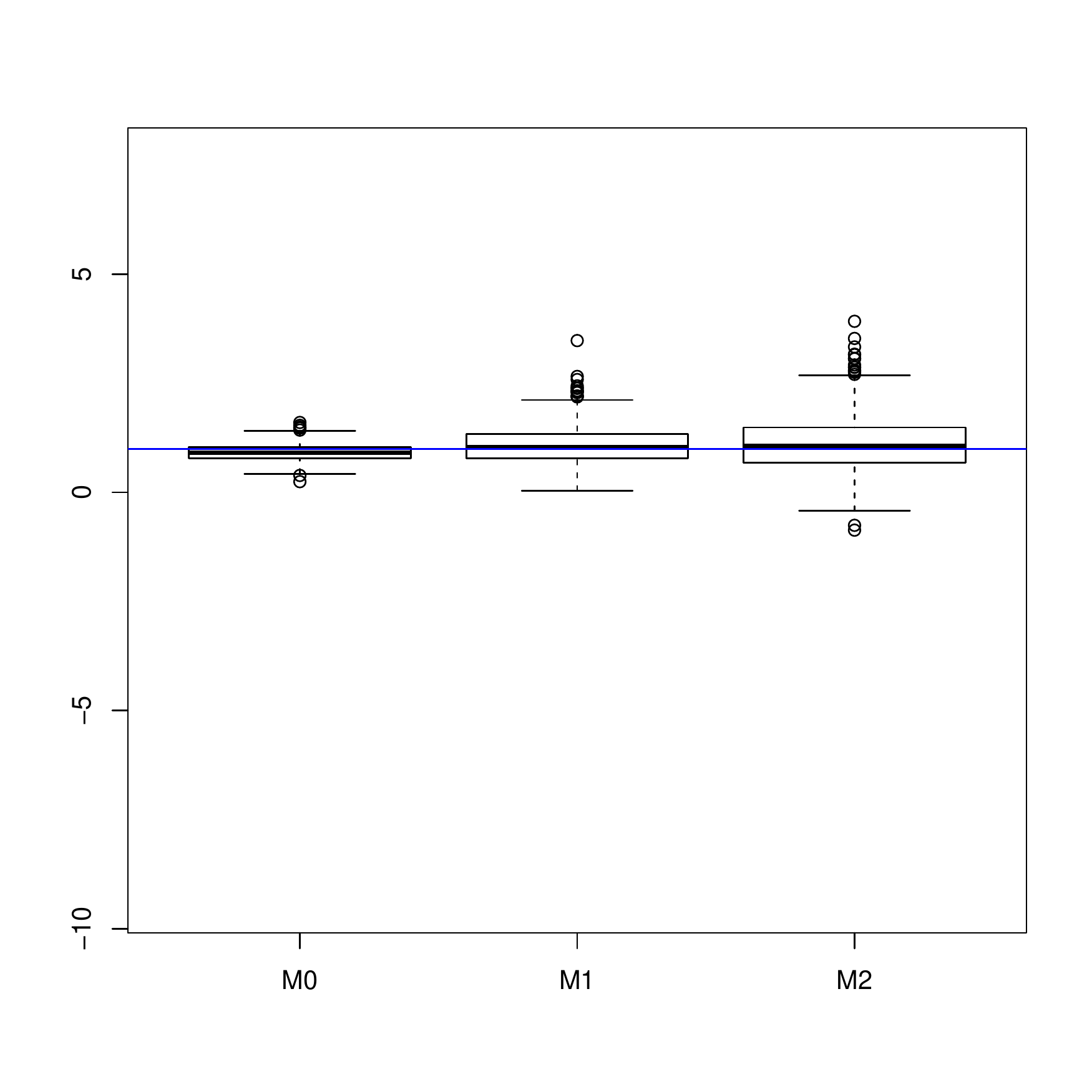}
&\includegraphics[height=2cm, width=3.5cm]{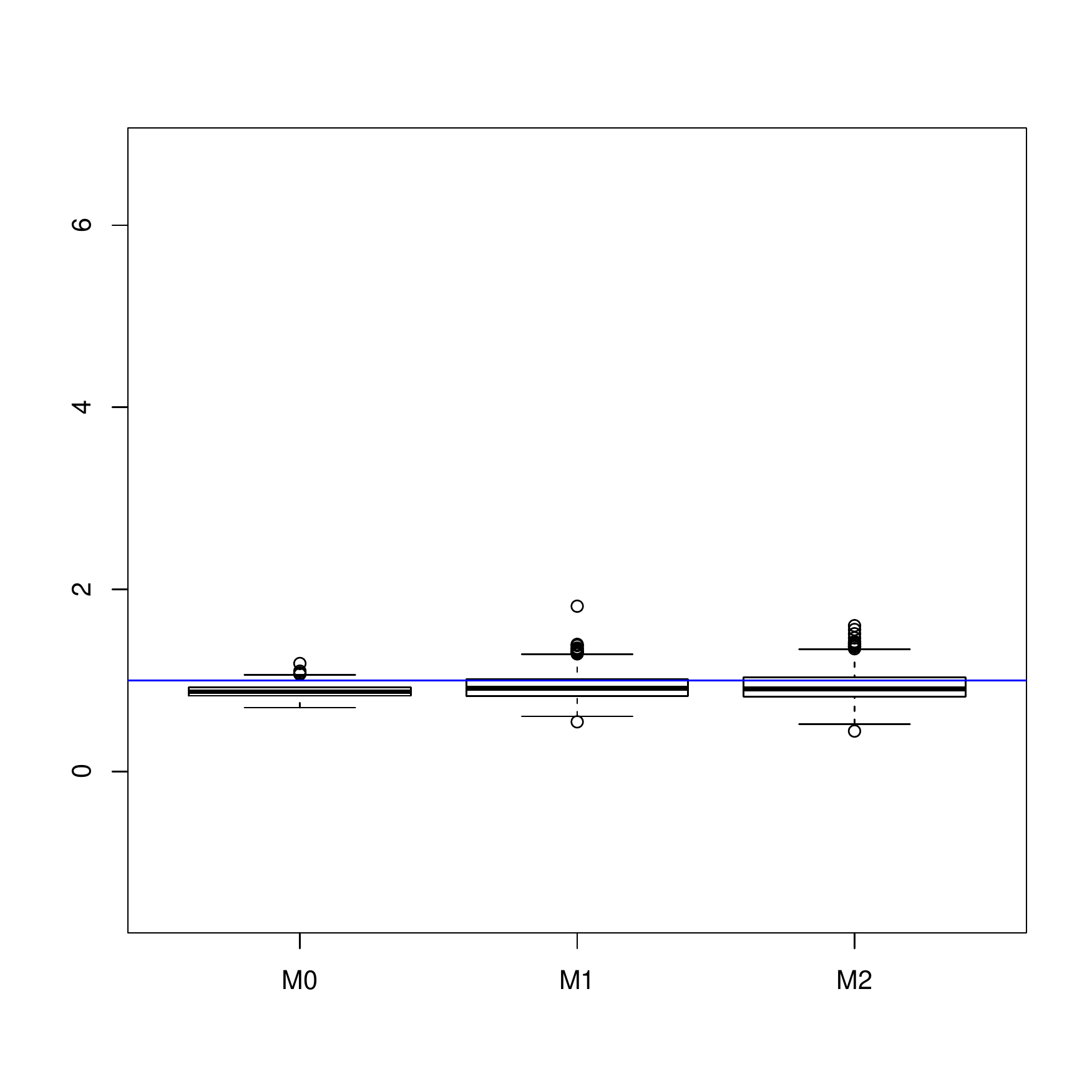}&\includegraphics[height=2cm, width=3.5cm]{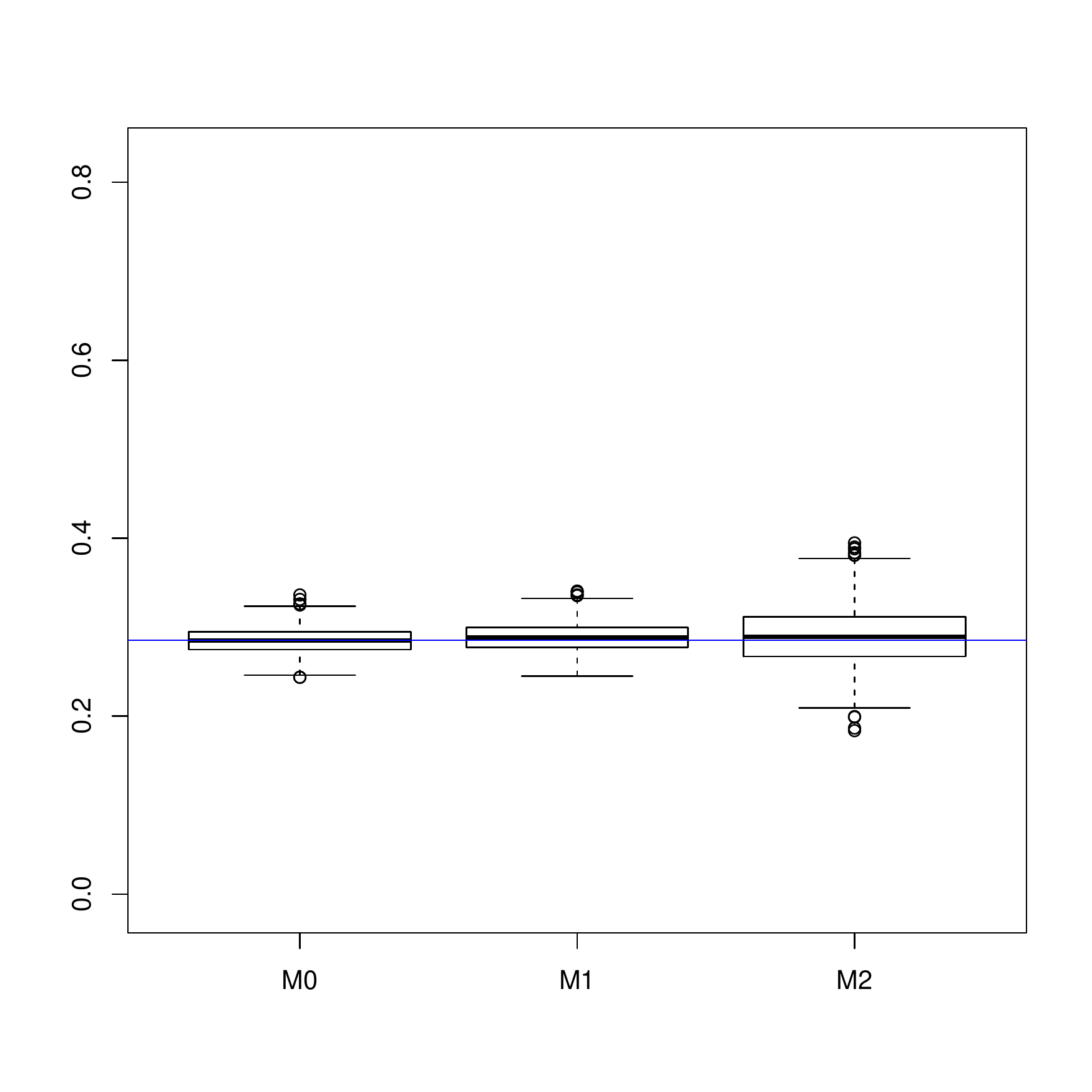}\\
\midrule
\raisebox{1cm}{1500}&\includegraphics[height=2cm, width=3.5cm]{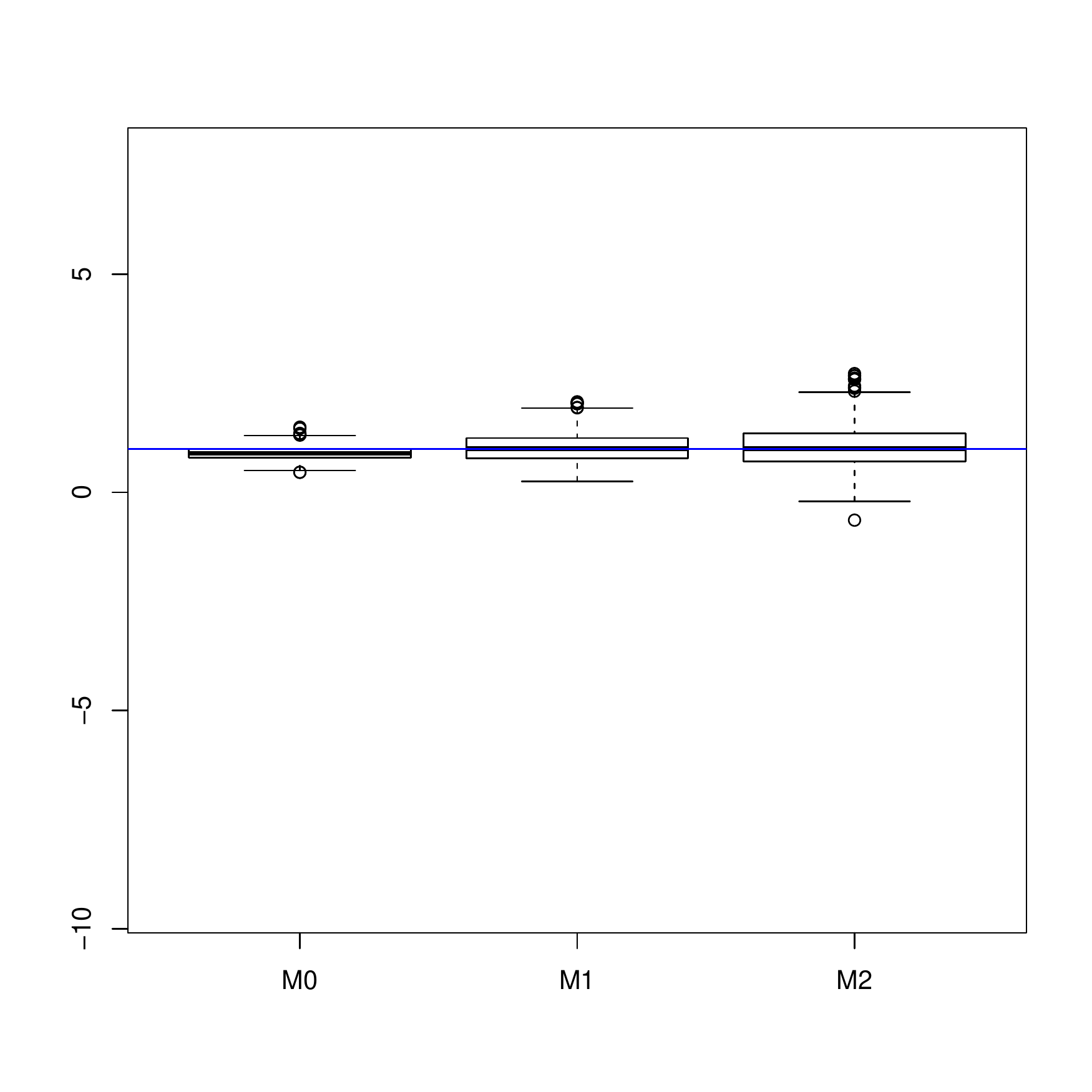}
&\includegraphics[height=2cm, width=3.5cm]{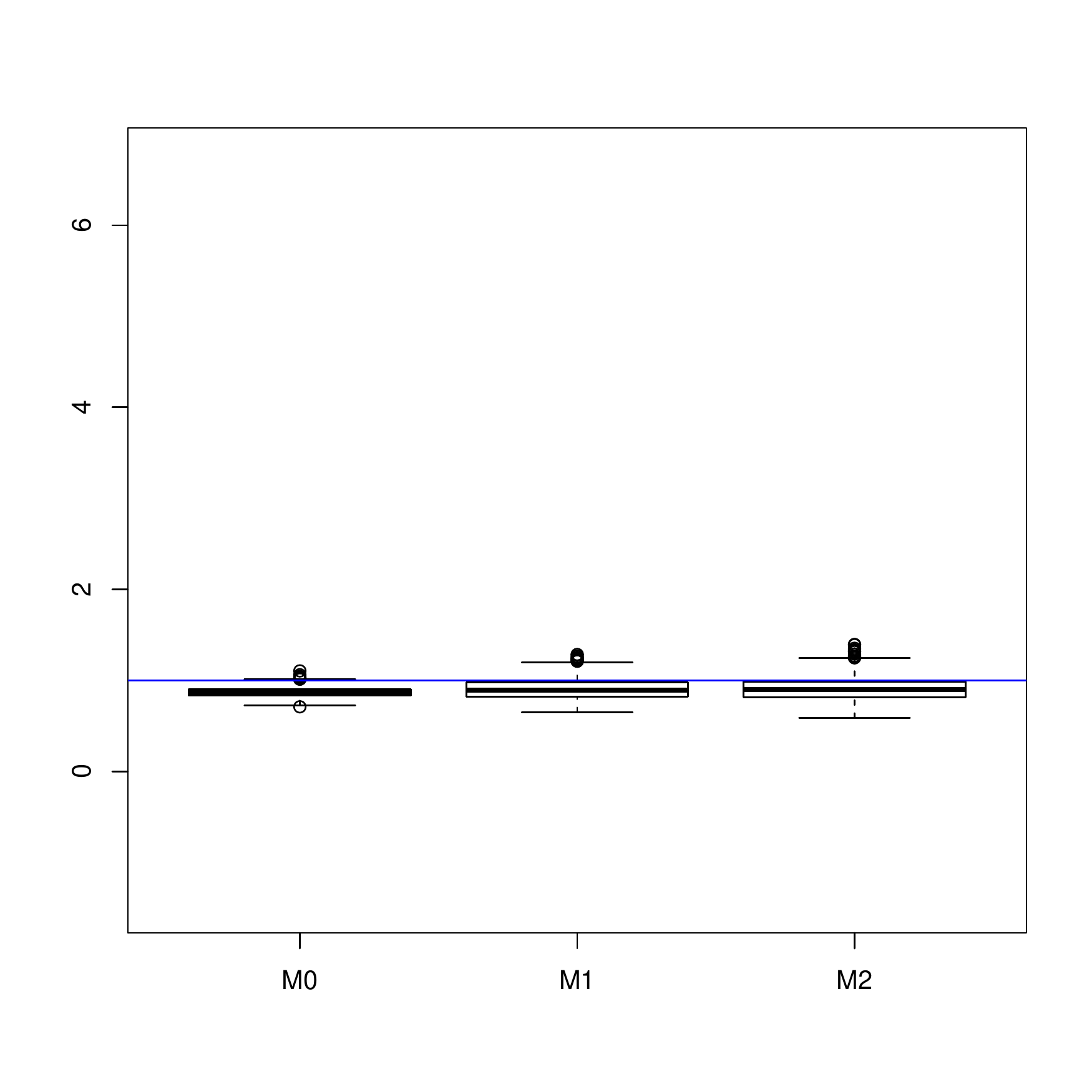}&\includegraphics[height=2cm, width=3.5cm]{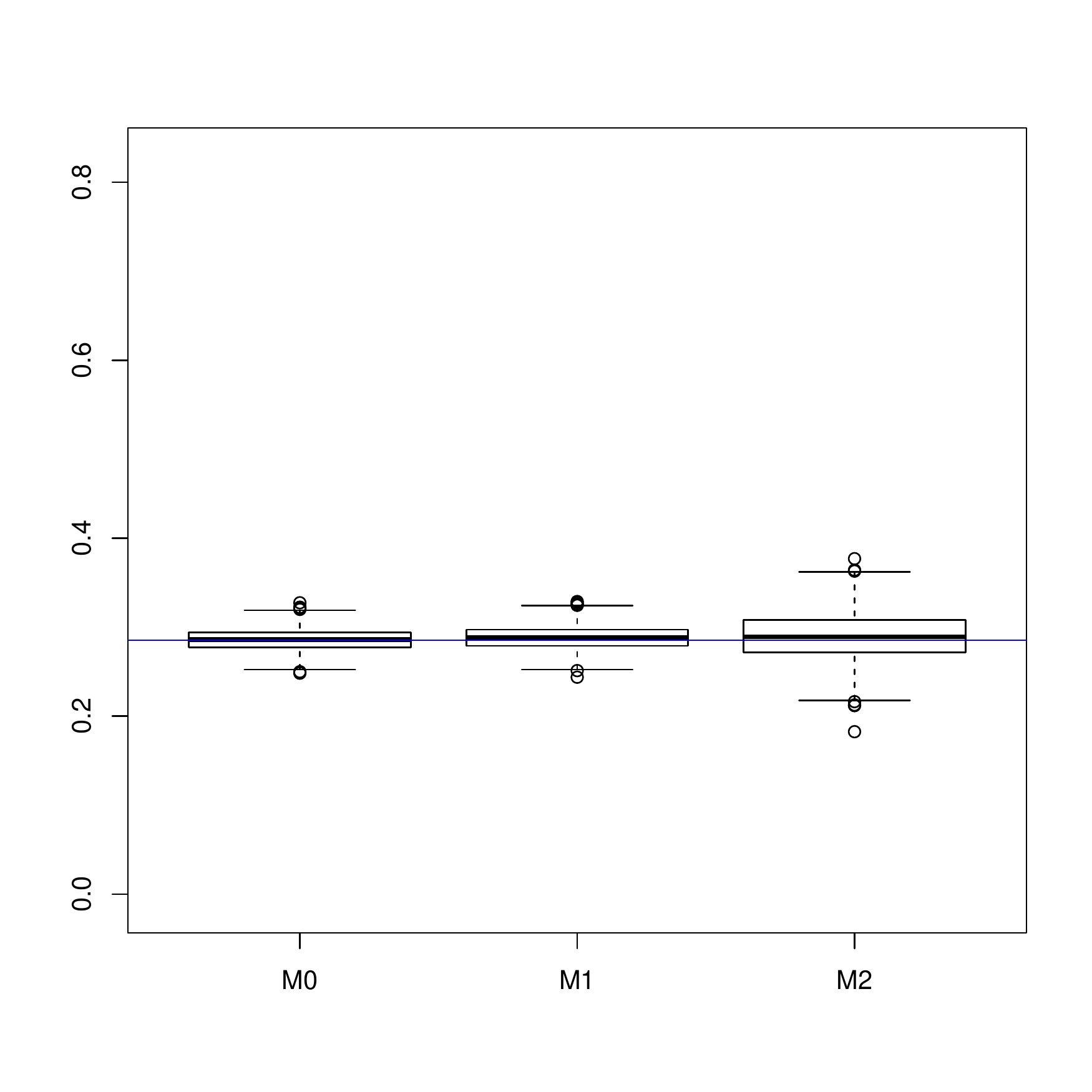}\\
\midrule
\raisebox{1cm}{2000}&\includegraphics[height=2cm, width=3.5cm]{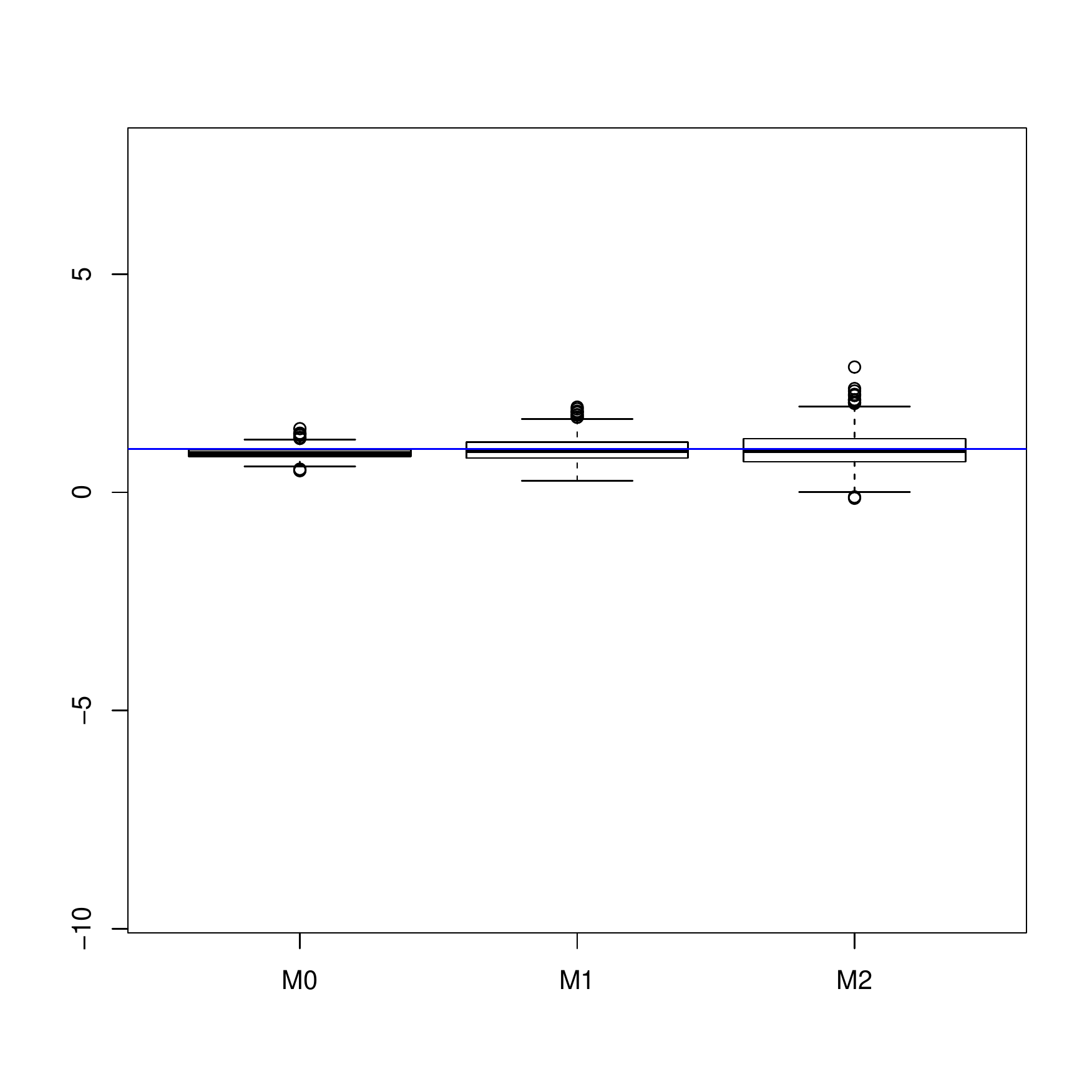}
&\includegraphics[height=2cm, width=3.5cm]{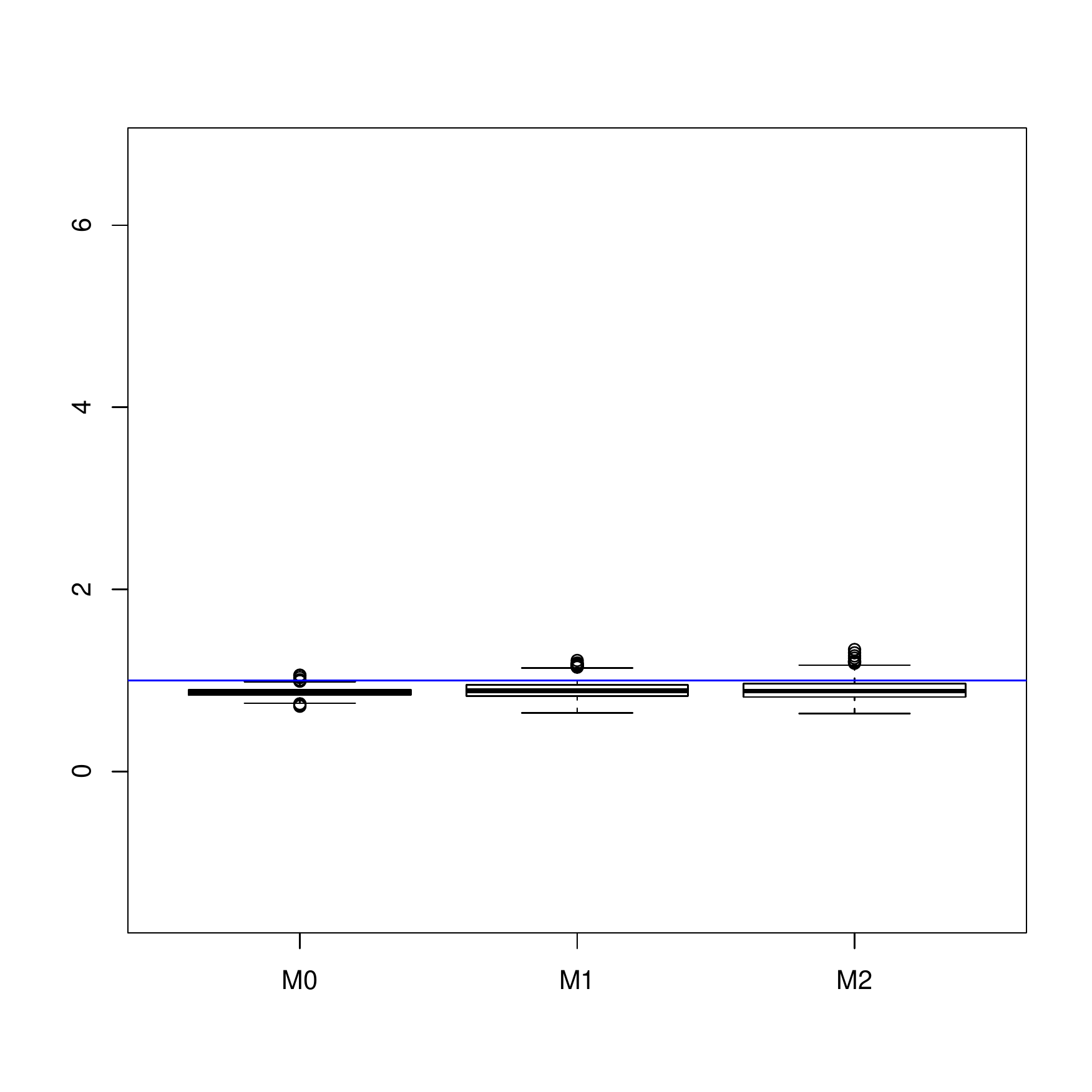}&\includegraphics[height=2cm, width=3.5cm]{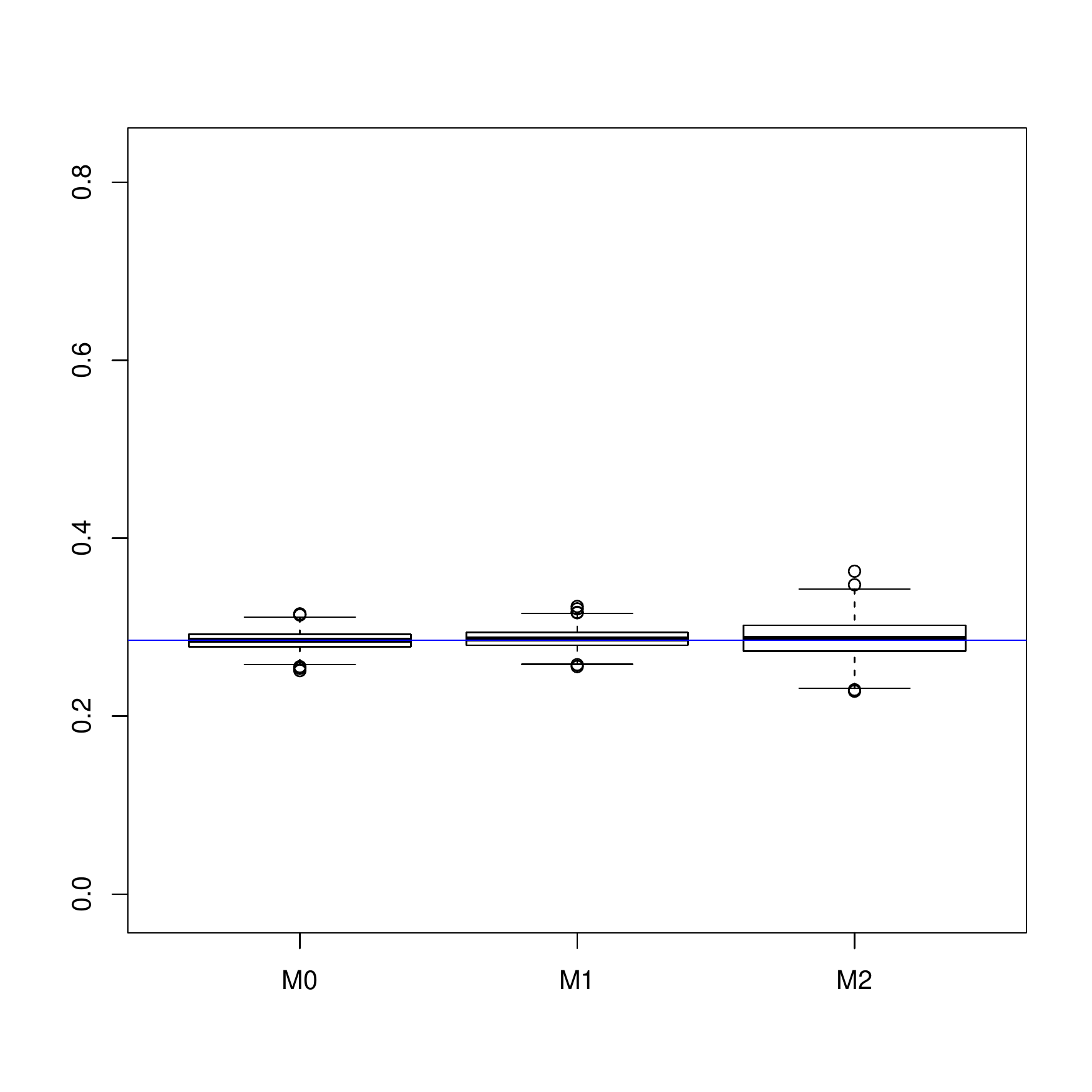}\\
\midrule
\raisebox{1cm}{3000}&\includegraphics[height=2cm, width=3.5cm]{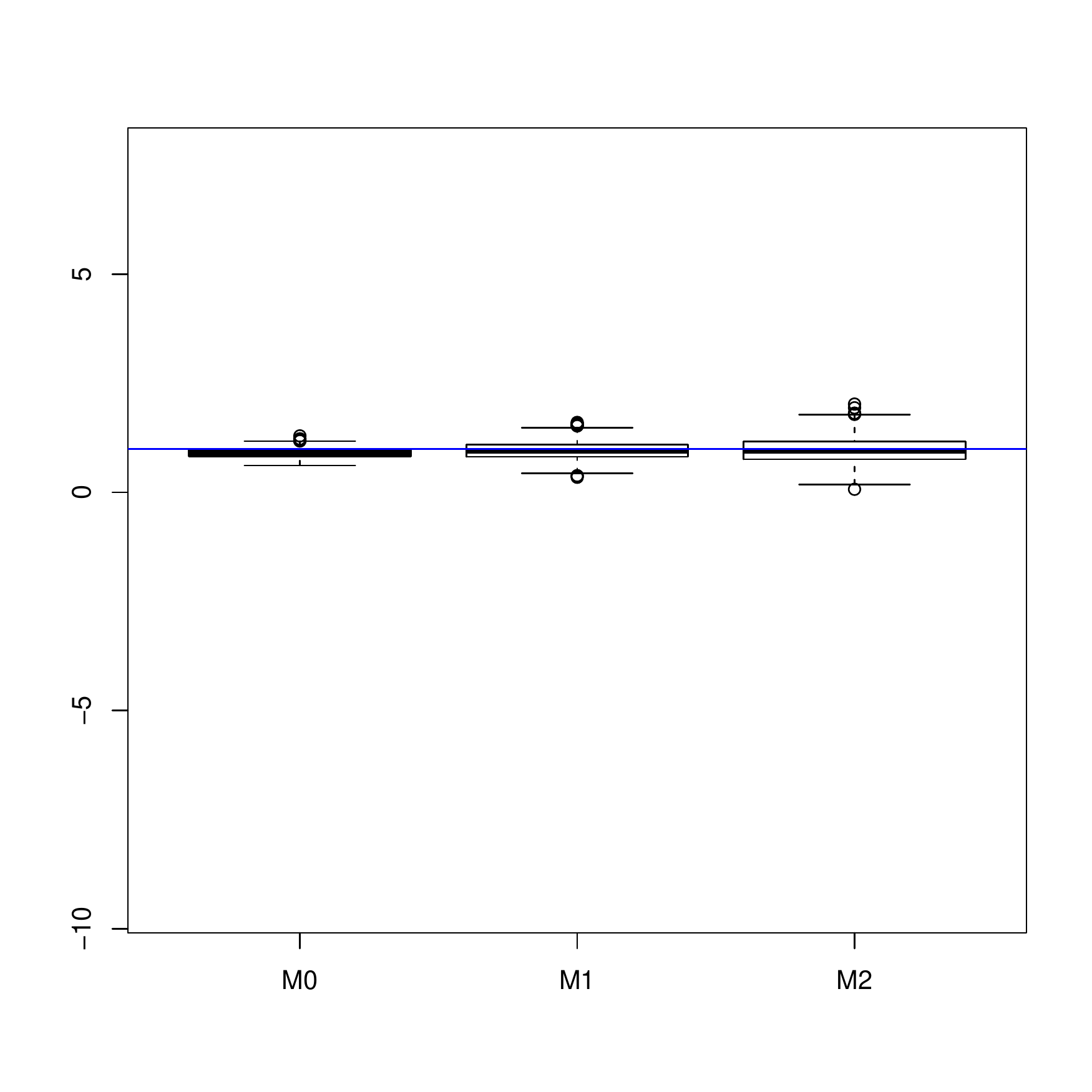}
&\includegraphics[height=2cm, width=3.5cm]{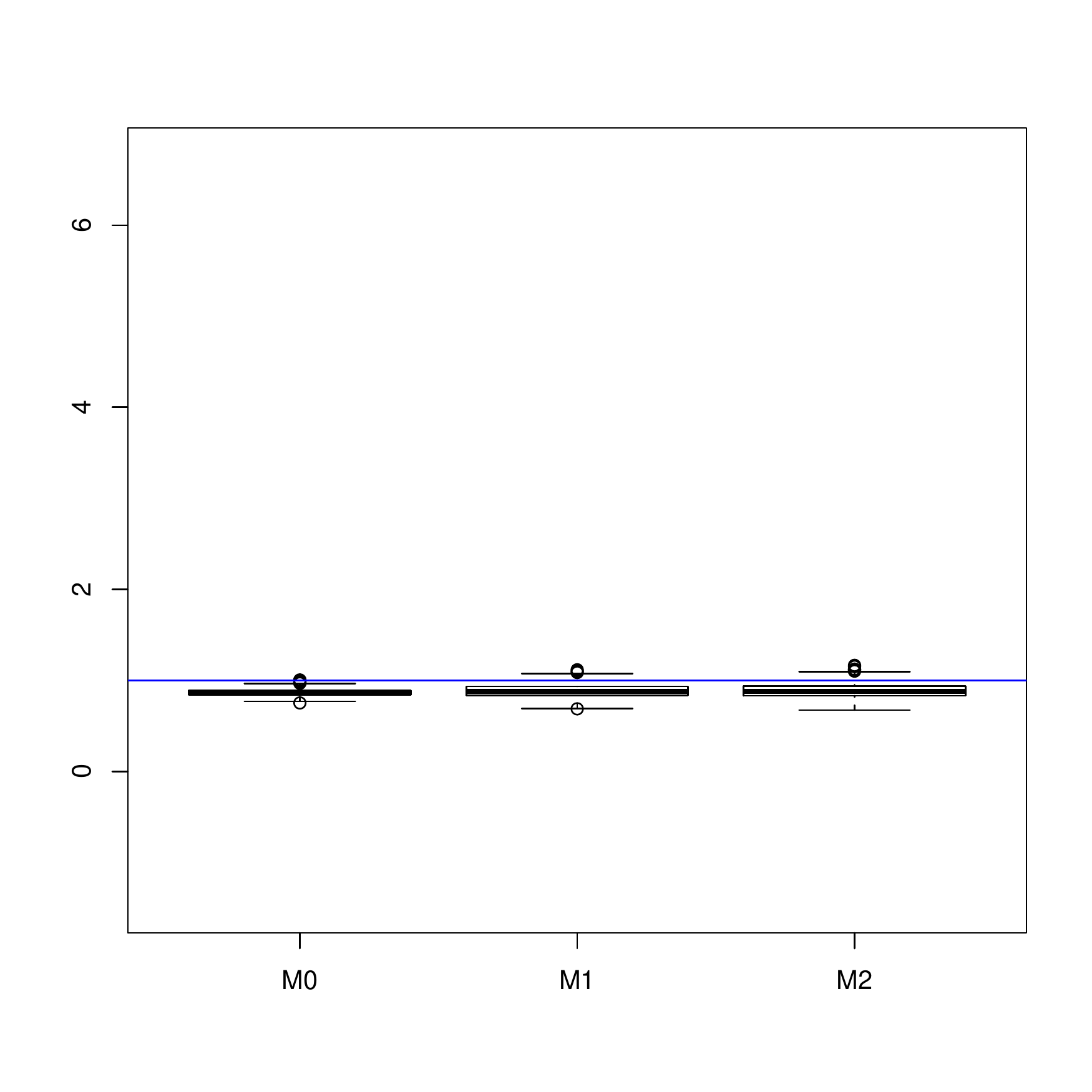}&\includegraphics[height=2cm, width=3.5cm]{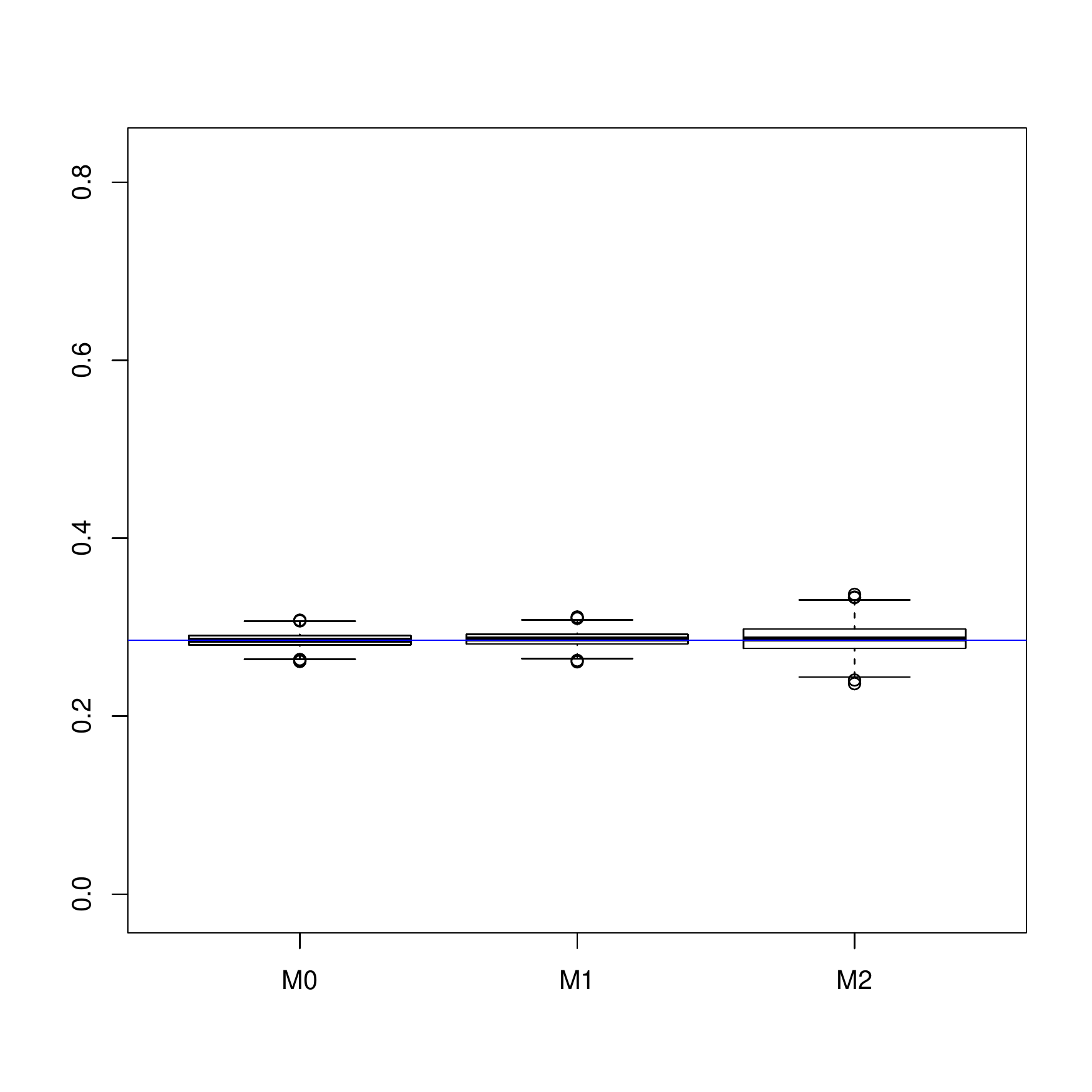}\\
\bottomrule
\end{tabular}
\caption{Scenario (iii): boxplots of simulations with increasing sample sizes and different models($M_0$,$M_1$ and $M_2$).}\label{fig:box}
\end{figure}

\begin{figure}[H]
\centering
\begin{tabular}{c}
\includegraphics[height=6cm, width=12cm]{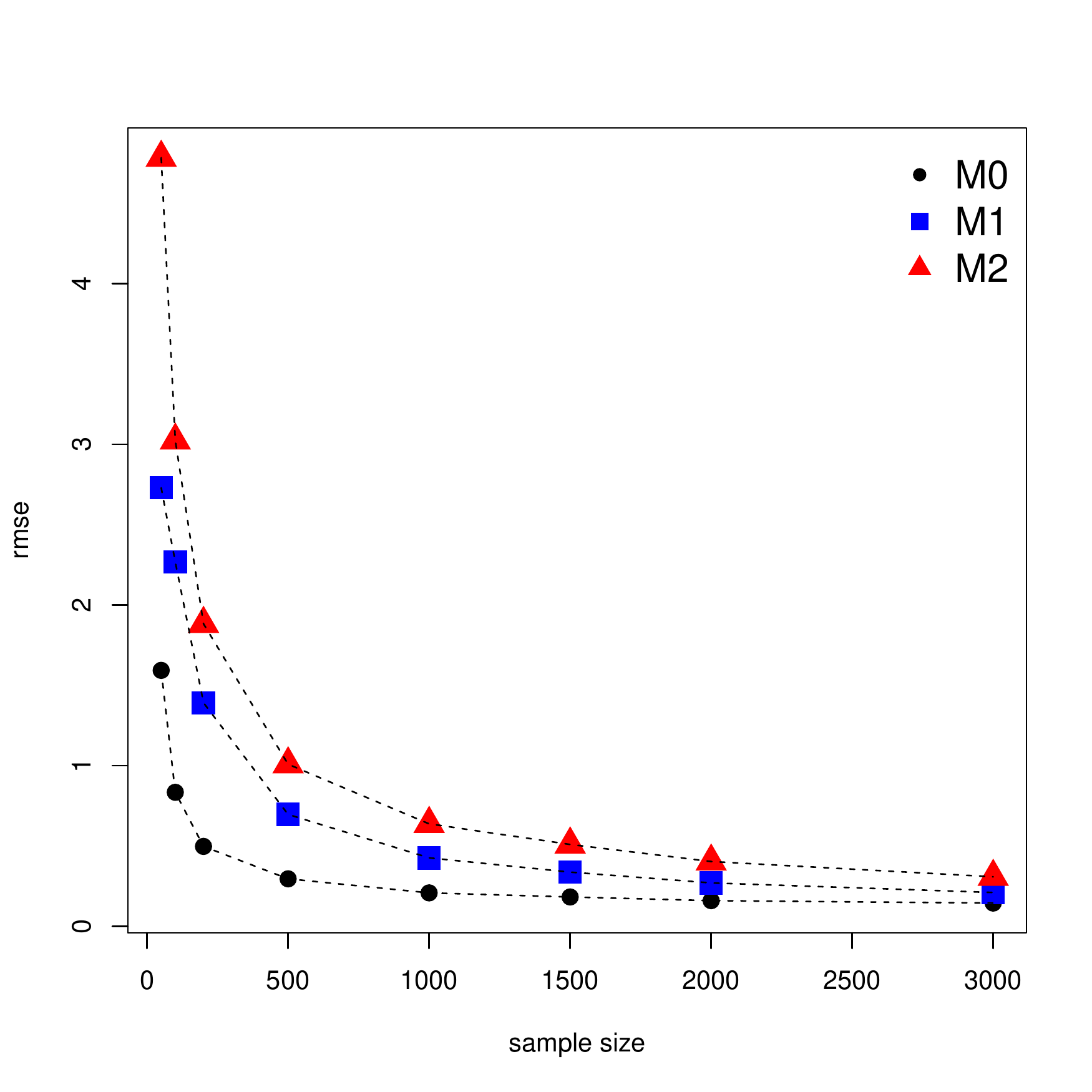} \\
(a)\\
\includegraphics[height=6cm, width=12cm]{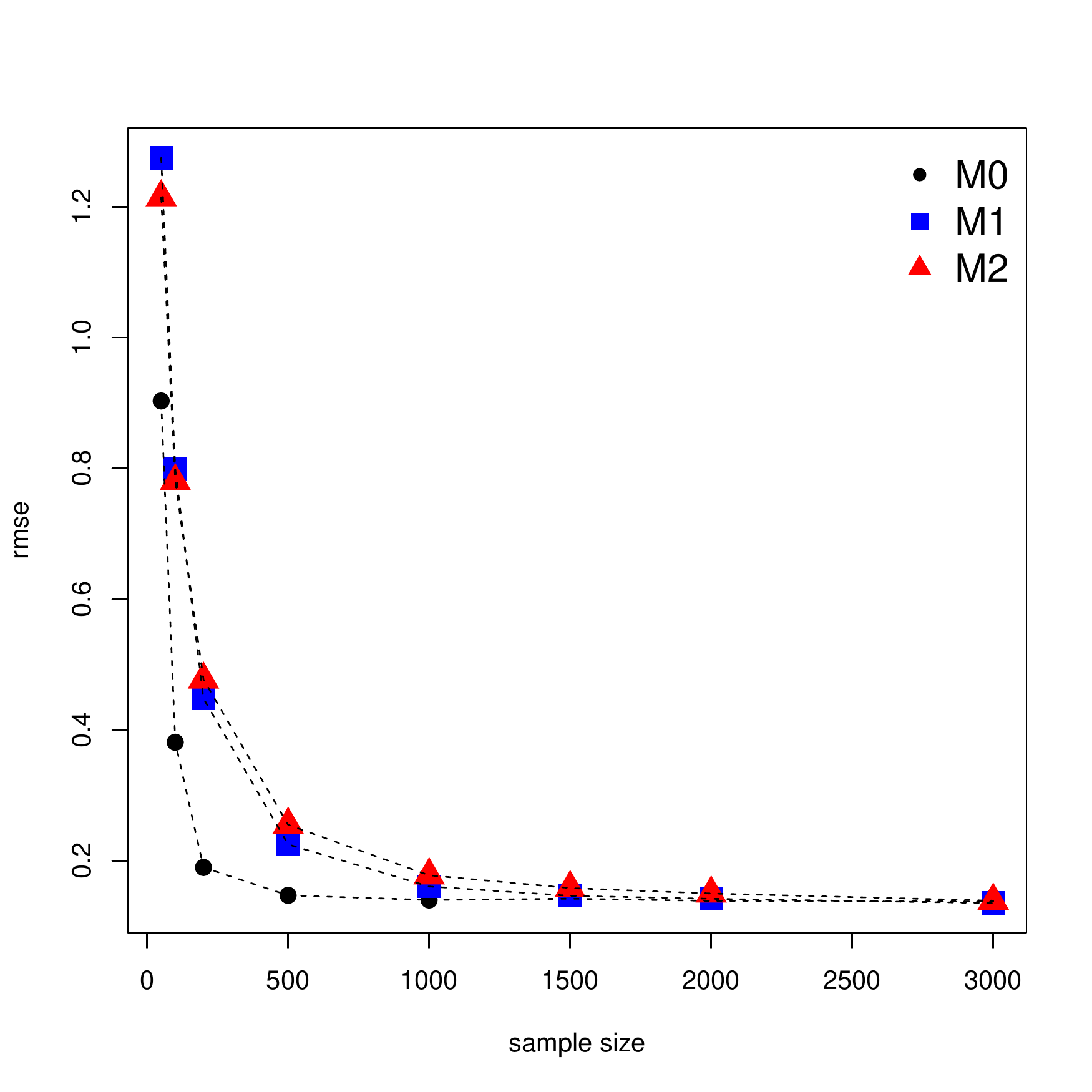} \\
(b)\\
\includegraphics[height=6cm, width=12cm]{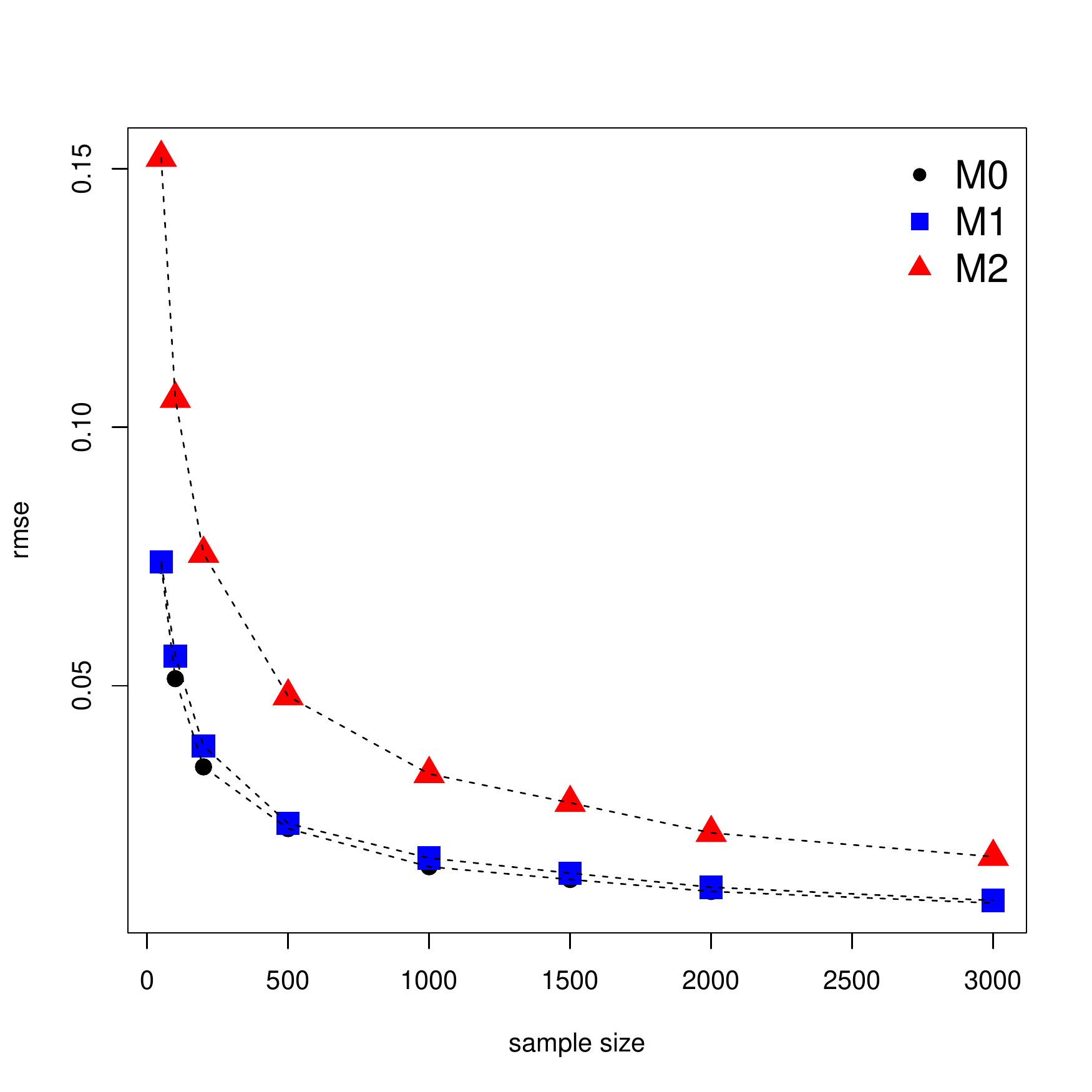} \\
(c)
\end{tabular}
\caption{Scenario (iii): root mean squared errors for different models ($M_0$,$M_1$ and $M_2$) over the 1000 replications, plots with increasing sample sizes for $\beta_0$ (a), $\beta_1$ (b) and $\pi$ (c). Dashed trajectories are reported to show the patterns.}\label{fig:rmse}
\end{figure}

\begin{figure}[H]
\centering
\begin{tabular}{cccc}
\toprule
$n$&$M_0$&$M_1$&$M_2$\\
\toprule
\raisebox{1cm}{50}&\includegraphics[height=2cm, width=3.5cm]{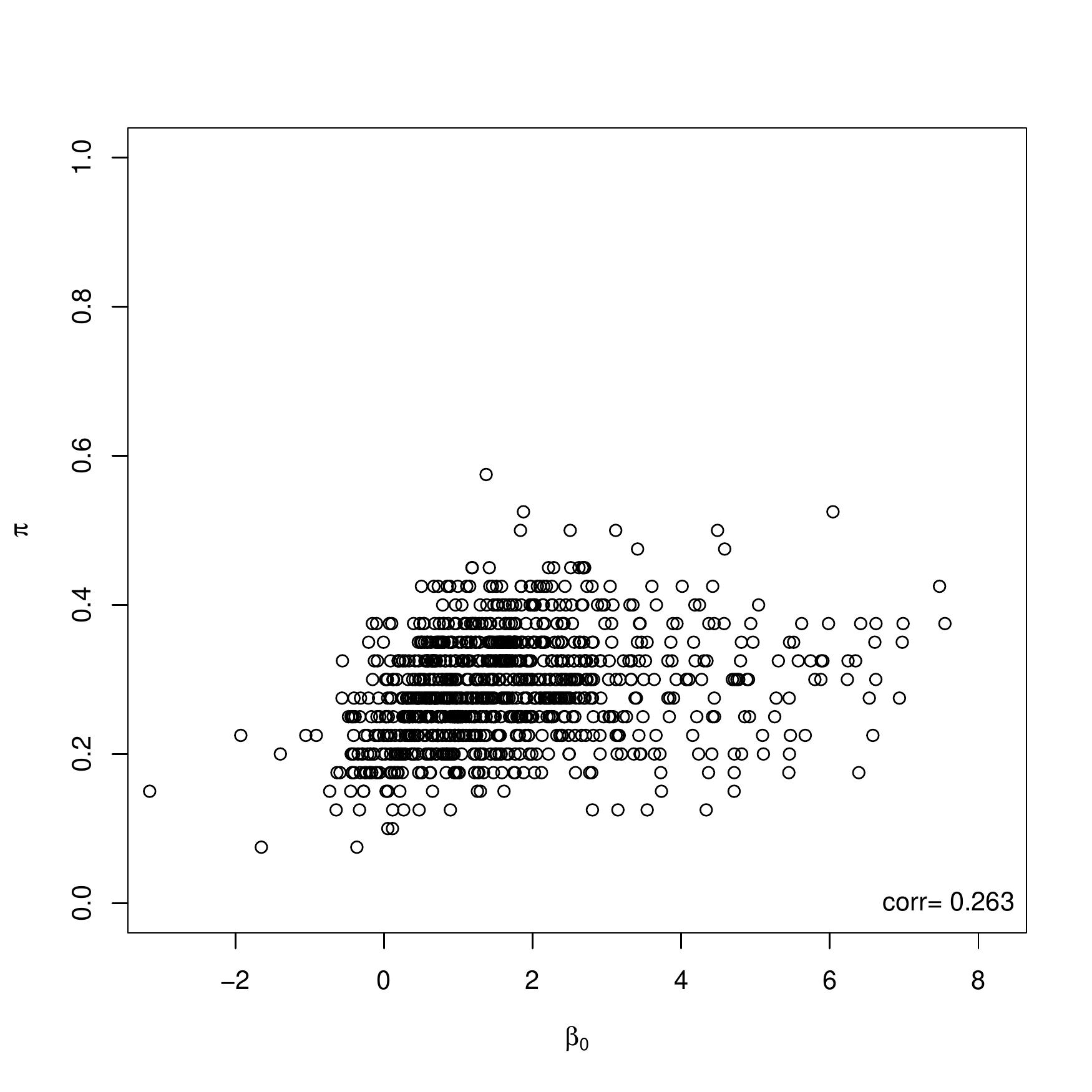}&\includegraphics[height=2cm, width=3.5cm]{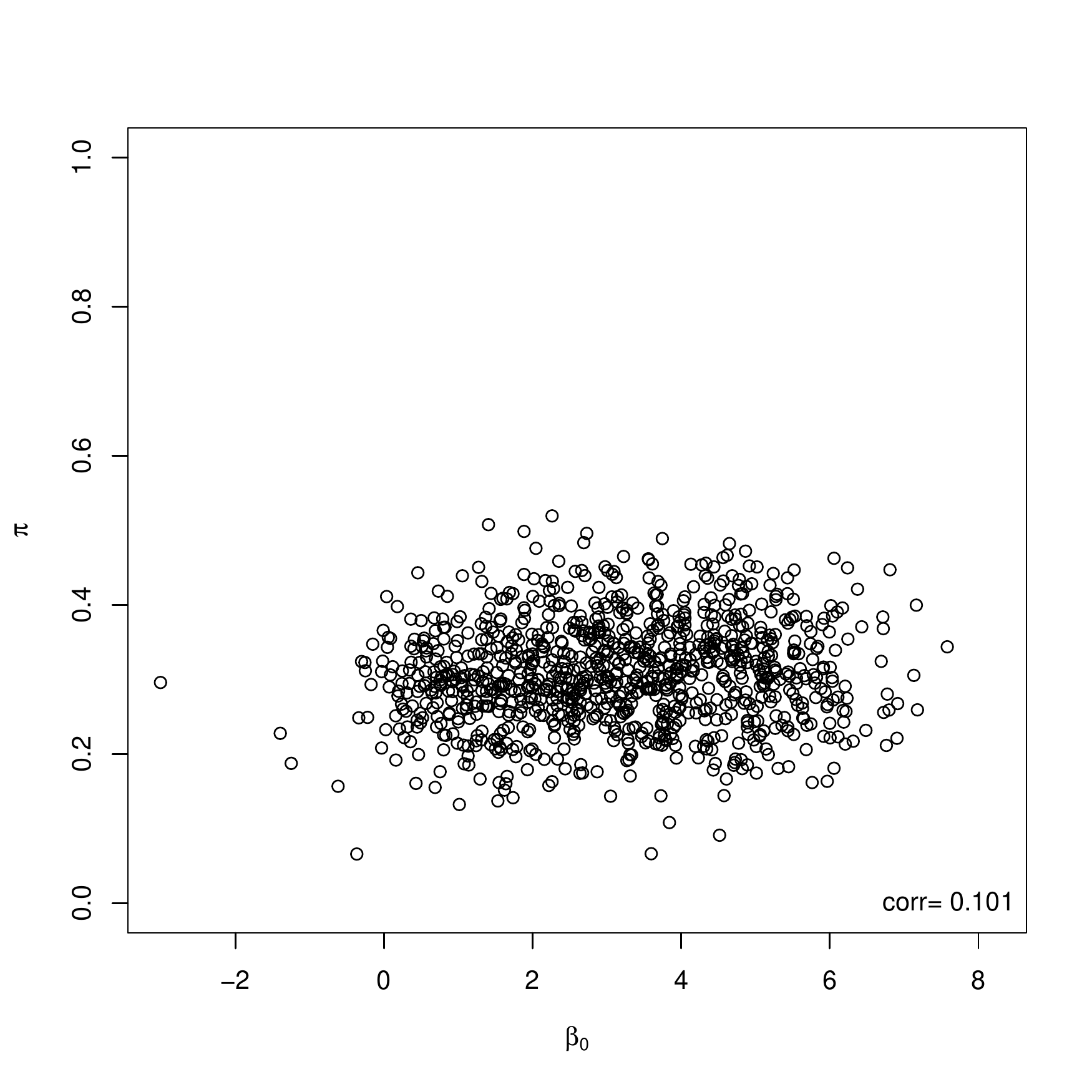}&\includegraphics[height=2cm, width=3.5cm]{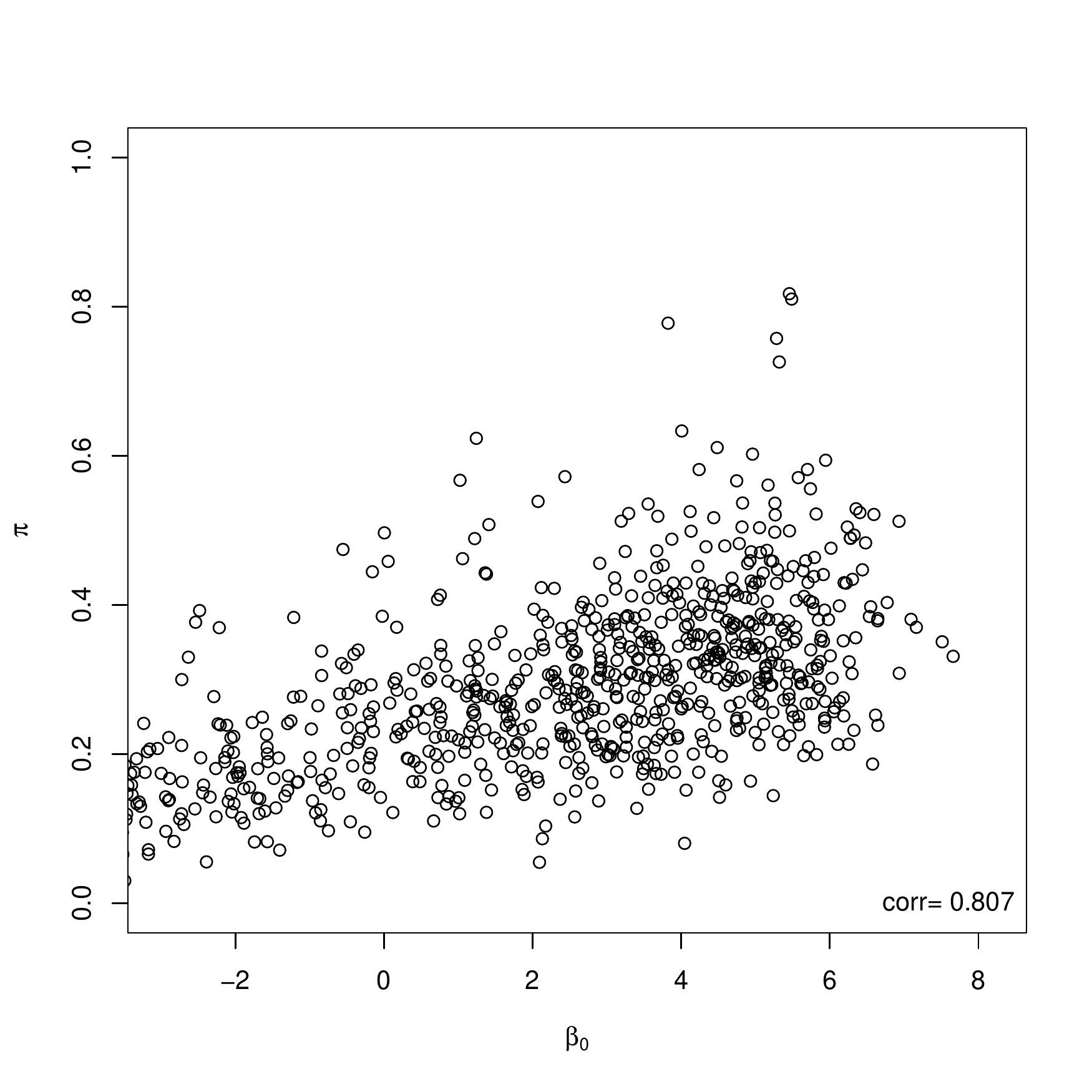}\\
\midrule
\raisebox{1cm}{100}&\includegraphics[height=2cm, width=3.5cm]{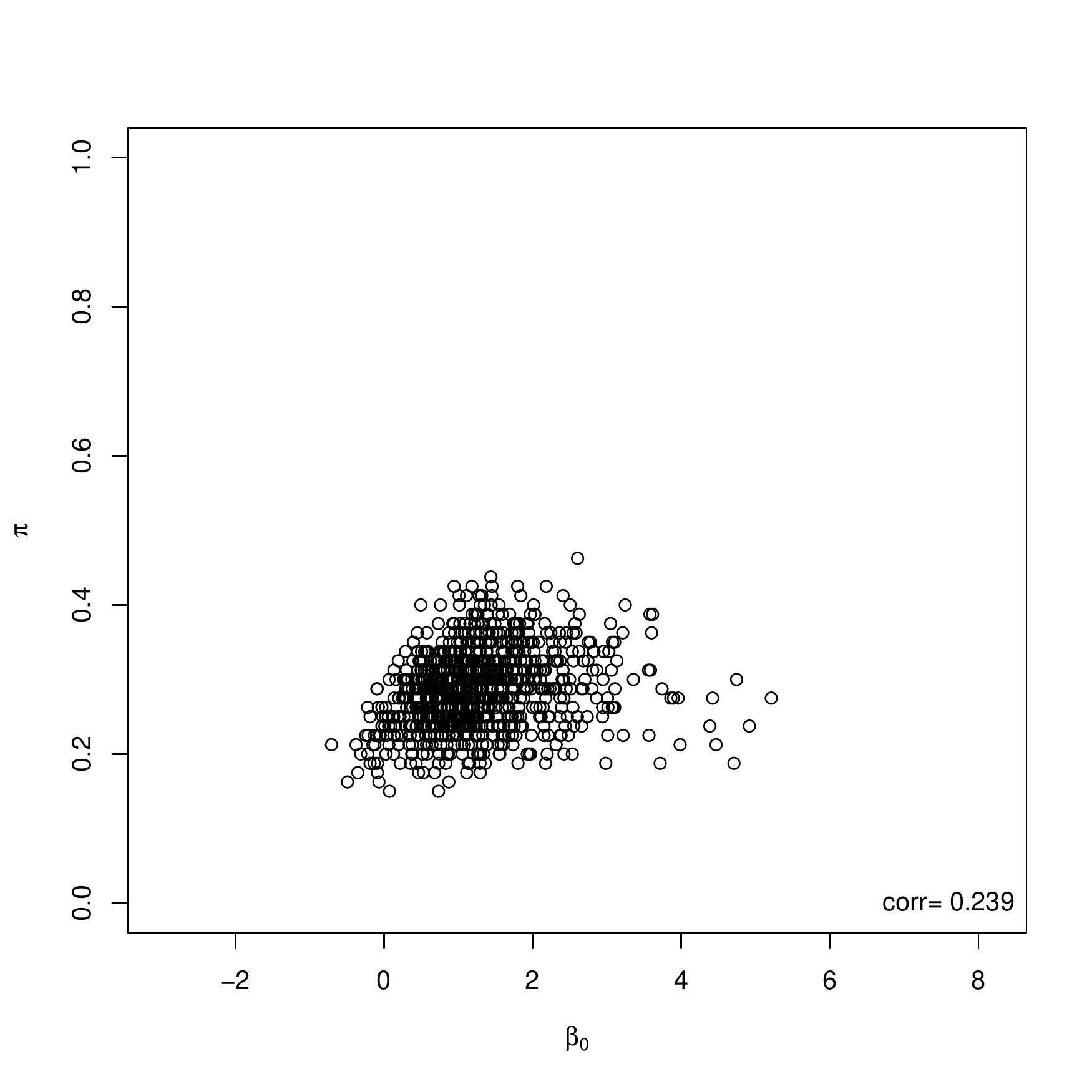}&\includegraphics[height=2cm, width=3.5cm]{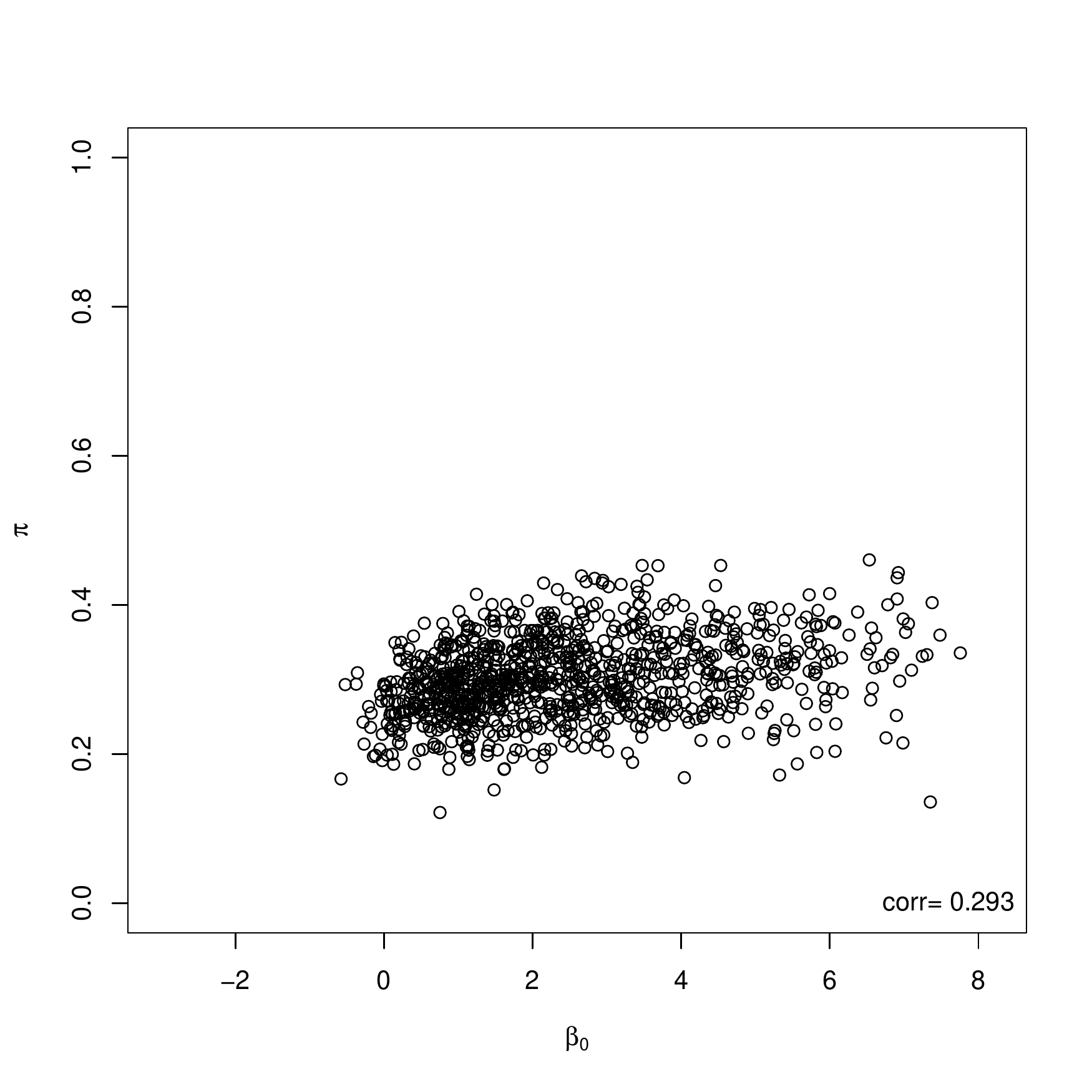}&\includegraphics[height=2cm, width=3.5cm]{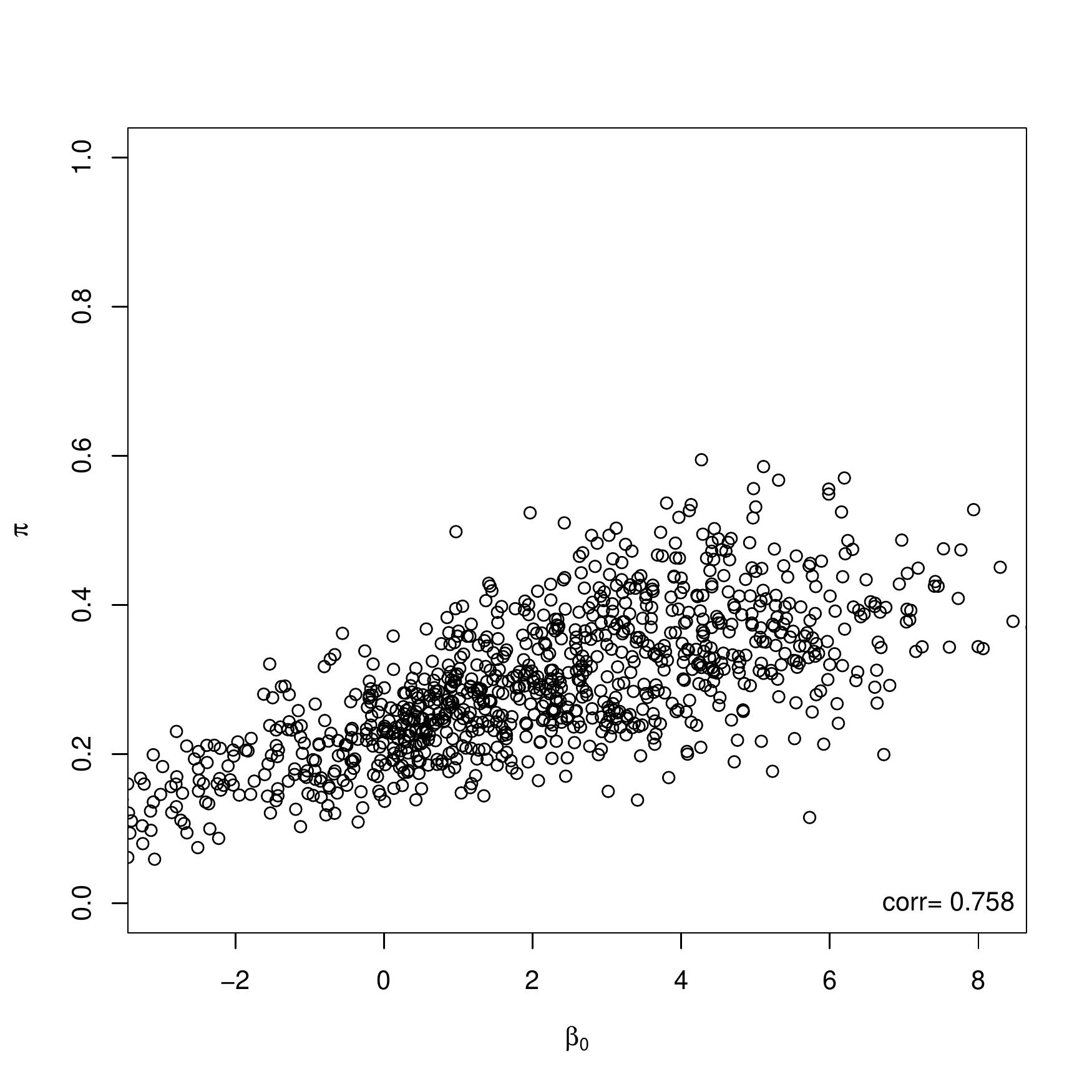}\\
\midrule
\raisebox{1cm}{200}&\includegraphics[height=2cm, width=3.5cm]{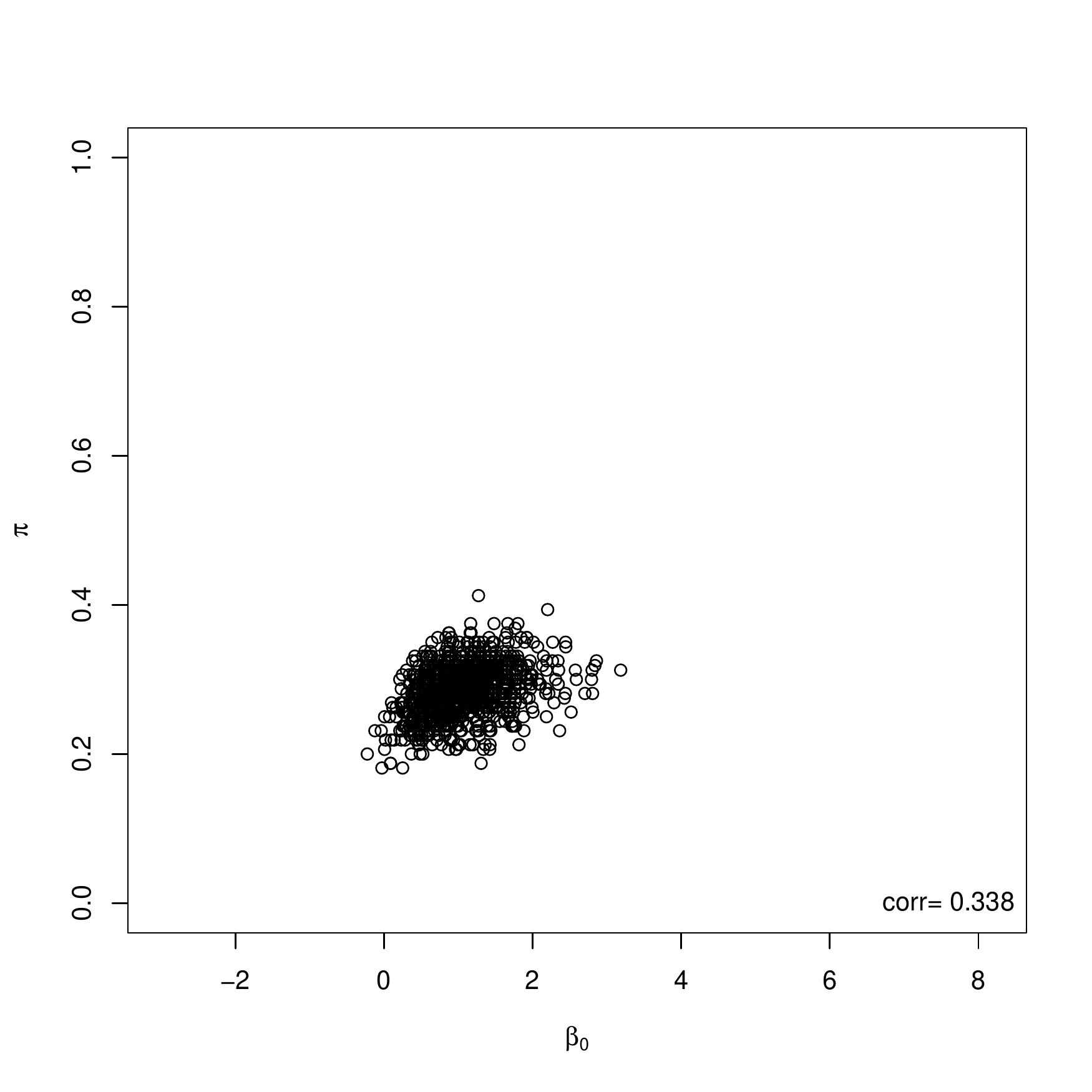}&\includegraphics[height=2cm, width=3.5cm]{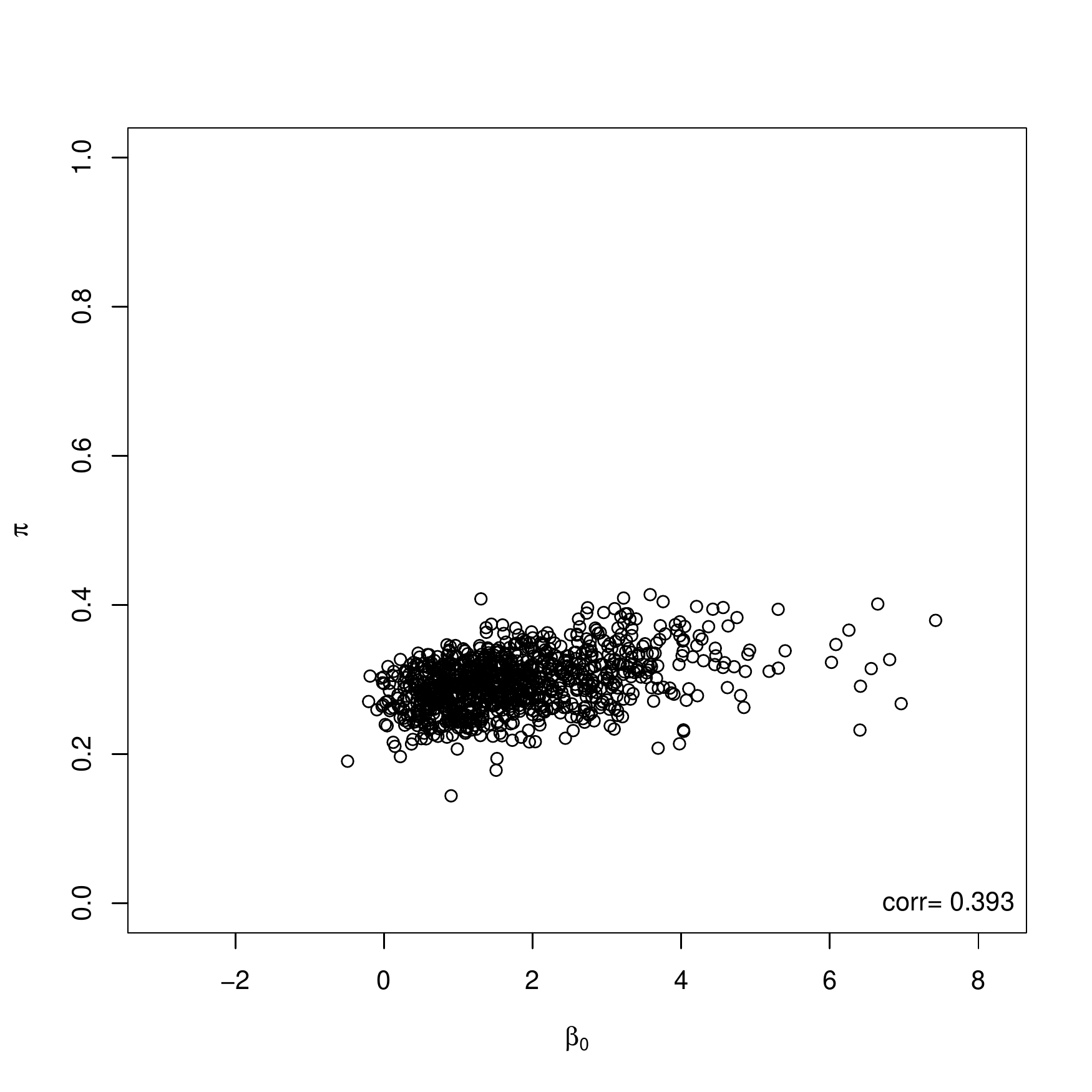}&\includegraphics[height=2cm, width=3.5cm]{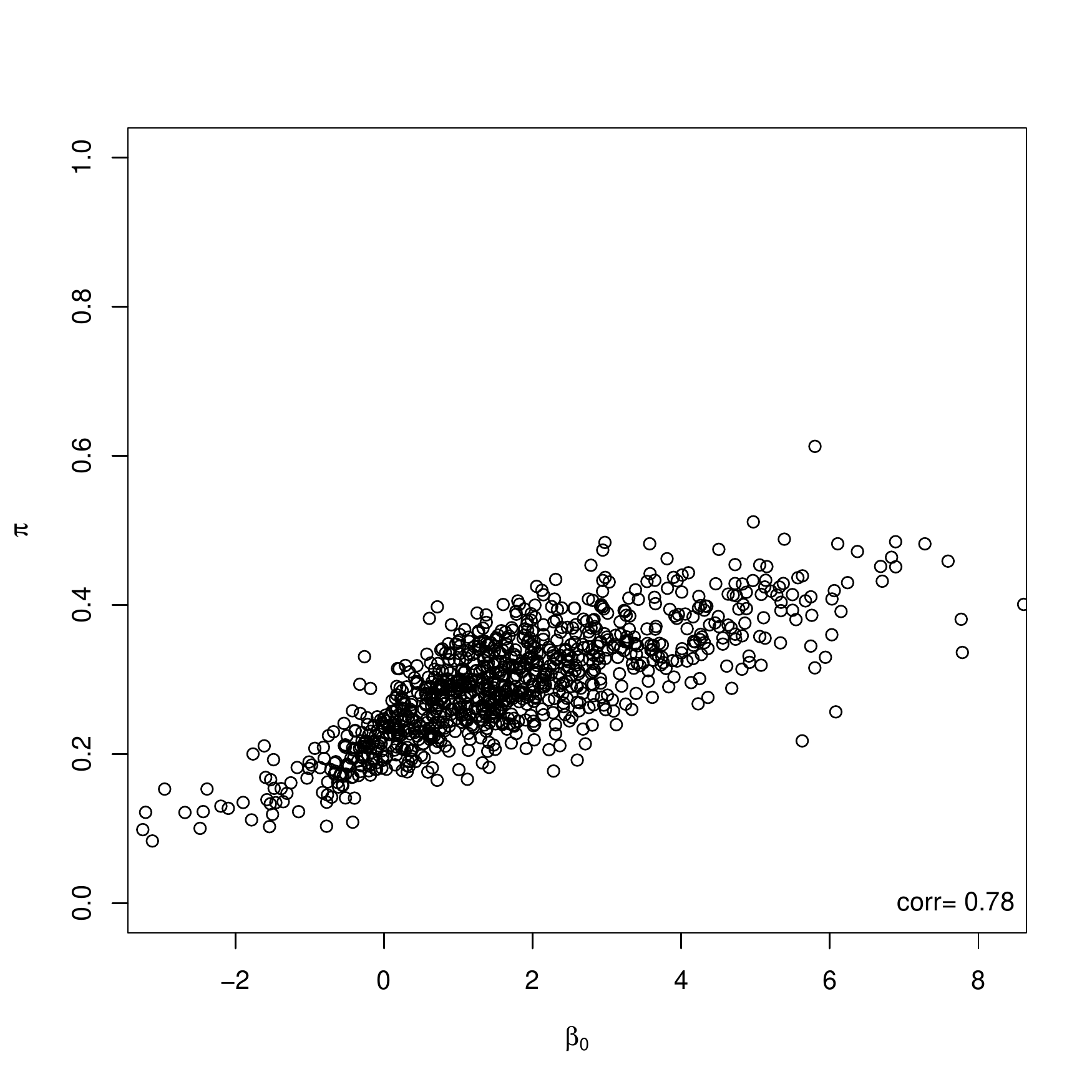}\\
\midrule
\raisebox{1cm}{500}&\includegraphics[height=2cm, width=3.5cm]{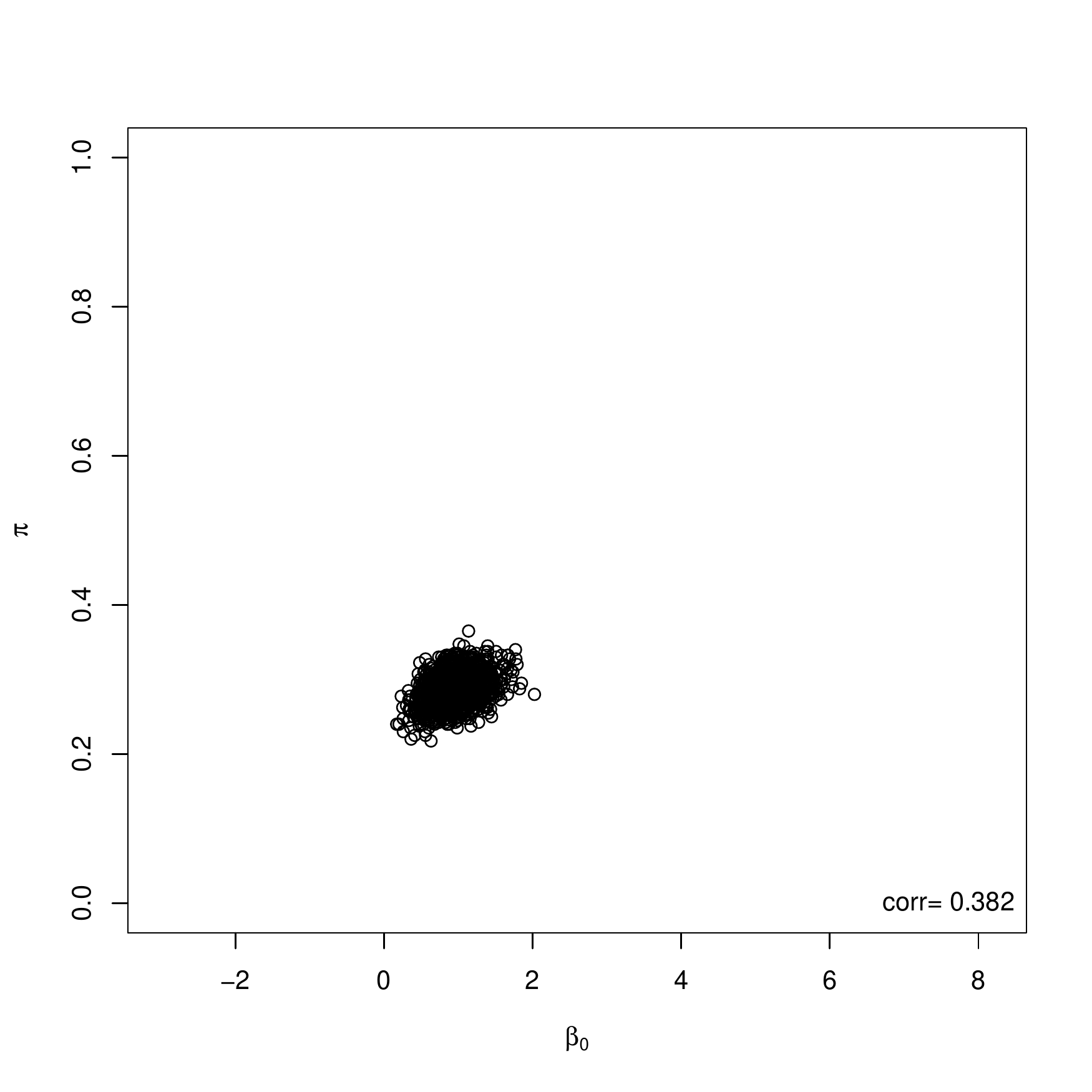}&\includegraphics[height=2cm, width=3.5cm]{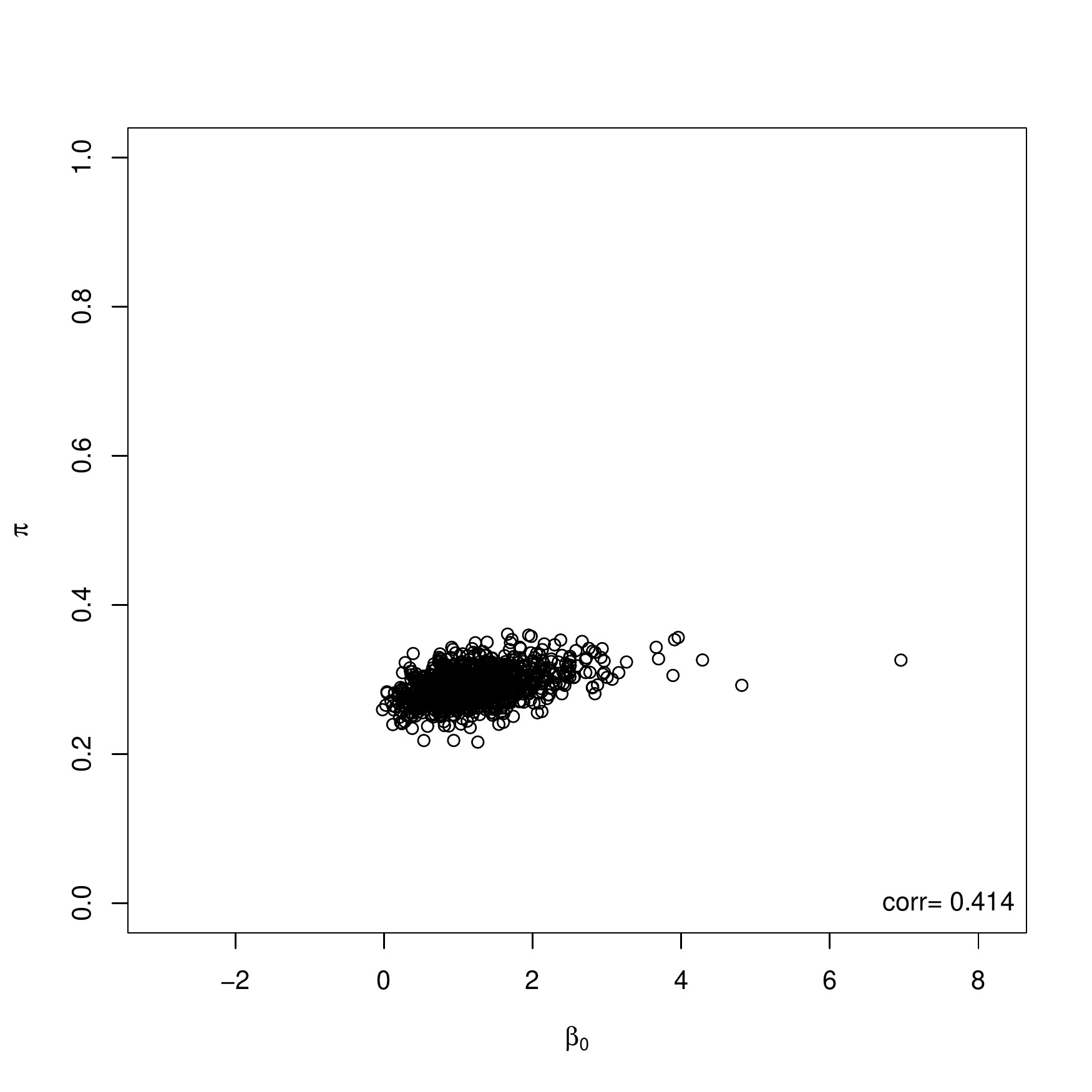}&\includegraphics[height=2cm, width=3.5cm]{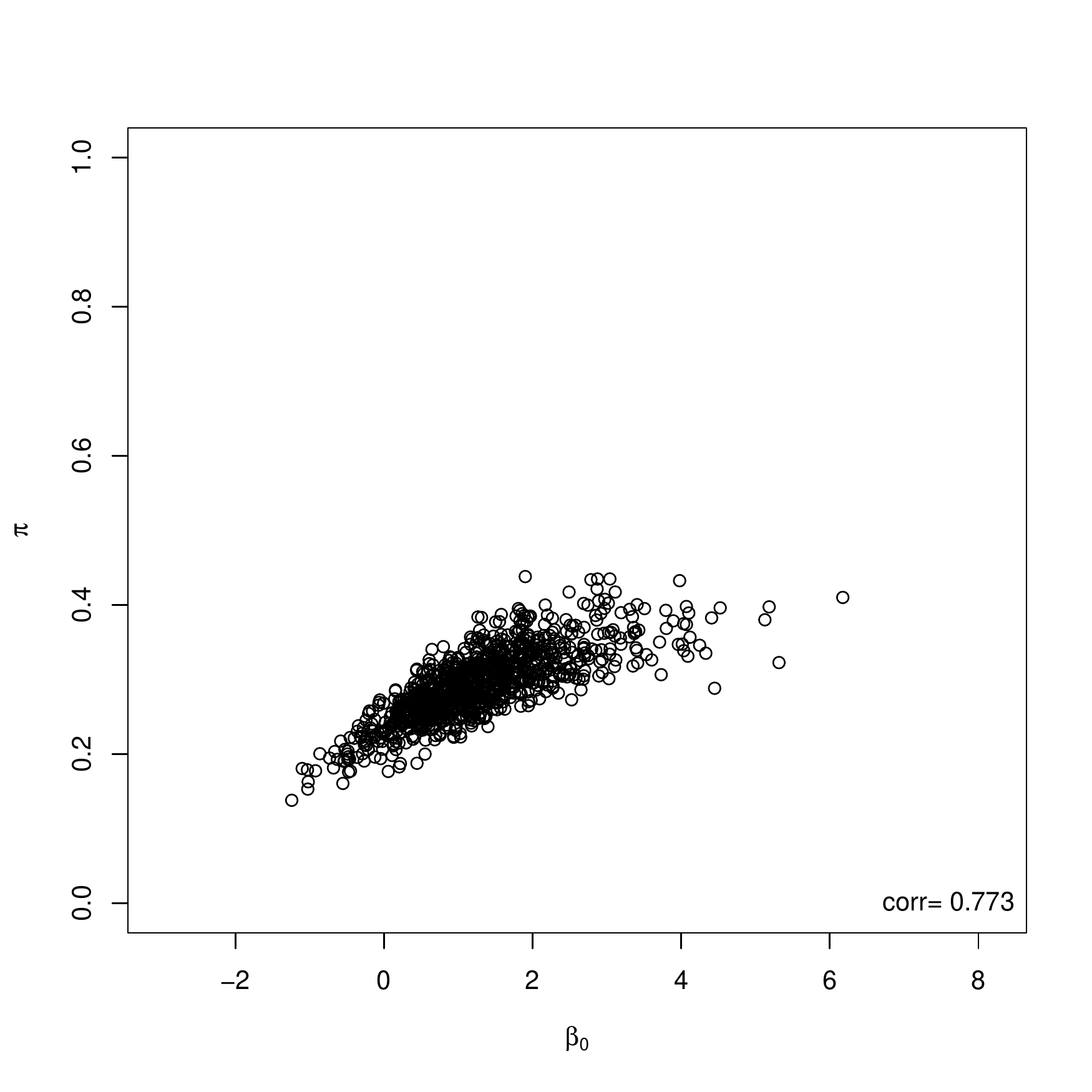}\\
\midrule
\raisebox{1cm}{1000}&\includegraphics[height=2cm, width=3.5cm]{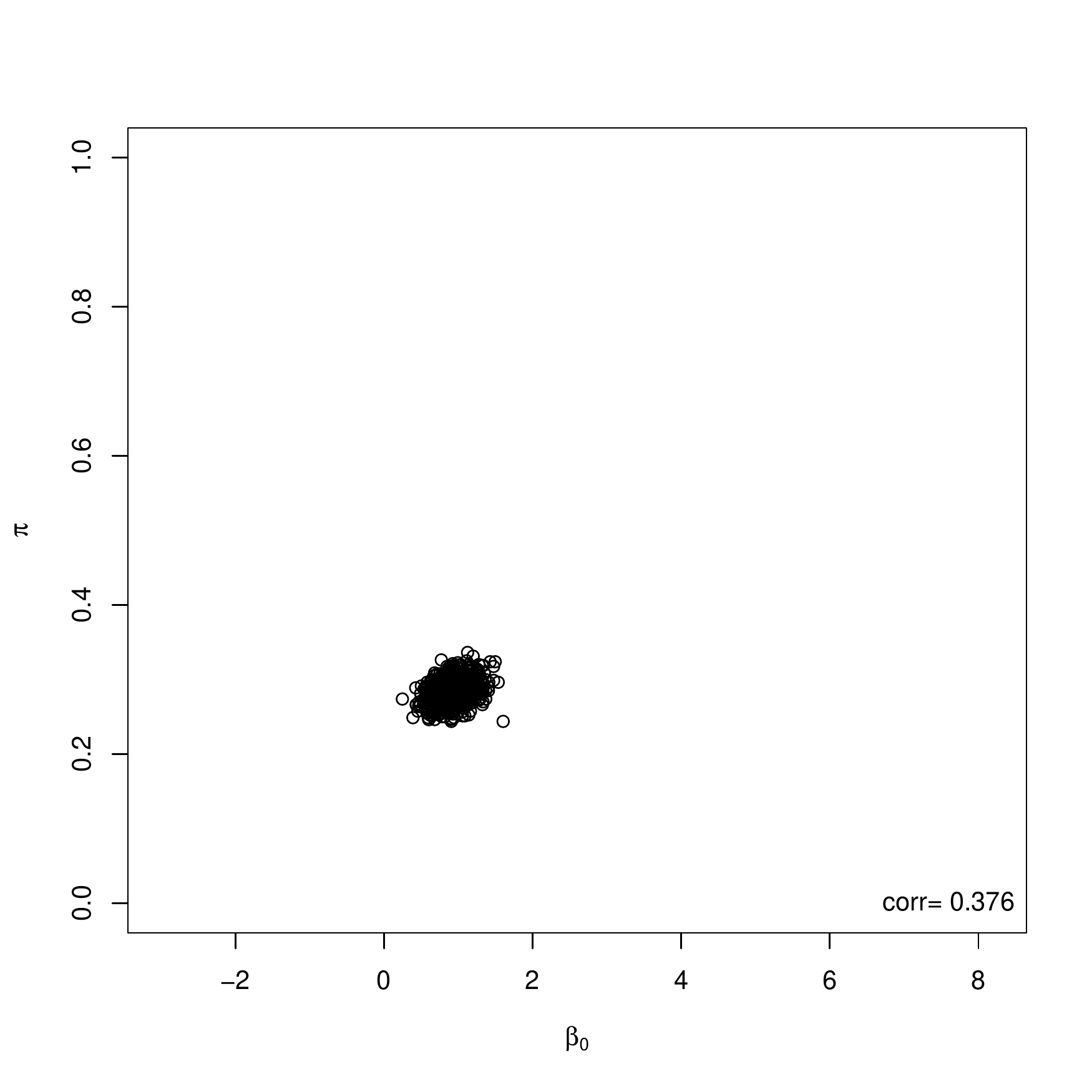}&\includegraphics[height=2cm, width=3.5cm]{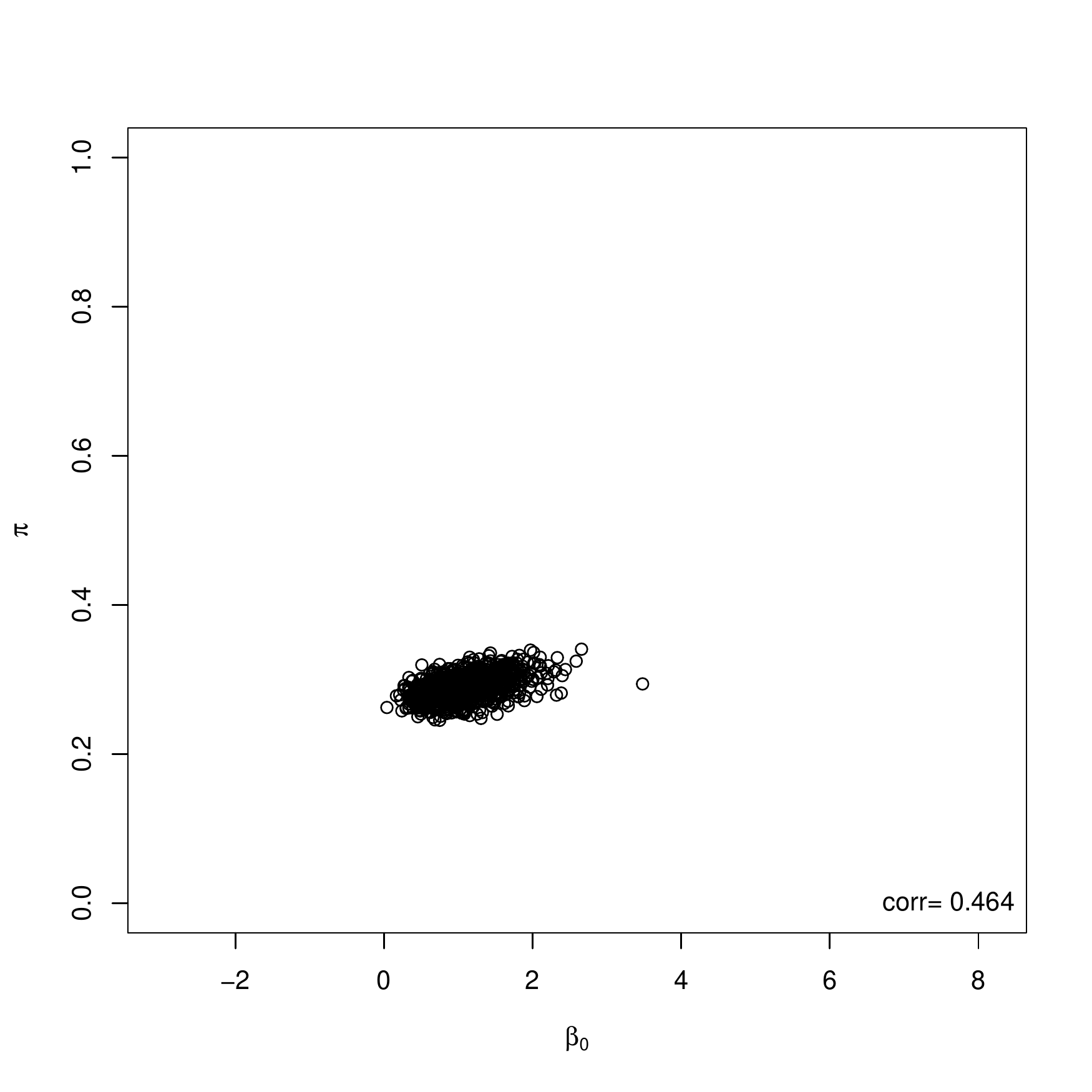}&\includegraphics[height=2cm, width=3.5cm]{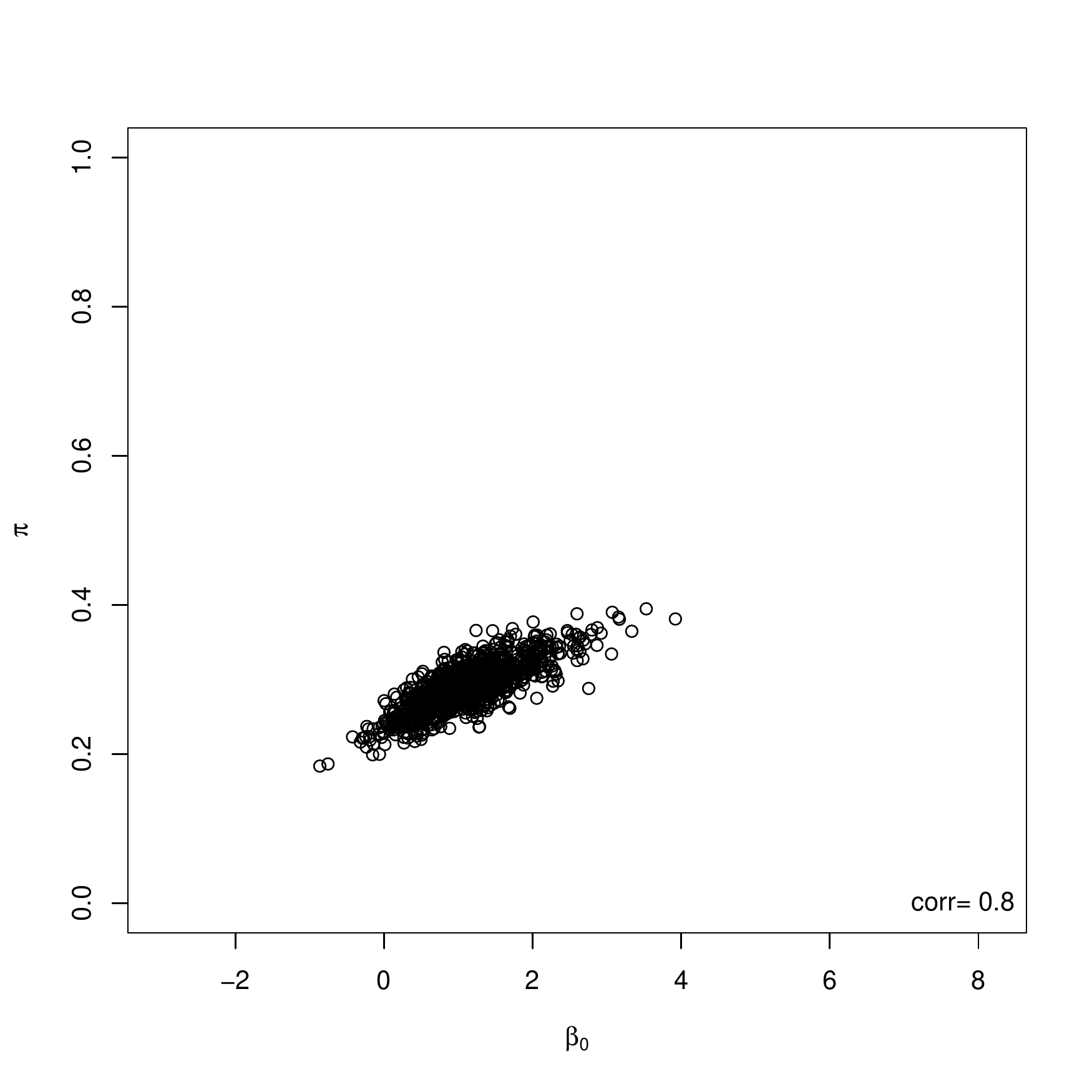}\\
\midrule
\raisebox{1cm}{1500}&\includegraphics[height=2cm, width=3.5cm]{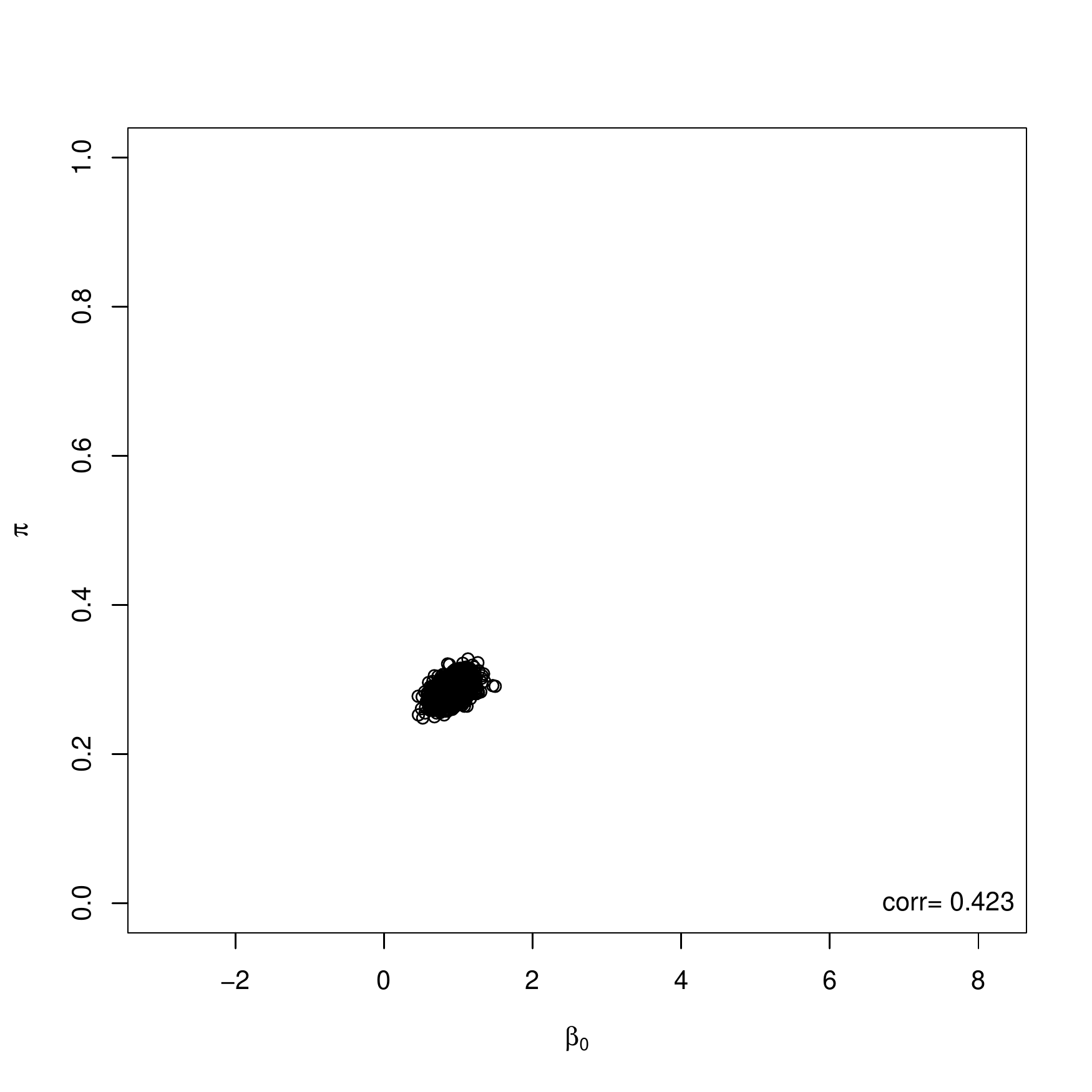}&\includegraphics[height=2cm, width=3.5cm]{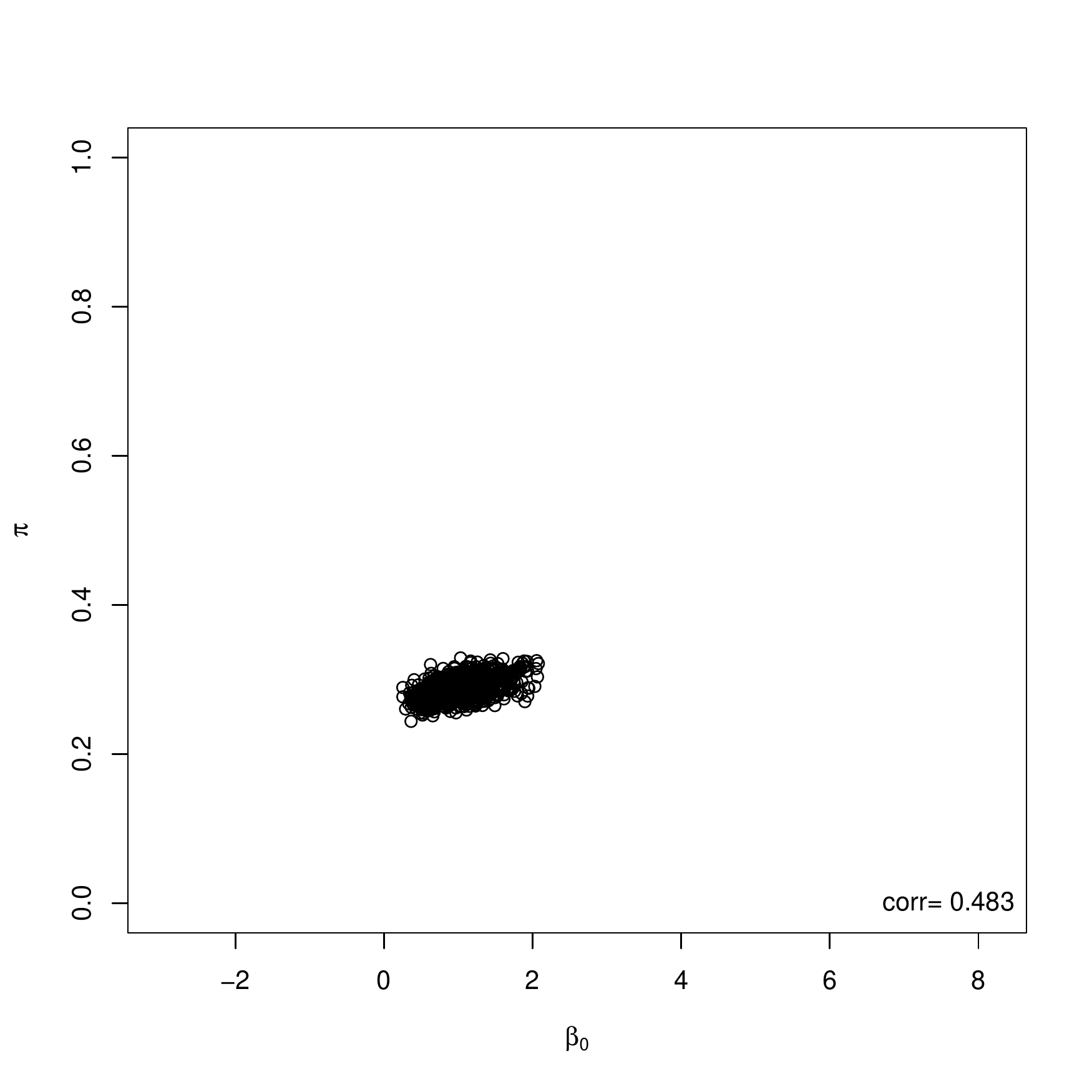}&\includegraphics[height=2cm, width=3.5cm]{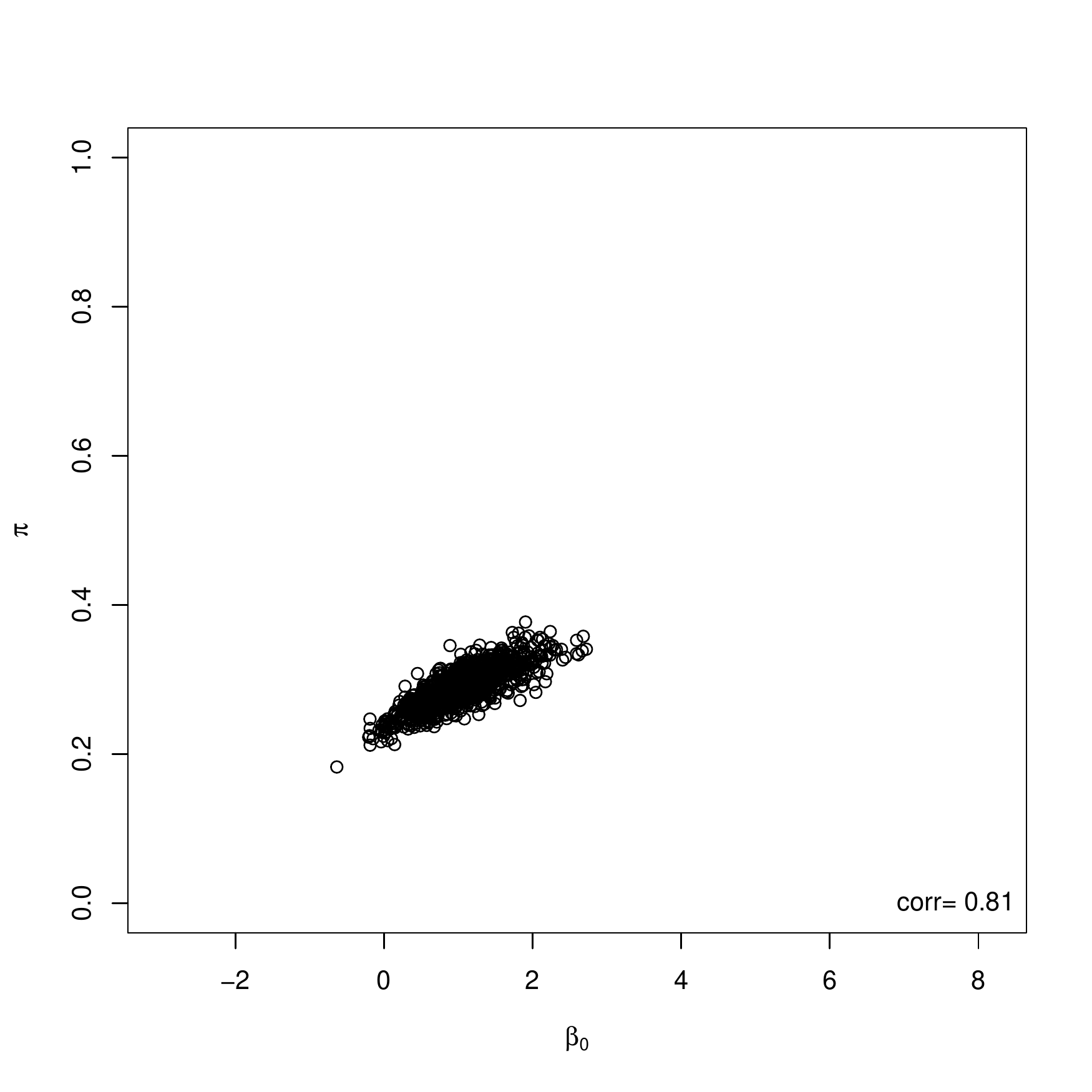}\\
\midrule
\raisebox{1cm}{2000}&\includegraphics[height=2cm, width=3.5cm]{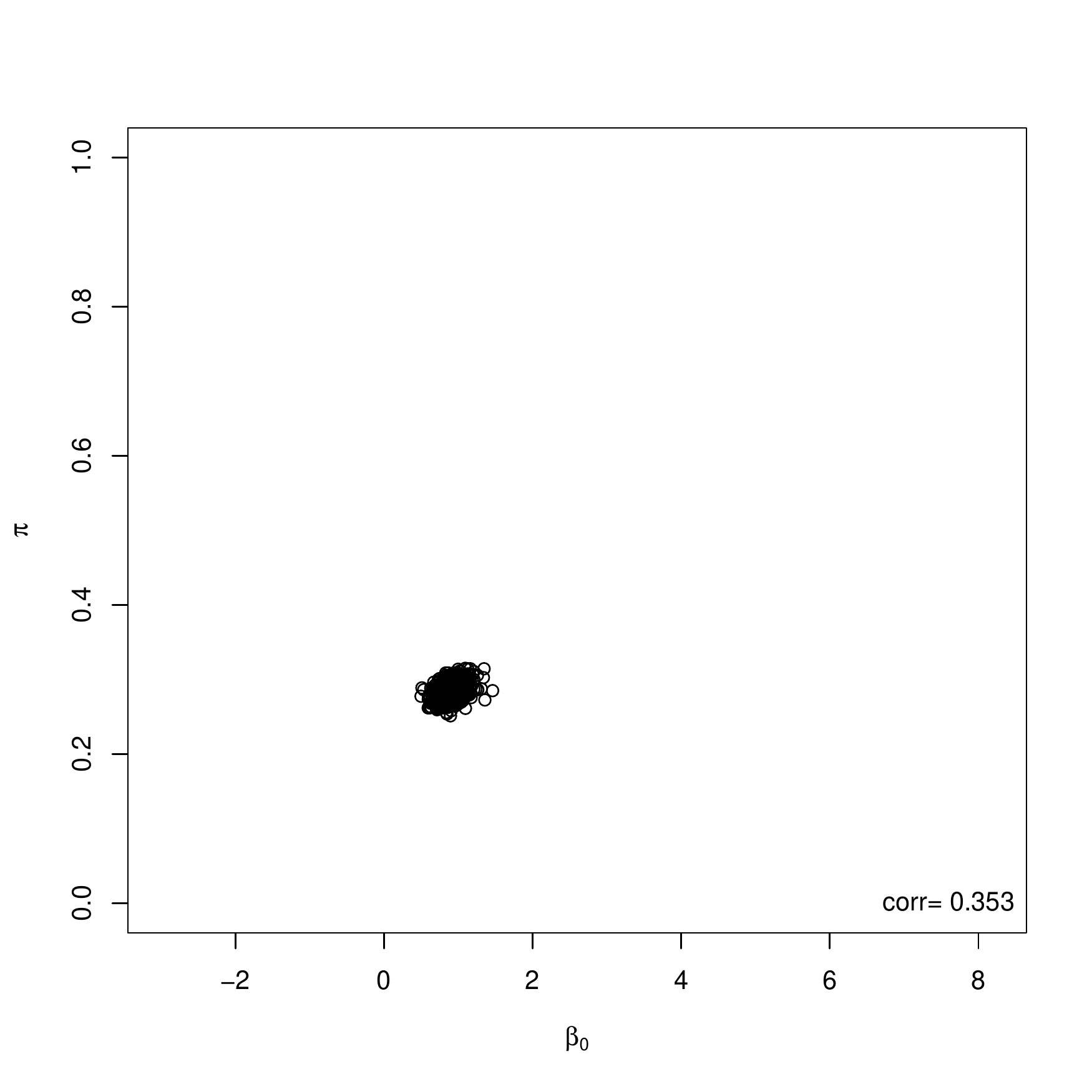}&\includegraphics[height=2cm, width=3.5cm]{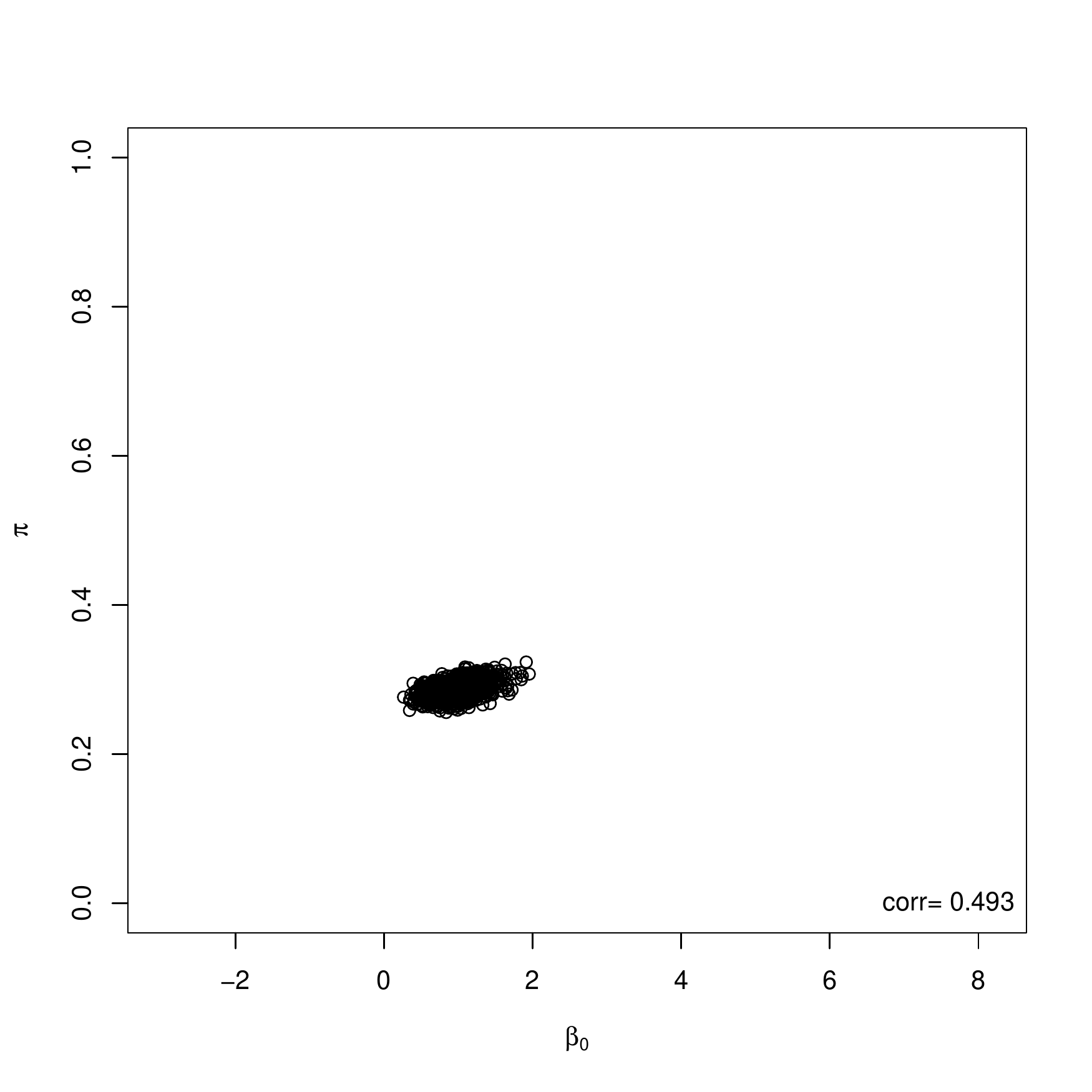}&\includegraphics[height=2cm, width=3.5cm]{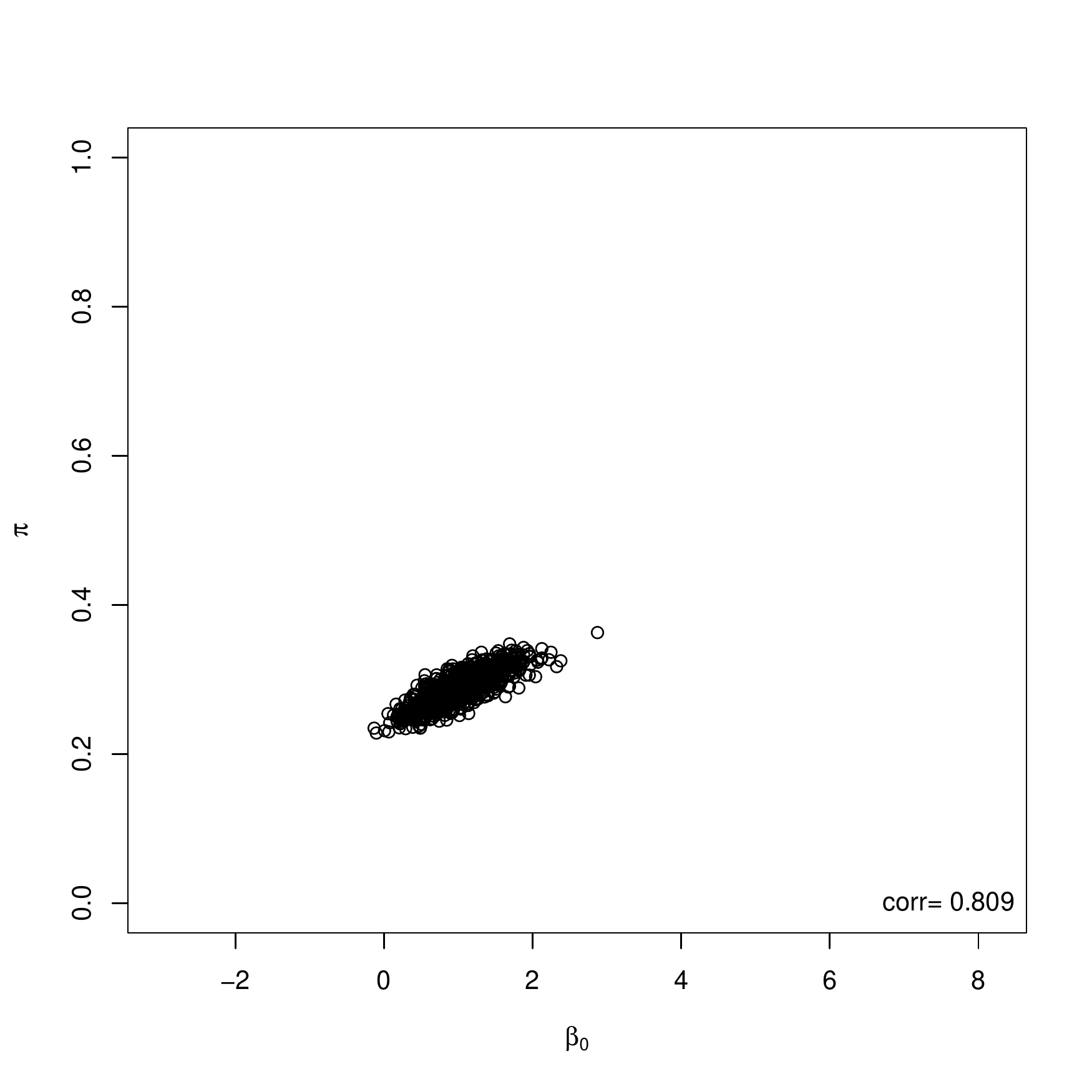}\\
\midrule
\raisebox{1cm}{3000}&\includegraphics[height=2cm, width=3.5cm]{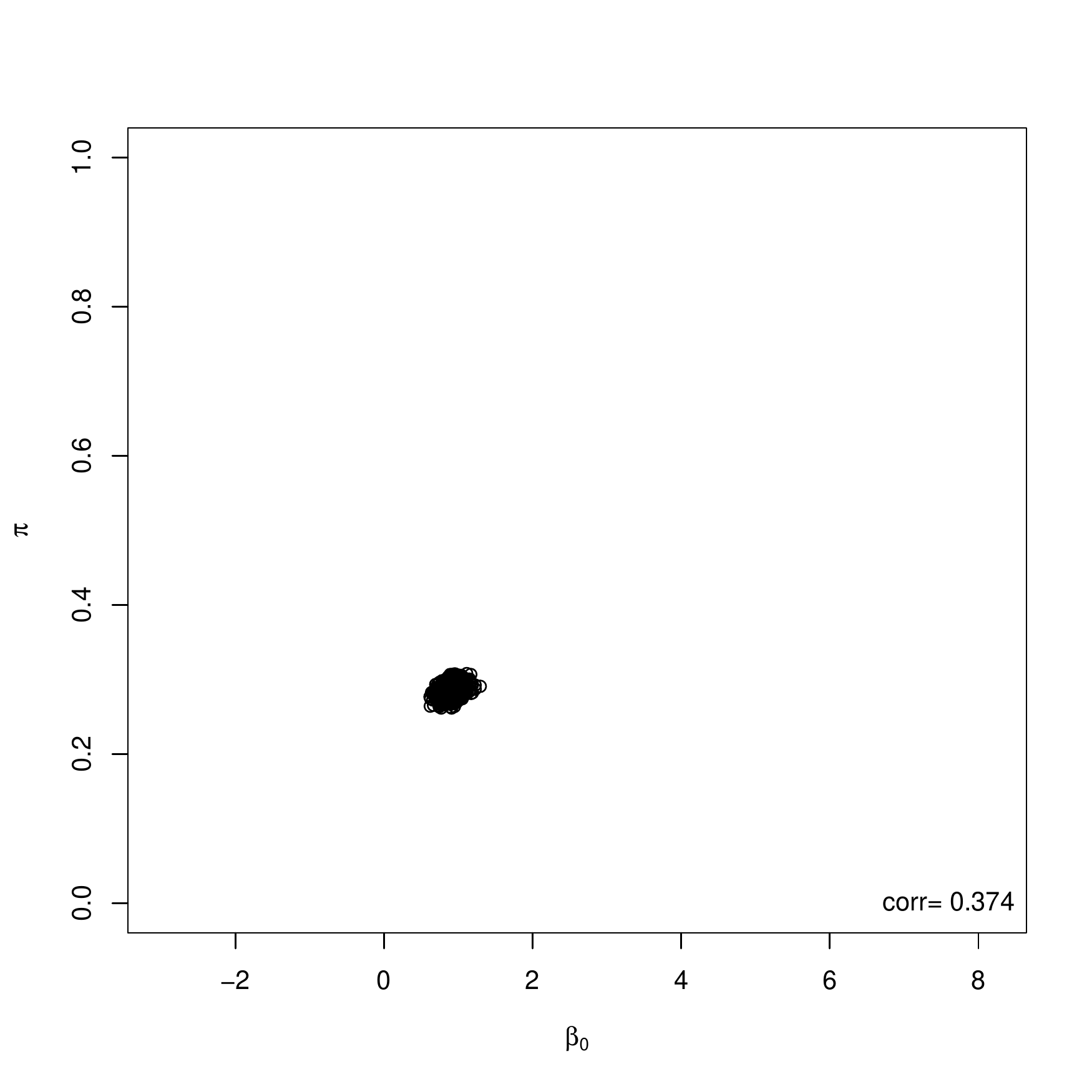}&\includegraphics[height=2cm, width=3.5cm]{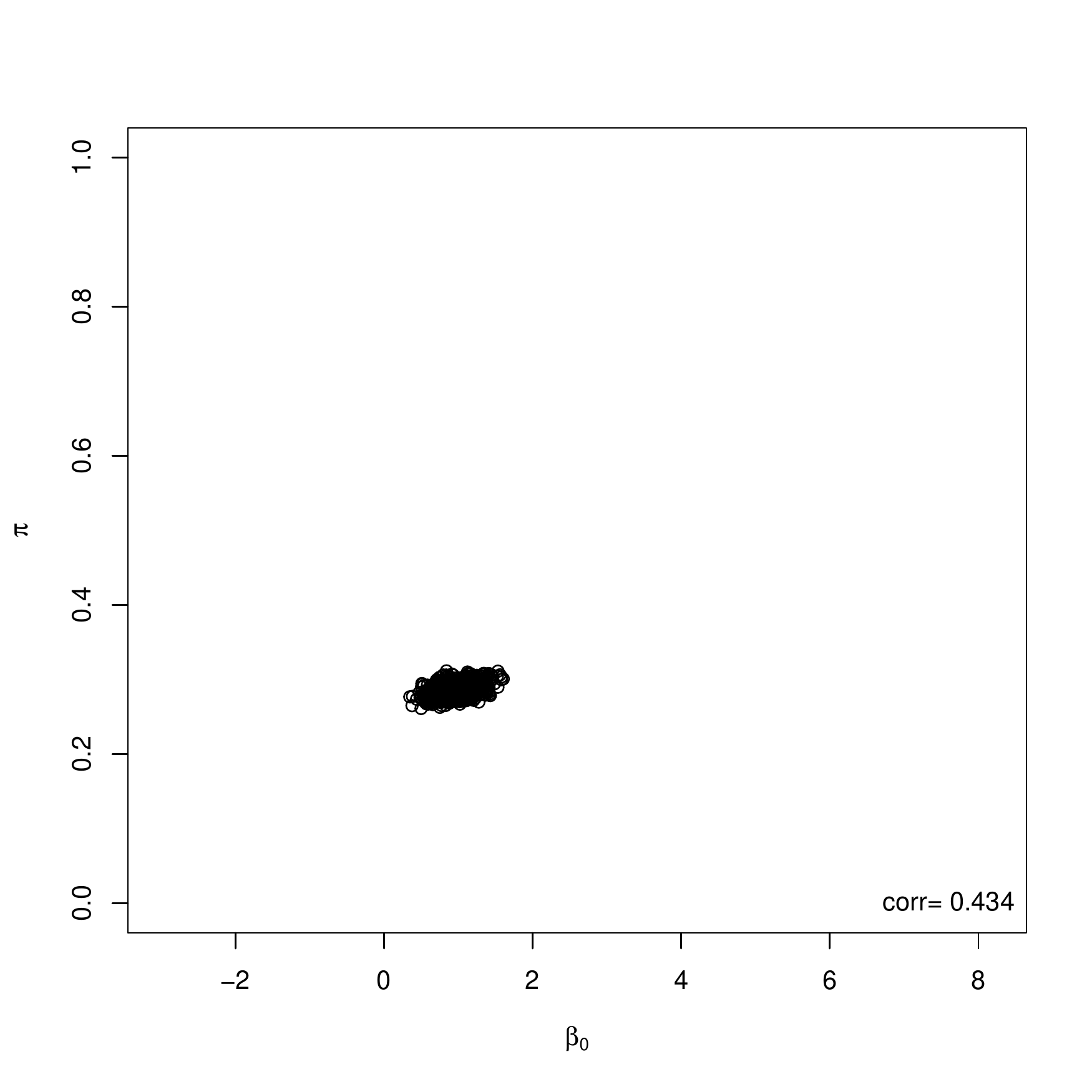}&\includegraphics[height=2cm, width=3.5cm]{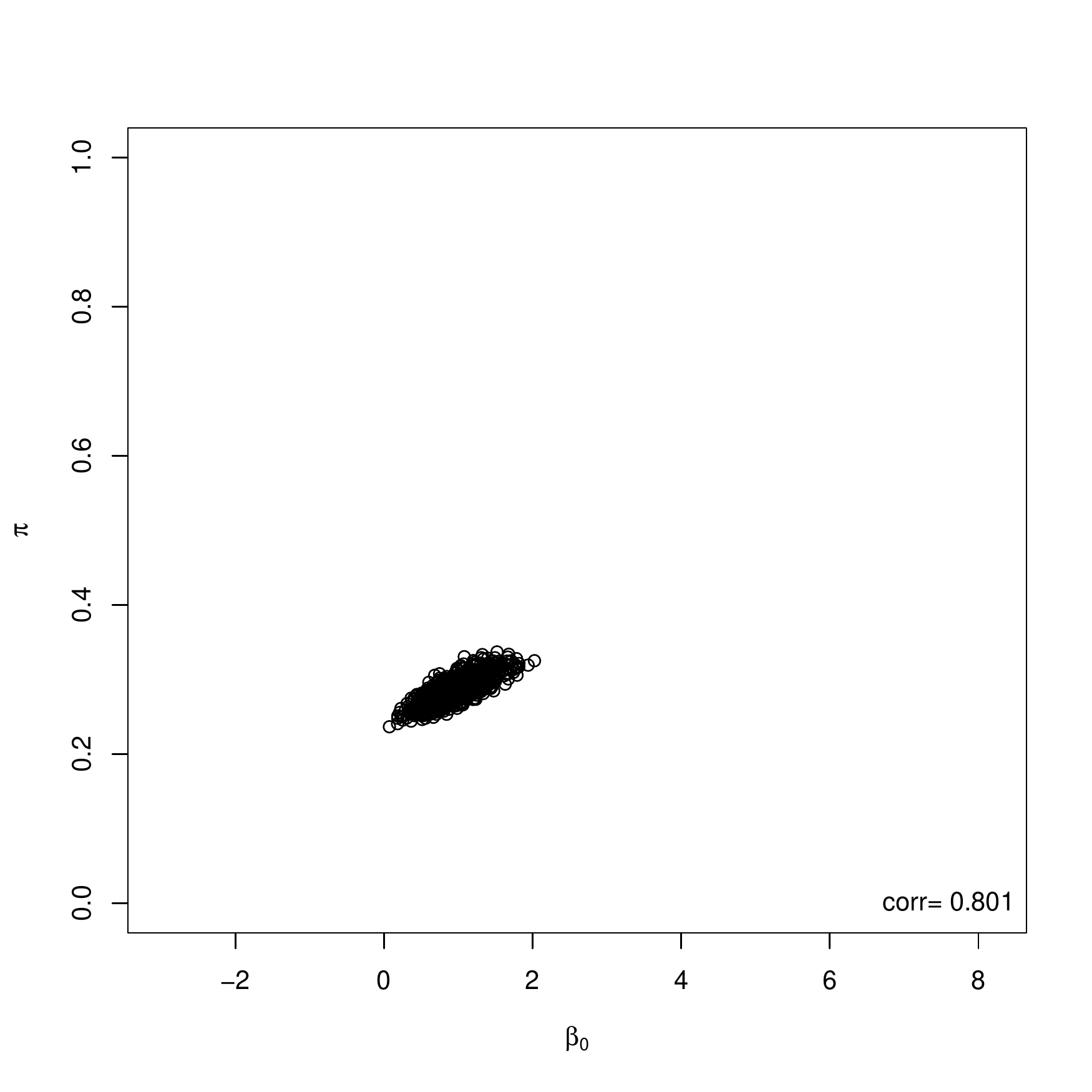}\\
\bottomrule
\end{tabular}
\caption{Scenario (iii): scatterplot of $\pi$ versus $\beta_0$ with increasing sample sizes and different models ($M_0$,$M_1$ and $M_2$).}\label{fig:scat}
\end{figure}

\begin{figure}[H]
\centering
\includegraphics[height=6cm, width=12cm]{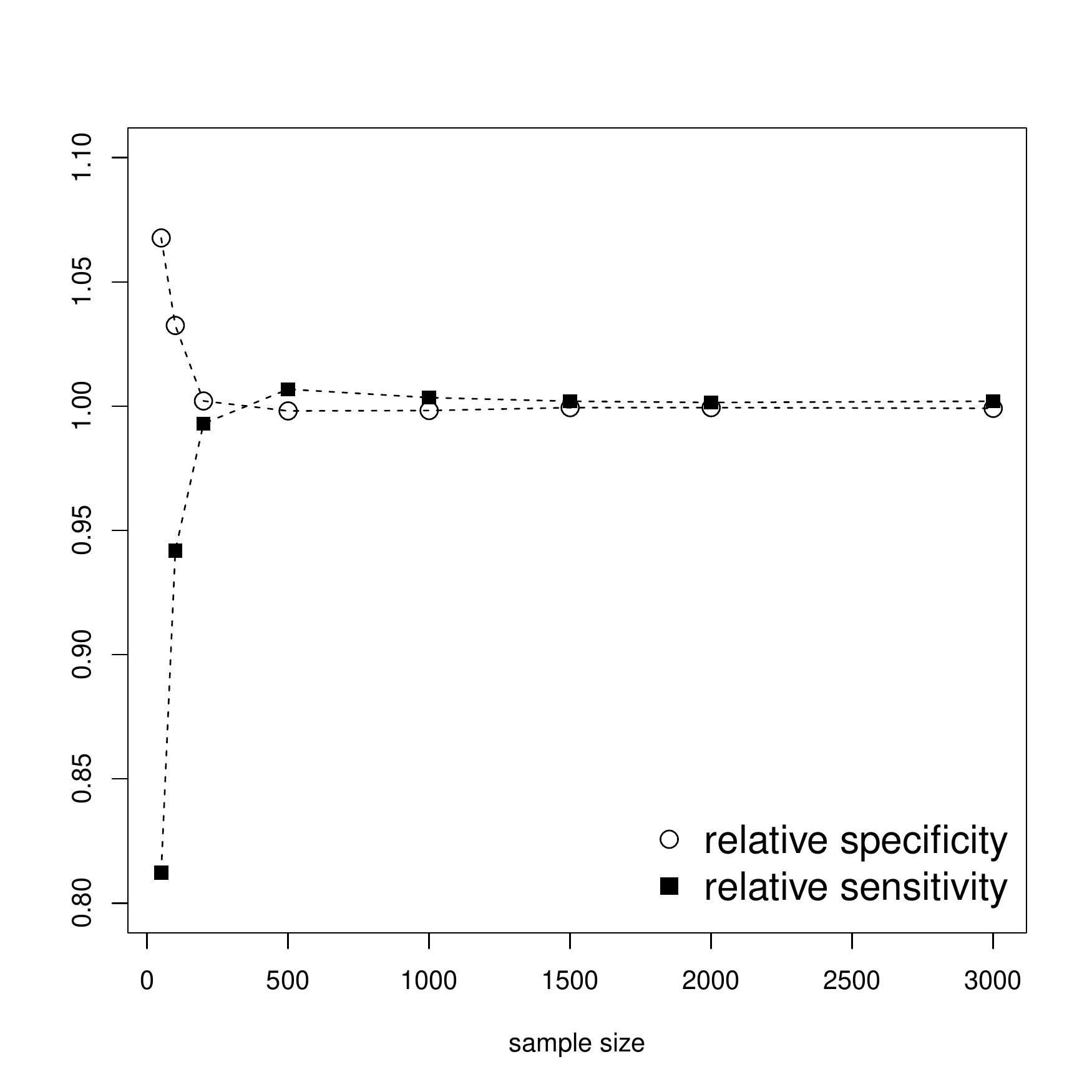}
\caption{Scenario (iii): relative specificity and sensitivity computed as ratios between $M_2$ and $M_1$ specificity and sensitivity measures with increasing sample sizes. Dashed trajectories are reported to show the patterns.}\label{fig:spec}
\end{figure}

\paragraph*{The case-control sampling.}
For each scenario, we sample under the case-control design with a ratio of presence/unobserved equal to $1:4$ and with respect to eight different sample sizes
\[n=50,100,200,500,1000,1500,2000,3000.\]
For example, if the sample size is  equal to $n=500$, we build the corresponding simulated experiment by extracting a random sample $S_{p}$ from $\mathcal{U}_{p}$ of $n_{p}=100$ presences and a random sample $S_{u}$ from $\mathcal{U}$ of $n_{u}=400$ unobserved values,  covariates are available for the whole sample $S$. We consider $k=1000$ independent replications of each experiment. In summary,  we generate a database of 24,000 datasets (8 sample sizes, 3 scenarios and 1000 replications).
With respect to the generating of the data, we considered a quite general framework since the contribution of an informative covariate was combined with a constant effect and a white \textit{Gaussian} noise. With respect to the three scenarios, we obtained empirical population prevalences respectively $\pi_{(i)}=0.215$, $\pi_{(ii)}=0.223$ and $\pi_{(iii)}=0.286$.
\paragraph*{The MCMC estimation.}
The estimation is performed in a Bayesian framework for all the models $M_{0}$, $M_{1}$ and $M_{2}$. The likelihood function we use in the estimation is based on a model that does not always replicate the model used to generate data. More precisely for all experiments (i), (ii) and (iii) the estimation model is:
\begin{equation}\label{eq:estmod}
\mbox{logit}(Pr(Y=1|X_1=x_1))=\beta_0+\beta_1x_1
\end{equation}
than with scenario (i) the model that generates the data and the one defining the likelihood are the same, whilst for scenarios (ii) and (iii) the likelihood model becomes increasingly different from the one that generates the data. Notice that we consider a simpler structure than the one shown in Figure (\ref{fig:graph}) as we choose a \textit{Gaussian} prior $N(0,25)$ for all regression parameters ($\beta_{0}$, $\beta_{1}$) and no hyper parameter is considered. Then, MCMC estimates are computed using 5000 runs after 10000 iterations of burn-in, no thinning is applied as samples autocorrelation is negligible. 
\paragraph*{Results.}
In what follows we report Figures and Tables built on scenario (iii) as it represents the most complex of the three alternatives and it is our ``worst'' case. In each replicate of an experiment, point estimates are computed as posterior means over 5000 iterations. In Figure \ref{fig:box} boxplots describing point estimates behaviour are reported, horizontal lines corresponding to the ``true'' values are drawn. The first box corresponds to procedure $M_0$, the second to $M_1$ and the third to our proposal $M_2$. In $M_0$ the prevalence $\pi$ is estimated as the ratio of the observed presences in $S_u$ to the sample size $n_u$.  In $M_1$, although $\pi$ is assumed known a priori, we consider its posterior prediction in $S_u$. Finally in $M_2$, the prevalence is obtained at each MCMC step as described in section \ref{sec:prev} and then the mean over 5000 runs is taken. In Table \ref{tab:tab3} further details of the point estimates are reported: the median and in parenthesis the first and third quartiles. From the Figures and the values we can see that the three procedure lead to ``comparable'' values with the obvious reduction of variability when $n$ increases. Remark that the estimates for $M_2$, although affected by a larger variability with small sample sizes, rapidly approaches $M_0$ and $M_1$ behaviour with increasing sample size. This can be seen more clearly in Figure \ref{fig:rmse} where rooted mean square errors (rmse) are reported. As far as $\beta_1$ is concerned the lack of knowledge on $X_2$ leads to biased point estimates regardless  the estimation procedure. Tables \ref{tab:tab4} and \ref{tab:tab5} in Appendix report point estimates for scenarios (i) and (ii).  For scenarios (i) unbiased estimates are obtained while (ii) is affected by the same distortion as (iii) but with smaller variability. \\
\begin{table}[H]\footnotesize
\centering
\begin{tabular}{ccccc}
\toprule
$n$  & Model  & $\beta_{0}$  & $\beta_{1}$  & $\pi$\tabularnewline
\toprule
\multirow{3}{*}{50}
 & $M_{0}$  & 1.42 (0.68 ; 2.33) & 1.15 (0.88 ; 1.55) & 0.28 (0.25 ; 0.35)\tabularnewline
 & $M_{1}$  & 3.13 (1.78 ; 4.46) & 1.69 (1.17 ; 2.28) & 0.31 (0.26 ; 0.35)\tabularnewline
 & $M_{2}$ & 1.79 (-3.38 ; 4.26) & 1.44 (0.76 ; 2.12) & 0.24 (0.13 ; 0.34)\tabularnewline
\midrule
\multirow{3}{*}{100}
 & $M_{0}$  & 1.14 (0.72 ; 1.62) & 1.00 (0.86 ; 1.22) & 0.29 (0.25 ; 0.33)\tabularnewline
 & $M_{1}$  & 2.12 (1.11 ; 3.51) & 1.30 (0.97 ; 1.80) & 0.30 (0.26 ; 0.34)\tabularnewline
 & $M_{2}$  & 1.92 (0.16 ; 3.87) & 1.24 (0.89 ; 1.78) & 0.28 (0.21 ; 0.36)\tabularnewline
\midrule
\multirow{3}{*}{200} 
 & $M_{0}$  & 1.01 (0.72 ; 1.36) & 0.94 (0.83 : 1.06) & 0.29 (0.26 ; 0.31)\tabularnewline
 & $M_{1}$  & 1.53 (0.89 ; 2.39) & 1.08 (0.87 ; 1.37) & 0.29 (0.27 ; 0.32)\tabularnewline
 & $M_{2}$  & 1.49 (0.59 ; 2.62) & 1.07 (0.83 ; 1.37) & 0.29 (0.24 ; 0.34)\tabularnewline
\midrule
\multirow{3}{*}{500}
 & $M_{0}$  & 0.94 (0.75 ; 1.15) & 0.89 (0.82 ; 0.96) & 0.29 (0.27 ; 0.30)\tabularnewline
 & $M_{1}$  & 1.12 (0.78 ; 1.57) & 0.94 (0.82 ; 1.10) & 0.29 (0.28 ; 0.31)\tabularnewline
 & $M_{2}$  & 1.17 (0.62 ; 1.82) & 0.94 (0.80 ; 1.12) & 0.29 (0.26 ; 0.32)\tabularnewline
\midrule
\multirow{3}{*}{1000}
 & $M_{0}$  & 0.91 (0.78 ; 1.04) & 0.88 (0.83 ; 0.92) & 0.28 (0.28 ; 0.30)\tabularnewline
 & $M_{1}$  & 1.03 (0.79 ; 1.34) & 0.91 (0.83 ; 1.01) & 0.29 (0.28 ; 0.30)\tabularnewline
 & $M_{2}$  & 1.05 (0.68 ; 1.49) & 0.91 (0.82 ; 1.03) & 0.29 (0.27 ; 0.31)\tabularnewline
\midrule
\multirow{3}{*}{1500}
 & $M_{0}$  & 0.89 (0.80 ; 1.00) & 0.86 (0.83 ; 0.91) & 0.29 (0.28 ; 0.29)\tabularnewline
 & $M_{1}$  & 1.00 (0.78 ; 1.24) & 0.89 (0.82 ; 0.98) & 0.29 (0.28 ; 0.30)\tabularnewline
 & $M_{2}$  & 1.01 (0.71 ; 1.35) & 0.90 (0.82 ; 0.99) & 0.29 (0.27 ; 0.31)\tabularnewline
\midrule
\multirow{3}{*}{2000} 
 & $M_{0}$  & 0.89 (0.82 ; 0.98) & 0.87 (0.84 ; 0.90) & 0.29 (0.28 ; 0.29)\tabularnewline
 & $M_{1}$  & 0.96 (0.79 ; 1.15) & 0.89 (0.83 ; 0.95) & 0.29 (0.28 ; 0.29)\tabularnewline
 & $M_{2}$  & 0.96 (0.71 ; 1.23) & 0.88 (0.82 ; 0.96) & 0.29 (0.27 ; 0.30)\tabularnewline
\midrule
\multirow{3}{*}{3000}
 & $M_{0}$  & 0.90 (0.83 ; 0.97) & 0.87 (0.84 ; 0.89) & 0.29 (0.28 ; 0.29)\tabularnewline
 & $M_{1}$  & 0.94 (0.82 ; 1.09) & 0.88 (0.84 ; 0.93) & 0.29 (0.28 ; 0.29)\tabularnewline
 & $M_{2}$  & 0.95 (0.76 ; 1.17) & 0.88 (0.83 ; 0.94) & 0.29 (0.28 ; 0.30)\tabularnewline
\bottomrule
\end{tabular}
\caption{Scenario (iii): point estimates of regression parameters and prevalence  computed as medians over 1000 replicates with increasing sample sizes and different models ($M_0$,$M_1$ and $M_2$). In parenthesis distributions quartiles are reported.}\label{tab:tab3}
\end{table}
From \citet{ward:al:2009} we know that pairwise correlation between parameters is present. In Table \ref{tab:tab6} we report the empirical pairwise correlation measures, obtained as the averages with respect to the 1000 samples, with increasing sample sizes across the different models. No significant differences in the pattern of correlation $(\beta_0;\beta_1)$ between the models $M_1$ and $M_2$ while the correlation $(\beta_1,\pi)$ has a general weaker pattern in $M_1$ than $M_2$. With respect to the correlation $(\beta_0,\pi)$ more significant difference are present between $M_1$ and $M_2$. In Figure \ref{fig:scat} scatterplots of $\beta_0$ versus $\pi$ in the 1000 replicates are plotted with equal axis across estimation procedures and sample sizes. These pictures help us to understand how this correlation  evolves with increasing sample sizes. $M_2$ produces the most positive correlated point estimates this being an advantage whenever the model is properly specified.
\begin{table}[H]\footnotesize
\centering
\begin{tabular}{cccccccccc}
\toprule
Model & \multicolumn{3}{c}{$M_{0}$} & \multicolumn{3}{c}{$M_{1}$} & \multicolumn{3}{c}{$M_{2}$}\tabularnewline
\toprule
$n$ & $\beta_{0};\beta_{1}$ & $\beta_{0};\pi$ & $\beta_{1};\pi$ & $\beta_{0};\beta_{1}$ & $\beta_{0};\pi$ &
$\beta_{1};\pi$ & $\beta_{0};\beta_{1}$ & $\beta_{0};\pi$ & $\beta_{1};\pi$\tabularnewline
\toprule
50   & 0.65 & 0.26 & -0.09 & 0.59 & 0.10 & -0.30 & 0.68 & 0.81 & 0.31\tabularnewline
100  & 0.75 & 0.24 & -0.12 & 0.89 & 0.29 &  0.02 & 0.82 & 0.76 & 0.37\tabularnewline
200  & 0.78 & 0.34 & -0.04 & 0.94 & 0.39 &  0.18 & 0.90 & 0.78 & 0.48\tabularnewline
500  & 0.79 & 0.38 &  0.00 & 0.95 & 0.41 &  0.24 & 0.92 & 0.77 & 0.51\tabularnewline
1000 & 0.77 & 0.38 & -0.01 & 0.94 & 0.46 &  0.27 & 0.91 & 0.81 & 0.54\tabularnewline
1500 & 0.78 & 0.42 &  0.00 & 0.95 & 0.48 &  0.28 & 0.92 & 0.81 & 0.55\tabularnewline
2000 & 0.77 & 0.35 & -0.06 & 0.94 & 0.49 &  0.30 & 0.92 & 0.81 & 0.55\tabularnewline
3000 & 0.81 & 0.37 & -0.01 & 0.95 & 0.43 &  0.23 & 0.91 & 0.80 & 0.52\tabularnewline
\bottomrule
\end{tabular}
\caption{Scenario (iii): pairwise parameters correlation (average over the 1000 replicates) with increasing sample sizes and different models ($M_0$,$M_1$ and $M_2$).}\label{tab:tab6}
\end{table}
To verify the predictive performance we considered relative measures of specificity and sensitivity \citep{fawcett:2006} build as the ratio of the same measures for $M_2$ (numerator) and for $M_1$ (denominator) respectively. In Figure \ref{fig:spec} the obtained values are reported versus sample sizes. Remark that $M_2$ rapidly reaches the same level of performance as $M_1$ with increasing sample size.
\section{Conclusions}\label{sec:conc}
In this work, we presented a Bayesian procedure to estimate the parameters of logistic regressions for presence-only data. The approach we proposed is based on a two levels scheme  where a generating probability law is combined with a case-control design adjusted for presence-only data. The new formalization allows to consider rigorously all the mathematical details of the model as for instance the approximation of the ratio \eqref{eq:ratio1} that represents the crucial point when modeling presence-only data in the finite population setting. We want to point out that our formalization is substantially different from the work by \cite{ward:al:2009}, although we end with the same statistical model. We concentrated on the case of the linear logistic regression because we were aware that  some care is necessary to handle the identifiability issues  present in the model.\\
The comparative simulation study considered three scenarios with different levels of complexity across increasing sample sizes. We presented detailed results with respect to the most difficult case where the contribution of an informative covariate was mixed with a constant effect and a white \textit{Gaussian} noise. In term of point estimation, the estimates based on our model were comparable to those obtained under the presence-only data benchmark in which the empirical population prevalence was assumed to be known. On the other hand, this lack of information on the population prevalence affected the efficiency of the estimates, that resulted smaller for our model than for the benchmark.  This difference was significant only when the sample size $n$ was smaller than 1000, i.e. when the number of observed presences $n_p$ was smaller than 200. From the  predictive point of view, our model performed as well as  the benchmark already for sample sizes about $n=200$, i.e. for a number of observed presences at least  $n_p=40$. Also the pairwise correlation between $\beta_0$ and $\pi$, that represents an important issue as pointed by \cite{ward:al:2009}, became negligible with increasing sample sizes.\\
From the computational point of view, the procedure were carried out through a MCMC scheme with data augmentation and implemented in Fortran codes.\\
Future work will investigate the possibility of adding dependence structures among the population units  into the model as, for instance, through the use of regression functions with structured random effects. 
\newpage
\section*{Appendix}
\paragraph*{Proposition 3.}
Under the assumption that, given $Y$, the inclusion into the sample ($C=1$) is independent from the covariates $X$, it results 
\[
Pr(Y=0|C=1,x)\,Pr(C=1|x)=\frac{1-\pi^*(x)}{1+\pi^*(x)}\,\rho_0 
\]
and
\[
Pr(Y=1|C=1,x)\,Pr(C=1|x)=\frac{2\pi^*(x)}{1+\pi^*(x)}\,\rho_1. 
\]
\paragraph*{Proof.}
In general we have that
\begin{equation}
Pr(Y|C=1,x)=\frac{Pr(C=1|Y,x)\,Pr(Y|x)}{Pr(C=1|x)} 
\label{eq:cor3.3}
\end{equation}
From the conditional independence between $C=1$ and $X$ given $Y$, the \eqref{eq:cor3.3} becomes
\[
Pr(Y|C=1,x)=\frac{Pr(C=1|Y)\,Pr(Y|x)}{Pr(C=1|x)}.
\]
Recalling that $Pr(Y=1|x)=\frac{2\pi^*(x)}{1+\pi^*(x)}$ and the definitions of $\rho_0=Pr(C=1|Y=0)$ and $\rho_1=Pr(C=1|Y=1)$ the proofs for $Y=0$ and $Y=1$ can be derive by simple algebra.

\newpage
\begin{table}[H]\footnotesize
\centering
\begin{tabular}{ccccc}
\toprule
$n$ & Model & $\beta_{0}$ & $\beta_{1}$ & $\pi$\tabularnewline
\toprule
\multirow{3}{*}{50} 
 & $M_{0}$  & 0.40 (-0.31 ; 1.45) & 1.56 (1.08 ; 2.72) & 0.20 (0.18 ; 0.25)\tabularnewline
 & $M_{1}$  & 2.19  (0.68 ; 3.57) & 2.19 (1.37 ; 3.74) & 0.23 (0.18 ; 0.27)\tabularnewline
 & $M_{2}$  & 1.03 (-2.51 ; 3.35) & 2.00 (1.07 ; 3.44) & 0.19 (0.12 ; 0.26)\tabularnewline
\midrule
\multirow{3}{*}{100}
 & $M_{0}$  & 0.31 (-0.18 ; 0.88) & 1.23 (0.99 ; 1.61) & 0.21 (0.19 ; 0.25)\tabularnewline
 & $M_{1}$  & 1.24  (0.29 ; 2.36) & 1.55 (1.12 ; 2.22) & 0.23 (0.19 ; 0.26)\tabularnewline
 & $M_{2}$  & 1.22 (-0.40 ; 2.69) & 1.50 (1.07 ; 2.22) & 0.16 (0.16 ; 0.27)\tabularnewline
\midrule
\multirow{3}{*}{200}
 & $M_{0}$  & 0.11 (-0.20 ; 0.46) & 1.08 (0.95 ; 1.28) & 0.22 (0.19 ; 0.24)\tabularnewline
 & $M_{1}$  & 0.46  (0.00 ; 1.27) & 1.23 (0.99 ; 1.56) & 0.22 (0.20 ; 0.24)\tabularnewline
 & $M_{2}$  & 0.48 (-0.24 ; 1.55) & 1.23 (0.96 ; 1.59) & 0.22 (0.19 ; 0.25)\tabularnewline
\midrule
\multirow{3}{*}{500}
 & $M_{0}$  & 0.06 (-0.10 ; 0.25) & 1.02 (0.94 ; 1.12) & 0.22 (0.20 ; 0.23)\tabularnewline
 & $M_{1}$  & 0.17 (-0.09 ; 0.47) & 1.04 (0.92 ; 1.19) & 0.22 (0.20 ; 0.23)\tabularnewline
 & $M_{2}$  & 0.14 (-0.26 ; 0.61) & 1.03 (0.89 ; 1.20) & 0.22 (0.19 ; 0.24)\tabularnewline
\midrule
\multirow{3}{*}{1000}
 & $M_{0}$  & 0.04 (-0.05 ; 0.17) & 1.01 (0.95 ; 1.07) & 0.22 (0.21 ; 0.22)\tabularnewline
 & $M_{1}$  & 0.04 (-0.13 ; 0.26) & 0.99 (0.91 ; 1.08) & 0.21 (0.21 ; 0.22)\tabularnewline
 & $M_{2}$  & 0.03 (-0.22 ; 0.34) & 0.98 (0.90 ; 1.09) & 0.21 (0.20 ; 0.23)\tabularnewline
\midrule
\multirow{3}{*}{1500}
 & $M_{0}$  & 0.05 (-0.04 ; 0.15) & 0.99 (0.95 ; 1.04) & 0.21 (0.21 ; 0.22)\tabularnewline
 & $M_{1}$  & 0.01 (-0.12 ; 0.18) & 0.97 (0.91 ; 1.05) & 0.21 (0.21 ; 0.22)\tabularnewline
 & $M_{2}$  & 0.00 (-0.24 ; 0.23) & 0.97 (0.90 ; 1.05) & 0.21 (0.20 ; 0.22)\tabularnewline
\midrule
\multirow{3}{*}{2000}
 & $M_{0}$  & 0.03 (-0.04 ; 0.10) & 0.99 (0.95 ; 1.03) & 0.21 (0.21 ; 0.22)\tabularnewline
 & $M_{1}$  & 0.00 (-0.12 ; 0.14) & 0.97 (0.92 ; 1.03) & 0.21 (0.21 ; 0.22)\tabularnewline
 & $M_{2}$  &-0.02 (-0.22 ; 0.14) & 0.96 (0.90 ; 1.02) & 0.21 (0.20 ; 0.22)\tabularnewline
\midrule
\multirow{3}{*}{3000}
 & $M_{0}$  & 0.03 (-0.02 ; 0.10) & 0.98 (0.96 ; 1.02) & 0.21 (0.21 ; 0.22)\tabularnewline
 & $M_{1}$  & 0.00 (-0.10 ; 0.09) & 0.96 (0.92 ; 1.00) & 0.21 (0.21 ; 0.22)\tabularnewline
 & $M_{2}$  &-0.03 (-0.18 ; 0.11) & 0.95 (0.91 ; 1.00) & 0.21 (0.20 ; 0.22)\tabularnewline
\bottomrule
\end{tabular}
\caption{Scenario (i): point estimates of regression parameters and prevalence  computed as medians over 1000 replicates with increasing sample sizes and different models ($M_0$,$M_1$ and $M_2$). In parenthesis distributions quartiles are reported.}\label{tab:tab4}
\end{table}
\begin{table}[H]\footnotesize
\centering
\begin{tabular}{ccccc}
\toprule
$n$  & Model  & $\beta_{0}$  & $\beta_{1}$  & $\pi$\tabularnewline
\toprule
\multirow{3}{*}{50}
 & $M_{0}$  & 0.42 (-0.33 ; 1.42) & 1.34 (0.94 ; 2.13) & 0.23 (0.18 ; 0.28)\tabularnewline
 & $M_{1}$  & 2.12  (0.78 ; 3.39) & 1.95 (1.25 ; 2.96) & 0.24 (0.20 ; 0.28)\tabularnewline
 & $M_{2}$  & 1.26 (-2.99 ; 3.33) & 1.75 (0.95 ; 2.79) & 0.20 (0.13 ; 0.28)\tabularnewline
\midrule
\multirow{3}{*}{100}
 & $M_{0}$  & 0.19 (-0.20 ; 0.73) & 1.07 (0.88 ; 1.35) & 0.23 (0.19 ; 0.25)\tabularnewline
 & $M_{1}$  & 1.13  (0.36 ; 2.40) & 1.39 (1.00 ; 1.96) & 0.23 (0.21 ; 0.26)\tabularnewline
 & $M_{2}$  & 1.03 (-0.40 ; 2.65) & 1.34 (0.96 ; 1.96) & 0.22 (0.17 ; 0.28)\tabularnewline
\midrule
\multirow{3}{*}{200}
 & $M_{0}$  & 0.13 (-0.18 ; 0.45) & 0.97 (0.84 ; 1.12) & 0.23 (0.20 ; 0.25)\tabularnewline
 & $M_{1}$  & 0.48  (0.02 ; 1.17) & 1.08 (0.88 ; 1.36) & 0.23 (0.21 ; 0.25)\tabularnewline
 & $M_{2}$  & 0.48 (-0.38 ; 1.56) & 1.07 (0.83 ; 1.41) & 0.23 (0.18 ; 0.27)\tabularnewline
\midrule
\multirow{3}{*}{500}
 & $M_{0}$  & 0.09 (-0.07 ; 0.27) & 0.92 (0.85 ; 1.00) & 0.22 (0.21 ; 0.24)\tabularnewline
 & $M_{1}$  & 0.22 (-0.04 ; 0.53) & 0.95 (0.84 ; 1.09) & 0.23 (0.21 ; 0.24)\tabularnewline
 & $M_{2}$  & 0.23 (-0.24 ; 0.69) & 0.95 (0.81 ; 1.09) & 0.22 (0.20 ; 0.25)\tabularnewline
\midrule
\multirow{3}{*}{1000}
 & $M_{0}$  & 0.08 (-0.02 ; 0.20) & 0.90 (0.86 ; 0.95) & 0.22 (0.21 ; 0.23)\tabularnewline 
 & $M_{1}$  & 0.09 (-0.07 ; 0.31) & 0.90 (0.83 ; 0.99) & 0.22 (0.21 ; 0.23)\tabularnewline
 & $M_{2}$  & 0.08 (-0.19 ; 0.38) & 0.89 (0.82 ; 1.00) & 0.22 (0.21 ; 0.24)\tabularnewline
\midrule
\multirow{3}{*}{1500}
 & $M_{0}$  & 0.08  (0.00 ; 0.18) & 0.90 (0.86 ; 0.94) & 0.22 (0.22 ; 0.23)\tabularnewline
 & $M_{1}$  & 0.08 (-0.05 ; 0.23) & 0.89 (0.84 ; 0.96) & 0.22 (0.22 ; 0.23)\tabularnewline
 & $M_{2}$  & 0.05 (-0.17 ; 0.30) & 0.89 (0.82 ; 0.96) & 0.22 (0.21 ; 0.23)\tabularnewline
\midrule
\multirow{3}{*}{2000}
 & $M_{0}$  & 0.07  (0.00 ; 0.15) & 0.89 (0.86 ; 0.92) & 0.22 (0.22 ; 0.23)\tabularnewline
 & $M_{1}$  & 0.06 (-0.06 ; 0.20) & 0.89 (0.84 ; 0.95) & 0.22 (0.22 ; 0.23)\tabularnewline
 & $M_{2}$  & 0.02 (-0.17 ; 0.24) & 0.88 (0.82 ; 0.94) & 0.22 (0.21 ; 0.23)\tabularnewline
\midrule
\multirow{3}{*}{3000}
 & $M_{0}$  & 0.07  (0.01 ; 0.13) & 0.89 (0.87 ; 0.91) & 0.22 (0.22 ; 0.23)\tabularnewline
 & $M_{1}$  & 0.04 (-0.04 ; 0.15) & 0.88 (0.84 ; 0.92) & 0.22 (0.22 ; 0.23)\tabularnewline
 & $M_{2}$  & 0.02 (-0.13 ; 0.17) & 0.87 (0.83 ; 0.92) & 0.22 (0.21 ; 0.23)\tabularnewline
\bottomrule
\end{tabular}
\caption{Scenario (ii): point estimates of regression parameters and prevalence  computed as medians over 1000 replicates with increasing sample sizes and different models ($M_0$,$M_1$ and $M_2$). In parenthesis distributions quartiles are reported.}\label{tab:tab5}
\end{table}

\newpage

\begin{thebibliography}{}

\bibitem[Ara\`ujo and Williams(2000)Ara\`ujo and Williams]{ara:wil:2000}
Ara\`ujo, M. and Williams, P. (2000).
\newblock Selecting areas for species persistence using occurrence data.
\newblock {\em Biological Conservation\/}, {\bf 96}, 331--345.

\bibitem[Armenian(2009)Armenian]{armenian:2009}
Armenian, H. (2009).
\newblock {\em The Case-Control Method: Design And Applications\/}.
\newblock Oxford University Press, New York, USA.

\bibitem[Breslow(2005)Breslow]{breslow:2005}
Breslow, N.~E. (2005).
\newblock {\em Handbook of Epidemiology\/}, chapter {6: Case-Control Studies},
  pages 287--319.
\newblock Springer, New York, USA.

\bibitem[Breslow and Dey(1980)Breslow and Dey]{bre:dey:1980}
Breslow, N.~E. and Dey, N.~E. (1980).
\newblock {\em Statistical Methods In Cancer Research, Volume 1 - The analysis
  of case-control studies\/}.
\newblock WHO International Agency for Research on Cancer, Lyon, France.

\bibitem[Chakraborty {\em et~al.}(2011)Chakraborty, Gelfand, Wilson, Latimer,
  and Silander]{chakraborty:al:2011}
Chakraborty, A., Gelfand, A.~E., Wilson, A.~M., Latimer, A.~M., and Silander,
  J.~A. (2011).
\newblock Point pattern modelling for degraded presence-only data over large
  regions.
\newblock {\em Journal of the Royal Statistical Society: Series C (Applied
  Statistics)\/}, {\bf 5}, 757--776.

\bibitem[Di~Lorenzo {\em et~al.}(2011)Di~Lorenzo, Farcomeni, and
  Golini]{dilorenzo:al:2011}
Di~Lorenzo, B., Farcomeni, A., and Golini, N. (2011).
\newblock A {B}ayesian model for presence-only semicontinuous data with
  application to prediction of abundance of {T}axus {B}accata in two {I}talian
  regions.
\newblock {\em Journal of Agricultural, Biological and Environmental
  Statistics\/}, {\bf 16}(3), 339--356.

\bibitem[Divino {\em et~al.}(2011)Divino, Golini, Jona~Lasinio, and
  Pettinen]{divino:al:2011}
Divino, F., Golini, N., Jona~Lasinio, G., and Pettinen, A. (2011).
\newblock Data augmentation approach in bayesian modelling of presence-only
  data.
\newblock {\em Procedia Environmental Sciences\/}, {\bf 7}, 38--43.

\bibitem[Dorazio(2012)Dorazio]{dorazio:2012}
Dorazio, R.~M. (2012).
\newblock Predicting the geographic distribution of a species from
  presence-only data subject to detection errors.
\newblock {\em Biometrics\/}, {\bf 68}, 1303--1312.

\bibitem[Elith and Leathwick(2009)Elith and Leathwick]{eli:lea:2009}
Elith, J. and Leathwick, J.~R. (2009).
\newblock Species distribution models: ecological explanation and prediction
  across space and time.
\newblock {\em Annual Review of Ecology, Evolution and Systematics\/}, {\bf
  40}, 677--697.

\bibitem[Elith {\em et~al.}(2006)Elith, Graham, Anderson, Dudik, Ferrier,
  Guisan, Hijmans, Huettmann, Leathwick, Lehmann, Li, Lohmann, Loiselle,
  Manion, Moritz, Nakamura, Nakazawa, Overton, Peterson, Phillips, Richardson,
  Scachetti-Pereira, Schapire, Soberon, Williams, Wisz, and
  Zimmermann]{elith:al:2006}
Elith, J., Graham, C.~H., Anderson, R.~P., Dudik, M., Ferrier, S., Guisan, A.,
  Hijmans, R.~J., Huettmann, F., Leathwick, J.~R., Lehmann, A., Li, J.,
  Lohmann, L.~G., Loiselle, B.~A., Manion, G., Moritz, C., Nakamura, M.,
  Nakazawa, Y., Overton, J.~M., Peterson, A.~T., Phillips, S.~J., Richardson,
  K.~S., Scachetti-Pereira, R., Schapire, R.~E., Soberon, J., Williams, S.,
  Wisz, M.~S., and Zimmermann, N.~E. (2006).
\newblock Novel methods improve prediction of species' distribution from
  occurence data.
\newblock {\em Ecography\/}, {\bf 29}, 129--151.

\bibitem[Elith {\em et~al.}(2011)Elith, Phillips, Hastie, Dud{\'\i}k, Chee, and
  Yates]{elith:al:2011}
Elith, J., Phillips, S.~J., Hastie, T., Dud{\'\i}k, M., Chee, Y.~E., and Yates,
  C.~J. (2011).
\newblock A statistical explanation of {M}ax{E}nt for ecologists.
\newblock {\em Diversity and Distributions\/}, {\bf 17}, 43--57.

\bibitem[Fawcett(2006)Fawcett]{fawcett:2006}
Fawcett, T. (2006).
\newblock An introduction to {ROC} analysis.
\newblock {\em Pattern Recognition Letter\/}, {\bf 27}, 861--874.

\bibitem[Franklin(2010)Franklin]{franklin:2010}
Franklin, J. (2010).
\newblock {\em Mapping Species Distributions: Spatial Inference And
  Prediction\/}.
\newblock Cambridge University Press, Cambridge, UK.

\bibitem[Jaynes(1957)Jaynes]{jaynes:1957}
Jaynes, E.~T. (1957).
\newblock Information theory and statistical mechanics.
\newblock {\em The Physical Review\/}, {\bf 106}(4), 620--630.

\bibitem[Keating and Cherry(2004)Keating and Cherry]{kea:che:2004}
Keating, K.~A. and Cherry, S. (2004).
\newblock Use and interpretation of logistic regression in habitat-selection
  studies.
\newblock {\em Journal of Wildlife Management\/}, {\bf 68}, 774--789.

\bibitem[Lancaster and Imbens(1996)Lancaster and Imbens]{lanc:imb:1996}
Lancaster, T. and Imbens, G. (1996).
\newblock Case-control studies with contaminated controls.
\newblock {\em Journal of Econometrics\/}, {\bf 71}, 145--160.

\bibitem[Little and Rubin(1987)Little and Rubin]{lit:rub:1987}
Little, R. J.~A. and Rubin, D.~B. (1987).
\newblock {\em Statistical Analysis With Missing Data\/}.
\newblock John Wiley \& Sons, New York, USA.

\bibitem[Liu(2008)Liu]{liu:2008}
Liu, J.~S. (2008).
\newblock {\em Monte Carlo Strategies In Scientific Computing\/}.
\newblock Springer, New York, USA.

\bibitem[Liu and Wu(1999)Liu and Wu]{liu:wu:1999}
Liu, S.~Y. and Wu, Y.~N. (1999).
\newblock Parameter expansion for data augmentation.
\newblock {\em Journal of American Statistical Association\/}, {\bf 94},
  1264--1274.

\bibitem[Pearce and Boyce(2006)Pearce and Boyce]{pea:boy:2006}
Pearce, J.~L. and Boyce, M.~S. (2006).
\newblock Modelling distribution and abundance with presence-only data.
\newblock {\em Journal of Applied Ecology\/}, {\bf 43}, 405--412.

\bibitem[Phillips {\em et~al.}(2006)Phillips, Anderson, and
  Schapire]{philips:al:2006}
Phillips, S.~J., Anderson, R.~P., and Schapire, R.~E. (2006).
\newblock Maximum entropy modeling of species geographic distributions.
\newblock {\em Ecological Modelling\/}, {\bf 190}, 231--259.

\bibitem[Robert and Casella(2004)Robert and Casella]{rob:cas:2004}
Robert, C.~P. and Casella, G. (2004).
\newblock {\em Monte Carlo Statistical Metods\/}.
\newblock Springer, New York, USA.

\bibitem[Rubin(1976)Rubin]{rubin:1976}
Rubin, D.~B. (1976).
\newblock Inference and missing data.
\newblock {\em Biometrika\/}, {\bf 63}(3), 581--592.

\bibitem[S\"arndal(1978)S\"arndal]{sarndal:1978}
S\"arndal, C.~E. (1978).
\newblock Design-based and model-based inference in survey sampling.
\newblock {\em Scandinavian Journal of Statistics\/}, {\bf 5}, 27--52.

\bibitem[Tanner(1996)Tanner]{tanner:1996}
Tanner, M. (1996).
\newblock {\em Tools for Statistical Inference: Observed Data And Data
  Augmentation\/}.
\newblock Springer, New York, USA.

\bibitem[Tanner and Wong(1987)Tanner and Wong]{tan:won:1987}
Tanner, M. and Wong, W. (1987).
\newblock The calculation of posterior distribution by data augmentation.
\newblock {\em Journal of American Statistical Association\/}, {\bf 82},
  528--550.

\bibitem[Ward {\em et~al.}(2009)Ward, Hastie, Barry, Elith, and
  Leathwick]{ward:al:2009}
Ward, G., Hastie, T., Barry, S., Elith, J., and Leathwick, A. (2009).
\newblock Presence-only data and the {EM} algorithm.
\newblock {\em Biometrics\/}, {\bf 65}, 554--563.

\bibitem[Warton and Shepherd(2010)Warton and Shepherd]{war:she:2010}
Warton, D.~I. and Shepherd, L. (2010).
\newblock Poisson point porcess models solve the ``pseudo-absence problem'' for
  presence-only data in ecology.
\newblock {\em Annals of Applied Statistics\/}, {\bf 4}(3), 1383--1402.

\bibitem[Woodward(2005)Woodward]{woodward:2005}
Woodward, M. (2005).
\newblock {\em Epidemiology: Study Design And Data Analysis\/}.
\newblock Chapman \& Hall, New York, USA.

\bibitem[Zaniewski {\em et~al.}(2002)Zaniewski, Lehmann, and
  Overton]{zaniewski:al:2002}
Zaniewski, A.~E., Lehmann, A., and Overton, J.~M. (2002).
\newblock Prediction species spatial distributions using presence-only data: a
  case study of native {N}ew {Z}eland ferns.
\newblock {\em Ecological Modelling\/}, {\bf 157}, 261--280.

\end{thebibliography}
%

\end{document}